\documentclass[12pt,a4paper]{book}

\usepackage{amsmath}

\usepackage{amsfonts}

\usepackage{amssymb}

\usepackage{graphicx}

\usepackage{listings}
\newcommand{\code}[2][C++]{\lstset{language=#1, basicstyle=\sffamily}\lstinline{#2}}
\lstnewenvironment{codeblock}[1][C++]
    {\lstset{basicstyle=\small \sffamily}}
    {}
    
\usepackage{commands} 

\begin{document}    

\frontmatter
    \begin{titlepage}
	\title{Numerical Experimentation within \grwb}
	\author{
	\textbf{Andrew Moylan} \\
	\\
	14 November, 2003 \\
	\\
	\\
	\includegraphics[width = 5 cm]{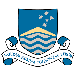} \\
	\\
	\small A thesis submitted in partial fulfillment of the requirements for \\
	\small the degree of Bachelor of Science with Honours in Theoretical Physics \\
	\small at The Australian National University \\
	\\
	\textbf{Supervisor} \\
	Dr Susan Scott \\
	\\
	\textbf{Advisor} \\
	Antony Searle \\
	}
	\date{}
	\maketitle	
\end{titlepage} \clearemptydoublepage
    \chapter{Acknowledgements}

    \begin{quote}
        ``I see that your wisdom has outstripped my own \ldots \\
        You must be killed!'' \\
        \hspace*{24 pt} \emph{-Antony `Surly' Searle}
    \end{quote}

    Thanks are due to my supervisor, Dr Susan Scott, and my advisor, Antony Searle (whose wisdom I have not outstripped), for their excellent guidance and advice, and for making this year very interesting and enjoyable.  Thanks also to Ben Cusack and Ingrid Irmer and my fellow Honours students for many interesting discussions.

    Thanks to my father Ron for additional help in proofreading this thesis, and to Katie and my mother Jenny for their loving support throughout the year. \clearemptydoublepage
    \chapter{Declaration}
    
    This thesis is an account of research undertaken between February 2003 and November 2003 at The Department of Physics, Faculty of Science, The Australian National University, Canberra, Australia.  Except where acknowledged, the material presented is, to the best of my knowledge, original, and has not been submitted for a degree at any university. \\
    \\
    \\
    \\
    \noindent Andrew Moylan \\
    14 November, 2003
\clearemptydoublepage
    \chapter{Abstract}

    The software tool \grwb\ is an ongoing project in visual, numerical General Relativity at The Australian National University.  This year, \grwb\ has been significantly extended to facilitate numerical experimentation.  The numerical differential geometric engine has been rewritten using functional programming techniques, enabling fundamental concepts to be directly represented as variables in the C++ code of \grwb.  Sophisticated general numerical methods have replaced simpler specialised algorithms.  Various tools for numerical experimentation have been implemented, allowing for the simulation of complex physical situations.

	A recent claim, that the mass of the Milky Way can be measured using a small interferometer located on the surface of the Earth, has been investigated, and found to be an artifact of the approximations employed in the analysis.  This difficulty is symptomatic of the limitations of traditional pen-and-paper analysis in General Relativity, which was the motivation behind the original development of \grwb.  The physical situation pertaining to the claim has been modelled in a numerical experiment in \grwb, without the necessity of making any simplifying assumptions, and an accurate estimate of the effect has been obtained. \clearemptydoublepage
    \tableofcontents\clearemptydoublepage

\mainmatter
    \chapter{Introduction}
    
    \grwb\ is a numerical, visual tool for exploring analytic space-times in General Relativity.  This year, the numerical differential geometric engine of \grwb\ has been rewritten using functional programming techniques, with the objective of creating a general platform in which complex physical situations can be simulated in \emph{numerical experiments}.  New tools for modelling physical systems were implemented within the functional framework.  A recently proposed experiment, to determine the mass of the Milky Way, was analysed, and then investigated numerically in \grwb, demonstrating the applicability of the new techniques for numerical experimentation.

    \section{Summary of thesis}

        \grwb\ arose from work in visual numerical relativity by S.\,M. Scott, B.\,J.\,K. Evans, and A.\,C. Searle, at The Australian National University.  Most recently, A.\,C. Searle implemented a numerical differential geometric engine, and improved 3-D visualisation \cite{R:searle}.  The efficacy of the differential geometric engine, and the utility of \grwb\ as an intuitive visualisation tool, has been demonstrated \cite{R:grwb-grossman, R:grwb-grg}.  Chapter~\ref{C:grwb} presents an overview of the \grwb\ project.

        In order to facilitate the creation of a general system for numerical experimentation in analytic space-times, the numerical and differential geometric aspects of \grwb\ have, this year, been rewritten using functional programming techniques.  Functional programming allows functions, like normal data, to be stored in program variables and manipulated by other functions.  Important concepts in differential geometry, which are naturally thought of as functions, can thus be directly represented in the C++ code of \grwb.  The functional programming methods employed in \grwb\ are introduced in Chapter~\ref{C:functional}.

        Some of the numerical methods previously employed by \grwb\ were found to be too inflexible or inaccurate to be applied to potentially complex and computationally intensive numerical experiments.  Sophisticated new algorithms have been implemented this year for key numerical operations including differentiation, integration, and minimisation; these operations act directly on functions, using the new functional framework of \grwb.  A general notion of approximate equality permits the numerical methods to be implemented in a consistent and elegant way.  Numerical methods are the topic of Chapter~\ref{C:numerical}.
    
        Appendix~\ref{A:grwb-code} lists the C++ code, written by the author, for the new numerical algorithms discussed in Chapter~\ref{C:numerical}.

        The differential geometric engine of \grwb, which relies on numerical methods for operations such as the transformation of tangent vector components between coordinate systems, has been rewritten within the functional framework, to interact cleanly with the numerical engine of \grwb.  Abstract notions such as points and tangent vectors are represented by C++ classes, which provide routines to obtain the coordinates of the objects in any coordinate system.  The functional numerical differential geometric engine is described in Chapter~\ref{C:differential-geometry}.
    
        Physical situations in numerical experiments are modelled in terms of important objects in differential geometry, particularly points, tangent vectors, and geodesics.  The key operation of geodesic tracing from initial data has been re-implemented using the new functional numerical engine. New methods for locating geodesics that are implicitly defined by boundary conditions have been developed using the function minimisation algorithms.  These tools facilitating numerical experimentation in \grwb\ are the topic of Chapter~\ref{C:numerical-experiments}.  Appendix~\ref{A:karim-code} lists the C++ code, written by the author, for a numerical experiment described in Chapter~\ref{C:karim-grwb}.

        An analysis of a recent claim by Karim \etal\ \cite{R:karim}, that the mass of the Milky Way can be determined using a small interferometer located on the surface of the Earth, is presented in Chapter~\ref{C:karim}.  Properties of the interferometer model employed in the calculation of Karim \etal\ are investigated.  The claimed size of the effect is found to be due to the coordinate-dependent definition of the interferometer employed, and not to the effects of space-time curvature.  A more physically motivated interferometer model (`geodesic-defined interferometer') is proposed, and its properties are investigated.

        The interferometer model of Karim \etal\ and the geodesic-defined interferometer were each simulated in \grwb.  The results of these numerical experiments are presented in Chapter~\ref{C:karim-grwb}.  The analysis by Karim \etal\ of their proposed interferometer was found to be in agreement with the results of the \grwb\ simulations of that interferometer.  The behaviour of the geodesic-defined interferometer was characterised, and used to obtain a new, more accurate, estimate on the size of the effect described in \cite{R:karim}.  The effect was found to be too small to detect with an interferometer on Earth.  We conclude that the proposed experiment, to measure the mass of the Milky Way using an interferometer located on Earth, is not currently technically feasible. \clearemptydoublepage
    \chapter{\grwb}
    \label{C:grwb}
    
    \grwb\ is a software tool for visualising numerical operations on analytically defined space-times in General Relativity.  It has arisen from work in visual numerical relativity by S.\,M. Scott, B.\,J.\,K. Evans, and, most recently, A.\,C. Searle.  In this chapter we give an overview of the motivation behind, and history of, the \grwb\ project.

    \section{Motivation}
    
        Analytic results in General Relativity are, in general, difficult to obtain.  Exact solutions of the Einstein field equation are rare, and some physically important exact solutions are sufficiently complicated to be difficult to work with algebraically.  It is usually necessary to make approximations if algebraic results are desired; this is exemplified by the claim analysed in Chapter~\ref{C:karim}.

        Computational methods have been applied to the solution of the Einstein field equation for various boundary conditions, most famously to the currently unsolved problem of two in-spiralling black holes.  Symbolic algebra software such as \mathematica, as well as specialised packages, such as \emph{GRTensorII} and \emph{Sheep}, are used to manipulate the tensor equations of General Relativity.

        Computational methods have also been used to explore the physical properties of analytic solutions to the Einstein equation, through numerical operations such as geodesic tracing.  Traditionally, such simulations were performed using specialised codes as required.

        Visualisation in General Relativity is intrinsically difficult because space-times are 4-dimensional and curved, whereas computer monitors (and most other visualisation devices) are 2-dimensional and flat.  Traditionally, visualisation is performed by choosing a coordinate system, suppressing 1 coordinate, and plotting the remaining three coordinates via a projection from 3 dimensions to 2 dimensions.

        The goal of the \grwb\ project is to create a visual software tool for numerical General Relativity, in which a point-and-click interface encourages the user to explore freely in a space-time.  Such a tool would, for the first time, allow experimental techniques to be applied to problems in General Relativity in an intuitive, visual environment.

    \section{\grwb}
    
        Working with S.\,M. Scott and B.\,J.\,K. Evans, A.\,C. Searle implemented a new version of \grwb\ in 1999 \cite{R:searle}.  It featured an imbedded platform-independent \textsc{gui} (Graphical User Interface), a novel numerical differential geometric engine, and a flexible visualisation system, as well as being easy to extend with additional space-time definitions.
       
        The differential geometric engine of \grwb\ allowed for abstract objects, such as points and tangent vectors, to have multiple numerical representations, corresponding to different coordinate charts.  \grwb\ was informed, through the space-time definitions, of the maps between the various charts.  Numerical operations, such as geodesic tracing, are performed on a single chart, until a chart boundary or other obstacle is encountered, at which point the algorithms are able to transform the data into another coordinate system and resume computation there.

        The components of the metric tensor on each coordinate chart, together with the maps between charts, define a space-time in \grwb.  For numerical operations which involve derivatives of the metric components, such as geodesic tracing, simple, robust numerical methods are employed to compute the derivatives.

        A highly general visualisation system was implemented in \grwb.  In the coordinate system of choice, space-times are visualised by transforming the 4 coordinates under arbitrary \emph{distortions} down to a 3-dimensional visualisation space, which is then rendered on the screen using the OpenGL graphics library.  Higher-dimensional structures (surfaces, volumes, hyper-volumes), such as the event horizon of a black hole, are also intelligently visualised under arbitrary distortions.
           
        \begin{figure}
            \begin{center}
                \includegraphics[width = 12cm]{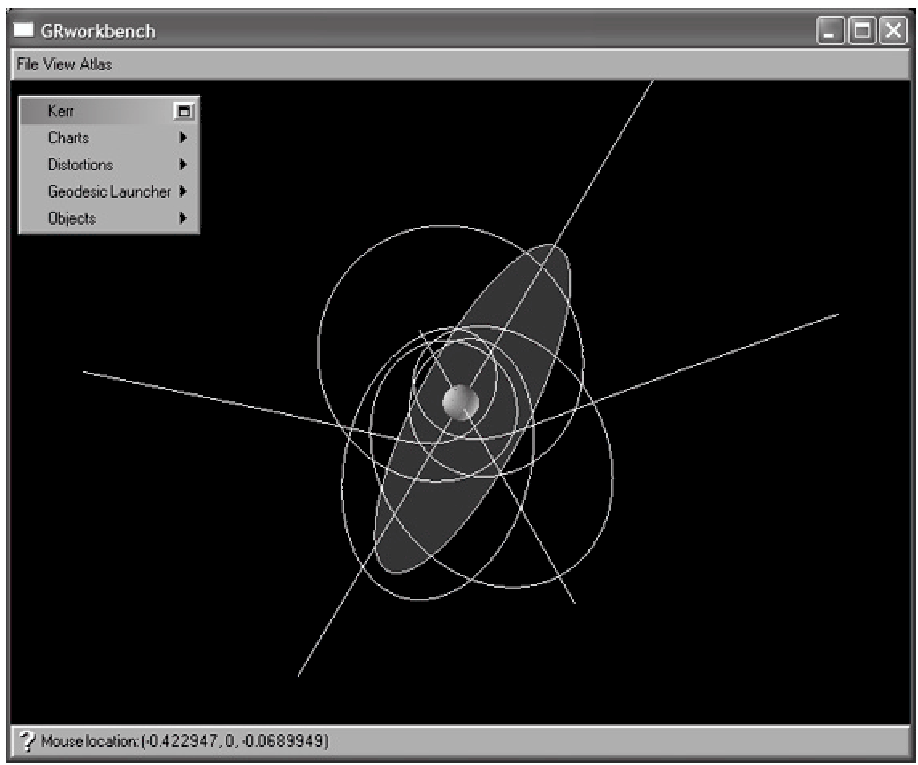}
            \end{center}
            \caption{\grwb\ screen-shot showing an interesting time-like geodesic in the Kerr rotating black hole space-time.}
            \label{F:grwb-geodesic}
        \end{figure}
        
        Figure~\ref{F:grwb-geodesic} is a screen-shot from \grwb\ showing a time-like geodesic in the Kerr space-time, which describes the gravitational field around a rotating black hole.  The geodesic represents the world-line of a particle falling into the near-field of the black hole, orbiting the event horizon several times, and then escaping in a different direction.  The spherical object in the centre of Figure~\ref{F:grwb-geodesic} is the event horizon of the black hole.  Elements of the \textsc{gui} are visible in the top-left corner.

        The interesting geodesic of Figure~\ref{F:grwb-geodesic} was obtained in a just a few minutes using the fast turn-around of real-time geodesic tracing and visualisation.  Other physically interesting situations can be explored visually in a similar way.  \grwb\ enables users to quickly get an intuitive `feel' for the properties of a space-time, and is thus also potentially useful as an educational tool.

    \section{Objective}
    
        Simple visual experiments have been performed in \grwb, demonstrating its utility.  However, the simulation of more complex physical situations was hindered by the numerical methods, which were not as efficient or flexible as they could be, and the differential geometric engine, which was not sufficiently general for rapid extension.  The modification of \grwb, with the aim of performing complex numerical experiments, is the topic of this thesis. \clearemptydoublepage
    \chapter{Functional programming}
    \label{C:functional}

    The numerical and differential geometric engine of \grwb\ has been rewritten during 2003 within the framework of functional programming.  An overview of C++ and functional programming is presented in this chapter.  The benefits for \grwb\ are discussed in Section~\ref{S:functional-benefits}.  Numerical methods and differential geometry within this functional framework are the topics of Chapters~\ref{C:numerical} and \ref{C:differential-geometry}, respectively.

    \section{Functions}
        \label{S:functional-functions}

        In the traditional programming languages of scientific computing, such as C, C++, and Fortran, a program typically consists of \emph{routines} which operate on data stored in program \emph{variables}.  Every variable in C++ has a \emph{type}, and there is a natural correspondence between C++ types and standard mathematical sets.  Table~\ref{T:functional-correspondence} lists the most important examples.
        
        \begin{table}
            \begin{center}
                \begin{tabular}{| c | c | l |}
                    \hline
                    Set & C++ type & Notes \\
                    \hline                        
                    $\mathbb{Z}$ & \code{int} & max.\ $\pm (2^{31} - 1)$ \\
                    $\mathbb{R}$ & \code{double} & max.\ $\sim \pm 10^{308}$, precision $\sim 10^{-15}$ \\
                    $\mathbb{R}^n$ & \code{nvector<double>} & (as for \code{double}) \\                    
                    $(A \to B)$ & \code{function<B (A)>} & see Section~\ref{SS:functional-data} \\
                    \hline
                \end{tabular}
                \caption{Correspondence between certain sets and C++ types in \grwb.}
                \label{T:functional-correspondence}
            \end{center}
        \end{table}
    
        The first two sets in Table~\ref{T:functional-correspondence}, $\mathbb{Z}$ and $\mathbb{R}$, are represented in some way or other in every language of scientific computing.  The type name \code{double} stands for `double-precision floating point number'.
    
        The \code{nvector<T>} type, written by Antony Searle, uses the C++ \emph{template} mechanism\footnote{See \cite{R:stroustrup}, page 327.} to provide a type representing $n$-tuples of any other type \code{T}.  The type \code{T} is called the \emph{template parameter}.  In the case of $\mathbb{R}^n$, \code{T} will be \code{double}.  The template parameter may itself be an \code{nvector}, as in \code{nvector<nvector<double>>}, which is a type representing the set of $m \times n$ matrices with real entries.
    
        The following is a routine in C++:
        \begin{codeblock}
double mean(double a, double b)
{
    return (a + b) / 2;
}
        \end{codeblock}
        The corresponding mathematical definition is
        \begin{align}
            \label{E:functional-mean}
            &\text{mean} \colon \mathbb{R} \times \mathbb{R} \to \mathbb{R}, \notag \\
            &\text{mean} (a, b) = \frac{a + b}{2}.
        \end{align}
        The first line of the routine conveys the same information as the first line of \eqref{E:functional-mean}: the routine \code{mean} takes two real numbers as arguments, and returns a real number.  The rest of the routine definition, enclosed in braces, encodes the second line of \eqref{E:functional-mean}.
    
        The \emph{signature} of a routine is obtained by taking the first line of a routine and removing the routine name and argument names, leaving only their types.  Thus the signature of the routine \code{mean} is \code{double (double, double)}, and the signature of a function $f \colon \mathbb{R}^n \times \mathbb{Z} \to \mathbb{R}$ would be \code{double (nvector<double>, int)}.
    
        We may define a \emph{function} as anything which behaves like the routine \code{mean} above, in the sense that it accepts zero or more arguments, and returns a value.  In traditional programming languages (C, Fortran) the only possible functions are routines, and so the terms `function' and `routine' are used interchangeably.  The key feature of functional programming is that there can be functions other than the routines typed in by the programmer---functions created while the program is running.  The mechanism to achieve this is introduced in Section~\ref{SS:functional-run-time}.  To create functions at run-time, we need to be able to store them in variables, which is the topic of the next section.

    \section{Functions as data}
        \label{SS:functional-data}          

        The capability to store functions in variables is not unique to functional programming.  Most languages used for scientific computation have some way to store a reference to a program routine; \grwb\ uses the \emph{Boost Function Library} \cite{R:boost-function}.  The Function Library provides the templatised type \code{function<T>} representing a function whose signature is \code{T}.  The following code fragment shows how the routine \code{mean} can thus be stored in a variable:\footnote{Anything after the characters \code{//} in a line of C++ code is a \emph{comment}, and is ignored by the compiler.}
        \begin{codeblock}
function<double (double, double)> f = mean;
// the following two lines are now equivalent
double x = f(1, 2);
double x = mean(1, 2);
        \end{codeblock}
        Observe from the last two lines that the variable \code{f} can be used just like the routine \code{mean}; they are both functions.

        In general, if we let $(A_1 \times \cdots \times A_n \to B)$ denote the set of functions from $A_1 \times \cdots \times A_n$ to $B$, then the corresponding C++ type is \code{function<B (A1, . . . , An)>}, where the sets $B, A_1, \ldots, A_n$ correspond to the types \code{B}, \code{A1}, \ldots, \code{An}.  The fourth row of Table~\ref{T:functional-correspondence} summarises this relationship.

        The most important consequence of the capability to store functions in variables is that functions can be arguments to other functions.  To illustrate this, consider the following routine, which approximates the derivative of a function $f$ at a point $x$:\footnote{This crude method for estimating the derivative is for illustrative purposes only; the differentiation algorithm employed by \grwb\ is described in Section~\ref{S:numerical-differentiation}.}
        \begin{codeblock}
double slope(function<double (double)> f, double x)
{
    double h = 0.1;
    return (f(x + h) - f(x - h)) / (2 * h);
}
        \end{codeblock}
        The corresponding mathematical definition is
        \begin{align}
            \label{E:functional-slope}
            &\text{slope} \colon (\mathbb{R} \to \mathbb{R}) \times \mathbb{R} \to \mathbb{R}, \notag \\
            &\text{slope} (f, x) = \frac{f (x + h) - f (x - h)}{2 h}, \quad h = 0.1.                
        \end{align}
        Again the first line of the routine definition encodes the same information as the first line of \eqref{E:functional-slope}, and the remainder of the routine definition, enclosed in braces, encodes the second line of \eqref{E:functional-slope}.

        In addition to differentiation, many other numerical algorithms naturally take a function as an argument.  Two classic examples are
        \begin{align}
            &\text{minimise} \colon (\mathbb{R} \to \mathbb{R}) \times \mathbb{R} \to \mathbb{R}, \notag \\
            &\text{minimise} (f, x) = (\text{a local minimum of $f$ near $x$}),
        \end{align}
        and
        \begin{align}
            &\text{integrate} \colon (\mathbb{R} \to \mathbb{R}) \times \mathbb{R} \times \mathbb{R} \to \mathbb{R}, \notag \\
            &\text{integrate} (f, a, b) = (\text{numerical estimate of $\int_a^b f(x) \, dx$}).
        \end{align}

        Finally, note that the signature of \code{slope} is \code{double (function<double (double)>, double)}, and so \code{slope} itself may be stored in a variable of type \code{function<double (function<double (double)>, double)>}.  Every function in C++ can be stored in a variable of type \code{function<T>}, where \code{T} is the signature of the function.
        
    \section{Creating functions at run-time}
        \label{SS:functional-run-time}
        
        Consider the following function, defined in terms of the $\text{slope}$ function \eqref{E:functional-slope}:
        \begin{align}
            \label{E:functional-derivative}
            &\text{derivative} \colon (\mathbb{R} \to \mathbb{R}) \to (\mathbb{R} \to \mathbb{R}), \notag \\
            &\text{derivative} (f) = g, \quad g \colon \mathbb{R} \to \mathbb{R}, \quad g (x) = \text{slope} (f, x).
        \end{align}
        For any function $f$, it returns \emph{the function which returns the slope of $f$ at its argument}.
        
        This expression of the operation of numerical differentiation as a mapping from functions to functions is more flexible than \code{slope}.  By recursively applying $\text{derivative}$, for example, we have $\text{derivative} (\text{derivative} (f))$, which is an approximation to the second derivative of $f$.  Using only the mechanisms introduced so far, however, we cannot encode \eqref{E:functional-derivative} in C++.

        \subsection{Functors}
            \label{S:functional-functors}
        
            New types are created in C++ by writing a \emph{class}.  A class may optionally define an \code{operator()}\footnote{(pronounced `operator parenthesis' or `the parenthesis operator')} routine, in which case it is called a \emph{functor class}.\footnote{The use of the term `functor' in category theory is not related to its use in this context.}  A variable whose type is a functor class is a function as defined in Section~\ref{S:functional-functions}.  To see this, consider the following functor class:
            \begin{codeblock}
class multiply_functor
{
public:
    // constructor (see the discussion, below)
    multiply_functor(double a_)
    {
        a = a_;
    }
    
    double operator()(double x)
    {
        return a * x;
    }

private:
    double a;
};
            \end{codeblock}
            It can be used in the following way:
            \begin{codeblock}
function<double (double)> f = multiply_functor(1.5);
double y = f(3);
            \end{codeblock}
            This code fragment sets \code{f} to the function which returns 1.5 times its argument, and thus it sets \code{y} to 4.5.
            
            A functor class represents the function encoded by its \code{operator()} routine, parameterised by the variables in its \code{private:} section.  The variables in the \code{private:} section are initialised by the \emph{constructor}, which always has the same name as the functor class.  In the code fragment, above, the line \code{a = a_;} initialises the \code{private:} variable \code{a} with the value of the variable \code{a_}, which was passed to the constructor of \code{multiply_functor}.

            Thus, \code{multiply_functor} represents the class of functions which multiply their argument by some constant $a \in \mathbb{R}$; the value of the parameter $a$ is the argument to the constructor.  We may even think of the constructor itself as a function:
            \begin{align}
                &\text{multiply\_functor} \colon \mathbb{R} \to (\mathbb{R} \to \mathbb{R}), \notag \\
                &\text{multiply\_functor} (a) = f, \quad f \colon \mathbb{R} \to \mathbb{R}, \quad f (x) = a x.
            \end{align}
            
            Using a functor class we can encode the $\text{derivative}$ function \eqref{E:functional-derivative} in C++:
            \begin{codeblock}
class derivative_functor
{
public:
    derivative_functor(function<double (double)> f_)
    {
        f = f_;
    }

    double operator()(double x)
    {
        return slope(f, x);
    }

private:
    function<double (double)> f;
};

function<double (double)> derivative(function<double (double)> f)
{
    return derivative_functor(f);
}
            \end{codeblock}                    
            If we were to replace the primitive \code{slope} routine with a more sophisticated algorithm for numerical differentiation, then this \code{derivative} routine would be a good approximation to the mathematical operation of differentiation.  For example, \code{derivative(sin)} would be a good approximation to the function \code{cos}.\footnote{The functions \code{sin} and \code{cos}, and many other standard functions, are built-in to C++.}

    \section{Applicability to \grwb}
        \label{S:functional-benefits}
        
        There are two reasons why functional programming is an ideal framework in which to implement the numerical and differential geometric aspects of \grwb.  Functional programming permits numerical operations, like \code{derivative}, to be expressed in a way which closely resembles the mathematical operation that they approximate; and many fundamental notions in differential geometry and general relativity, such as the action of the metric tensor, and particle world-lines, are functions.

        By elevating functions to the same level as traditional data types ($\mathbb{Z}$, $\mathbb{R}$), functional programming makes these notions directly representable as variables in C++ code.  As we shall see in Chapter~\ref{C:numerical-experiments}, this is invaluable in the construction of numerical experiments.

        \clearemptydoublepage
    \chapter{Numerical methods}
    \label{C:numerical} 
    
    The numerical engine of \grwb\ has been rewritten during 2003 within the framework of functional programming.  Functional algorithms have replaced third-party routines and inline implementations of simpler methods.  Some algorithms needed to be rewritten or added as part of the development of \grwb\ for numerical experiments, as described in Chapter~\ref{C:numerical-experiments}, while other changes were directed towards increasing robustness, accuracy, or speed of computation.

    A technique for scale-independent computation is described in Section~\ref{S:numerical-relative-difference}, and the method of Richardson extrapolation is introduced in Section~\ref{S:numerical-richardson}.  These tools are employed in new implementations for the operations of differentiation, integration of ordinary differential equations, and function minimisation, which are described in Sections~\ref{S:numerical-differentiation}, \ref{S:numerical-ode} and \ref{S:numerical-minimisation}, respectively.

    \section{Scale-independent computation}
        
        As mentioned in Section~\ref{S:functional-functions}, the name of the type \code{double}, which represents real numbers in \grwb, stands for `double-precision floating point number'.  The term `double-precision' arises from the size of the data type, 64 bits, being twice that of the smallest floating point data type in C++, which is called \code{float} and referred to as `single-precision'.  The term `floating point' refers to the particular way that numbers are encoded in the 64 bits.

        Floating point numbers are represented in \emph{mantissa-exponent form}, which is similar to standard scientific notation.  For example, the number $1.234 \times 10^{-56}$ is represented as a \code{double} by \code{1.234e-56}, where \code{1.234} is the \emph{mantissa}, which can contain up to 15 significant figures, and \code{-56} is the \emph{exponent}, which ranges from $-308$ to $+308$.\footnote{The mantissa is stored in 52 bits, so its precision is one part in $2^{52} \simeq 4.5 \times 10^{15}$.  The exponent is stored in 11 bits, so binary exponents up to $\pm 2^{10} = \pm 1024$ can be represented, corresponding to decimal exponents of $\pm \log_{10} 2^{1024} \simeq \pm 308$.}  These limitations of the \code{double} data type were summarised in Table~\ref{T:functional-correspondence}.

        The alternative to mantissa-exponent form is \emph{fixed-point form}, in which a certain number of bits (32 bits, say) store the part of the number to the left of the decimal point, and the remaining bits (31 bits, say) store the part of the number to the right of the decimal point, with 1 bit reserved to indicate the sign ($+$ or $-$) of the number.  In this form, the largest representable number is $\sim 2^{32}$, and the smallest (in magnitude) representable number is $\sim 2^{-31}$, so the example above, $1.234 \times 10^{-56}$, is not representable at all.  Mantissa-exponent form, offering a wider range of length scales, and the same precision at all length scales, is more suitable than fixed-point form for scientific computation.

        \subsection{Approximate equality}
            \label{S:numerical-relative-difference}

            In approximate methods, it is necessary to have a notion of two numbers being \emph{approximately equal}, to some relative precision $\epsilon$.  For example, suppose $\epsilon = 0.01$; then we want to consider $1.001 \times 10^{43}$ to be approximately equal to $1.002 \times 10^{43}$, because their difference, $10^{40}$, divided by either of their magnitudes, $\sim 10^{43}$, is $\sim 10^{-3} < \epsilon$.  On the other hand, we also want to consider $0$ to be approximately equal to $10^{-4}$, simply because $10^{-4} < \epsilon$.  We require a definition of approximate equality which satisfies both of these examples.

            A notion of approximate equality is also required for elements of other sets, most importantly $\mathbb{R}^n$, where there is an additional consideration.  Consider two vectors $\mathbf{v}_1, \mathbf{v}_2 \in \mathbb{R}^2$,
            \begin{equation}
                \label{E:numerical-vector-difference}
                \mathbf{v}_1 = \columnvector{10^5 \\ 1}, \quad \mathbf{v}_2 = \columnvector{10^5 \\ 2}.
            \end{equation}
            Denoting the standard Euclidean norm on $\mathbb{R}^2$ by $\| \cdot \|$, we have that $\| \mathbf{v}_1 \| \simeq \| \mathbf{v}_2 \| \gg 1$, $\| \mathbf{v}_2 - \mathbf{v}_1 \| = 1$, and
            \begin{equation}
                \label{E:numerical-vector-difference-problem}
                \frac{\| \mathbf{v}_2 - \mathbf{v}_1 \|}{\| \mathbf{v}_1 \|} < \epsilon.
            \end{equation}
            However, we may not want to consider the vectors $\mathbf{v}_1$ and $\mathbf{v}_2$ to be approximately equal, because their second components are not approximately equal, and the scale of interest of the first component may be different to that of the second component.

            In the literature, it is common for numerical algorithms to assume that the scale of interest is approximately unity, or at least that it is uniform for all components of a vector or matrix; for such algorithms, it is necessary to appropriately normalise input variables, and then apply the inverse transformation to the output of the algorithm.  Definitions like \code{double tiny = 1.0e-30;} are also common, where the variable \code{tiny} is intended to be smaller than any quantity that might otherwise arise, apart from zero.  Such a definition invalidates the routine for scales smaller than $10^{-30}$, which partially nullifies one of the main benefits of floating point arithmetic.  Whenever either of the two issues above was encountered while implementing the numerical methods of this chapter, it was found that, by rethinking the relevant parts of the algorithm in terms of a general notion of approximate equality, the problem could be avoided.

            In the redesigned numerical engine of \grwb, the notion of approximate equality for any set $S$ is represented by the function
            \begin{align}
                &\text{approx\_equal} \colon S \times S \times \mathbb{R} \to \{ \text{true}, \text{false} \}, \notag \\
                &\text{approx\_equal} (a, b, \epsilon) =
                    \begin{cases}
                        \text{true}, &\text{if $\text{relative\_difference} (a, b) < \epsilon$;} \\
                        \text{false}, &\text{otherwise,}
                    \end{cases}
            \end{align}
            where the function relative\_difference encodes, for each set $S$, a method to determine to what precision two given elements are equal.  The range of \text{approx\_equal}, $\{ \text{true}, \text{false} \}$, is represented by the type \code{bool} in C++.

            The default definition,\footnote{The C++ template mechanism allows for routines which have no particular type specified for one or more of their arguments; such a routine may be called with arguments of \emph{any} type for which the routine body makes sense.} for any set $S$ which has a norm\footnote{The norm on $\mathbb{R}$ is represented by the function \code{abs}, which is built-in to C++.  In \grwb\ the norm is defined for other types by specialising (\emph{overloading}) \code{abs} to take arguments of other types.} $\| \cdot \|$ and is closed under an addition operation, is
            \begin{align}
                \label{E:numerical-default-relative-difference}
                &\text{relative\_difference} \colon S \times S \to \mathbb{R}, \notag \\
                &\text{relative\_difference} (a, b) = \frac{\| a - b \|}{\max (\sqrt{\| a \| \| b \|}, 1)}.
            \end{align}
            Thus, the relative difference is the absolute difference divided by the geometric mean of the absolute values, unless the geometric mean is less than unity, in which case the relative difference is just the absolute difference.  Definition \eqref{E:numerical-default-relative-difference} is not the only conceivable default definition for \text{relative\_difference} that is suitable for $\mathbb{R}$ and that is easily generalisable to other sets with norms; but it is the definition employed in \grwb.  The code of the \code{relative_difference} routine is listed in Section~\ref{S:grwb-code-relative-difference}.

            The \text{relative\_difference} function is specialised for the case $S = \mathbb{R}^n$, to resolve the problem exemplified by \eqref{E:numerical-vector-difference-problem}:
            \begin{align}
                \label{E:numerical-relative-difference-r-n}
                &\text{relative\_difference} \colon \mathbb{R}^n \times \mathbb{R}^n \to \mathbb{R}, \notag \\
                &\text{relative\_difference} (\mathbf{a}, \mathbf{b}) = \sqrt{\sum_{i = 1}^n \text{relative\_difference} (a_i, b_i)^2},
            \end{align}
            where $\mathbf{a} = (a_1, \ldots, a_n)$ and $\mathbf{b} = (b_1, \ldots, b_n)$.  Thus, the square of the relative difference is the sum of the squares of the relative differences of the components.
    
            The specialisation of the \code{relative_difference} routine in \grwb\ has signature \code{double (nvector<T>, nvector<T>)}, where \code{T} is a template parameter.  As such, the componentwise definition \eqref{E:numerical-relative-difference-r-n} applies to $n$-tuples of any set.  In particular, recalling that matrices are represented by the type \code{nvector<nvector<double>>}, by recursively applying \eqref{E:numerical-relative-difference-r-n} we find that the square of the relative difference of two matrices is just the sum of the squares of the relative differences of their components, independent of their representation as vectors of vectors.

            More general than the notion of relative difference, as defined in \eqref{E:numerical-default-relative-difference} and \eqref{E:numerical-relative-difference-r-n}, is to associate with each set $S$ and norm $\| \cdot \|$ on $S$ not just a C++ type \code{S}, representing $S$, but also a function of signature \code{double (S)}, representing the norm $\| \cdot \|$.  The particular norm on $S$ will depend on what the elements of $S$ are being used to represent; multiple norms on $\mathbb{R}^n$, for example, could facilitate the correct definition of approximate equality for the two vectors in \eqref{E:numerical-vector-difference}, which will depend on the particular meaning of the various components of the vectors.

    \section{Evaluation of limits}
        \label{S:numerical-richardson}

        Two of the numerical methods presented in this chapter (that for differentiation and that for integration of ordinary differential equations) involve an algorithm $f (h)$ which approximates the desired solution as a function of a small parameter $h \in \mathbb{R}$, such that
        \begin{equation}
            \label{E:numerical-richardson-limit}
            \lim_{h \to 0} f (h) = \text{(the exact solution)},
        \end{equation}
        but such that $f (0)$ is not defined.  The limit \eqref{E:numerical-richardson-limit} must be estimated by evaluating $f (h)$ for a finite number of values of $h$.  For very large\footnote{(relative to the scale over which $f$ varies significantly)} values of $h$, $f (h)$ will be a poor estimate of the limit; but for very small values of $h$, \emph{roundoff error} in the floating point arithmetic will contribute significantly to the value of $f (h)$.

        To see the effect of roundoff error, let
        \begin{equation}
            f (h) = \frac{\sin{(\pi + h)} - \sin{\pi}}{h} = \frac{\sin{(\pi + h)}}{h},
        \end{equation}
        so that $\lim_{h \to 0} f (h) = -1$ is the derivative of $\sin{x}$ at $x = \pi$.  Now, $f (0.1) \simeq - 0.998$ equals the limit to 2 significant figures, and in general $f (10^{-n})$, $n \in \mathbb{N}$, equals the limit to $2 n$ significant figures, if we perform the computation to arbitrary precision.  However, if we evaluate, say, $f (10^{-12})$ using double precision floating point numbers, the result is approximately $- 0.99996$, accurate to only 4 significant figures.  Accuracy is lost because $\pi + h$ differs from $\pi$ only after 12 significant figures,\footnote{(out of the 15 or at most 16 significant figures representable in the \code{double} data type)} and so the computed quantity $\sin (\pi + h)$ is only accurate to 4 significant figures.

        \subsection{Richardson extrapolation}
            \label{S:numerical-richardson-in}

            The purpose of the technique called \emph{Richardson extrapolation} is to estimate the value of the limit \eqref{E:numerical-richardson-limit} from several values of $f (h)$, none of which may themselves be sufficiently accurate estimates.  The basic method is to construct a polynomial approximation to the function $f$, and evaluate it at $h = 0$.  That is, evaluate $p (0)$, where $p (h)$ is the unique polynomial of order $m$ fitting the $m$ known values $(h, f (h))$.
    
            Given the polynomial of order $m$ passing through $m$ known values, it is possible to efficiently determine the polynomial of order $m + 1$ passing through the $m + 1$ points consisting of the $m$ original points plus one additional point.  As such, if the estimate of the limit \eqref{E:numerical-richardson-limit} afforded by the first $m$ evaluations of $f (h)$ is not sufficiently accurate, another single evaluation can be made and a new estimate of the limit obtained.
    
            If the estimate after $m + 1$ function evaluations is approximately equal to the estimate after $m$ function evaluations, to within the desired relative precision $\epsilon$, in the sense defined in Section~\ref{S:numerical-relative-difference}, then no more function evaluations are made.  The most recent estimate, namely the estimate after $m + 1$ function evaluations, is then the output of the Richardson extrapolation process: an approximation of the limit \eqref{E:numerical-richardson-limit}.
    
            Richardson extrapolation is particularly useful when a power series of the function $f (h)$ about $h = 0$ is known to contain only even powers of $h$; this is the case for both of the applications of Richardson extrapolation in this chapter.  The power series may then be treated as a polynomial in $h^2$, rather than a polynomial in $h$.  The extrapolation polynomial is then $p (h^2)$, passing through known values $(h^2, f (h))$.  In evaluating the function $f$ at, say, half the previous value of $h$, a new polynomial fitting point is obtained which is four times closer to zero.
    
            In \grwb, the templatised class \code{richardson_extrapolation<T>}, whose code is listed in Section~\ref{S:grwb-code-richardson}, represents the operation of Richardson extrapolation on a function from $\mathbb{R}$ to the set represented by the type \code{T}; typically \code{T} is \code{double} or an \code{nvector} type.  The \code{refine} routine of the \code{richardson_extrapolation} class takes one argument of type \code{double} and one argument of type \code{T}, representing a new known value pair $(h, f (h))$; using the new values, and the values supplied in previous calls to the routine, \code{refine} computes a new estimate of the limit \eqref{E:numerical-richardson-limit}, and computes the difference between the new estimate and the previous estimate as an approximation of the error.  The most recent estimate and error are accessed, respectively, through the routines \code{limit} and \code{error} of the \code{richardson_extrapolation} class.

    \section{Differentiation}
        \label{S:numerical-differentiation}
        
        Numerical differentiation in \grwb\ is implemented in terms of the class \code{richardson_extrapolation} of Section~\ref{S:numerical-richardson-in}, exposing a functional interface similar to that developed for the \code{derivative} routine of Section~\ref{SS:functional-run-time}.  For a vector space $V$, numerical differentiation is encoded in a routine
        \begin{align}
            \label{E:numerical-derivative}
            &\text{derivative} \colon (\mathbb{R} \to V) \times \mathbb{R} \times \mathbb{R} \to (\mathbb{R} \to V), \notag \\
            &\text{derivative} (f, \mu, \epsilon) = g, \quad g \colon \mathbb{R} \to V, \notag \\
            &g (x) = \text{(the derivative of $f$ at $x$, to relative precision $\epsilon$)},
        \end{align}
        where the argument $\mu$ is a characteristic length scale over which the function $f$ varies significantly.  Depending on the choice of $\mu$, the routine may not successfully converge to an estimate of the derivative to relative precision $\epsilon$.  The code of the \code{derivative} routine is listed in Section~\ref{S:grwb-code-derivative}.

        The function $g$ in \eqref{E:numerical-derivative} employs Richardson extrapolation to estimate the value of
        \begin{equation}
            \label{E:numerical-centred-difference}
            \lim_{h \to 0} \frac{f (x + h) - f (x - h)}{2 h} = \lim_{h \to 0} d (h),
        \end{equation}
        which is the \emph{centred difference} approximation to the derivative of $f$ at $x$.  Observe that $d (h)$ is an even function of $h$; hence a power series expansion of $d (h)$ about $h = 0$ contains only even powers of $h$, and the Richardson extrapolation can be performed using the value pairs $(h^2, d (h))$, rather than the value pairs $(h, d (h))$, with the advantage described at the end of Section~\ref{S:numerical-richardson-in}.

        The first value of $h$ for which $d (h)$ is computed by the \code{derivative} routine is $h = \mu$, the characteristic length scale of the function $f$; the $n$th value of $h$ is $\mu / \sigma^{n - 1}$, where $\sigma = 1.7$ is a constant parameter of the algorithm.  At most $n_\text{max} = 13$ values of $h$ are processed, after which the algorithm terminates, and the derivative is undefined.  Thus, the algorithm explores the region around $x$ at length scales between $\mu / \sigma^{n_\text{max}} \simeq 10^{-3} \mu$ and $\mu$.  The particular values of the constants $\sigma$ and $n_\text{max}$ were empirically chosen to optimise computation speed for the applications of \grwb\ discussed in this thesis.

        Previously in \grwb, numerical differentiation was accomplished by, where an algorithm required it, evaluating $d (h)$ at progressively smaller values of $h$, until the difference between two successive evaluations was smaller than the desired precision.  The new implementation, employing Richardson extrapolation and the C++ template mechanism, converges faster and more accurately, and its interface is more general, in that functions from $\mathbb{R}$ to any sensible set can be differentiated.

        \subsection{Gradient}
        
            The gradient of a function of $\mathbb{R}^n$ is defined in terms of \code{derivative}.  For any vector space $V$, the gradient is defined by
            \begin{align}
                \label{E:numerical-gradient}
                &\text{gradient} \colon (\mathbb{R}^n \to V) \to (\mathbb{R}^n \to V^n), \notag \\
                &\text{gradient} (f) = g, \quad g \colon \mathbb{R}^n \to V^n, \notag \\
                &g (\mathbf{x}) = \text{(derivatives of $f$ at $\mathbf{x}$ with respect to the $n$ components)}.
            \end{align}
            The code of the \code{gradient} routine is listed in Section~\ref{S:grwb-code-gradient}.

            Like many routines in \grwb\ that employ \code{derivative}, \code{gradient} uses default values of $\mu = 1$ and $\epsilon = 10^{-9}$ for the arguments to \code{derivative}.  In general, these routines should be extended to accept these parameters as arguments, and to pass them on to all numerical routines which require them; the scale information $\mu$ in \grwb\ must originally be supplied with definitions of the metric.  For current applications, the metrics input to \grwb\ have unity as an appropriate length scale, and so this extension has not yet been performed.

            Previously in \grwb, the gradient of a field was computed by, where an algorithm required it, explicitly calculating the numerical derivatives with respect to the various components of the vector argument, and populating a vector with the results.  Like \code{derivative}, the new implementation employs the C++ template mechanism to create a more general algorithm, which can apply the definition \eqref{E:numerical-gradient} for any set $V$ for which it makes sense.

    \section{Integration of ordinary differential equations}
        \label{S:numerical-ode}
        
        Previously in \grwb, numerical integration of ordinary differential equations (\textsc{ode}s) was performed using the third-party \slatec\ \texttt{ddriv3} Runge-Kutta algorithm \cite{R:slatec}, originally written in Fortran, converted to C using a Fortran-to-C source code converter, and then adapted to the C++ code of \grwb.  During the course of the project, it was discovered that the \slatec\ algorithm was coded such that only one numerical integration can be in operation at any time.  Normally, this presents no problem; but in the case that the function $f$ which gives the derivatives in the initial value problem specification,
        \begin{align}
            \label{E:numerical-initial-value}
            \frac{d \mathbf{y}}{dx} &= f (\mathbf{y}, x), \notag \\
            \mathbf{y} (0) &= \mathbf{y}_0,
        \end{align}
        is defined in terms of the integration of another, separate \textsc{ode}, the \slatec\ algorithm is inadequate.
        
        It was decided that, rather than further adapting the \slatec\ algorithm, a general \textsc{ode} integrator should be directly implemented in the newly functional framework of \grwb.  The \emph{Bulirsch-Stoer} method, described in \cite{R:numerical-recipes}, pages 724--732, and \cite{R:bulirsch-stoer}, pages 484--486, was selected based on arguments in \cite{R:bulirsch-stoer}, pages 487--488, which recommend it for \textsc{ode}s whose derivative functions $f$ are smooth,\footnote{By `smooth' we mean not varying significantly on scales much smaller than the region of integration.} and for applications where high accuracy is required.  The Bulirsch-Stoer method is generally inferior to Runge-Kutta methods for \textsc{ode}s for which the derivative function $f$ contains discontinuities near the exact solution,\footnote{(because Bulirsch-Stoer steps are longer than Runge-Kutta steps, and are thus more likely to `accidentally' land on or near a discontinuity)} or for stiff \textsc{ode}s, but neither of these cases occur in the current applications of \grwb.

        The Bulirsch-Stoer method, as implemented in \grwb, applies Richardson extrapolation to a series of estimates obtained using the \emph{modified midpoint} method, from \cite{R:numerical-recipes}, pages 722--724.  The modified midpoint method is an algorithm for estimating $\mathbf{y} (H)$ from $\mathbf{y} (0)$, evaluating the derivative function $f$ at the initial point $\mathbf{y}_0$ and at $n$ other points, by the following process:
        \begin{align}
            h &= H / n, \notag \\
            \mathbf{z}_0 &= \mathbf{y}_0, \notag \\
            \mathbf{z}_1 &= \mathbf{z}_0 + h f (0, \mathbf{z}_0), \notag \\
            \mathbf{z}_{m + 1} &= \mathbf{z}_{m - 1} + 2 h f (m h, \mathbf{z}_m), \notag \\
            \mathbf{y} (H) &\simeq \frac{\mathbf{z}_n + \mathbf{z}_{n - 1} + h f (H, \mathbf{z}_n)}{2}.
        \end{align}
        It is a second-order method in $h$.

        The modified midpoint estimate of $\mathbf{y} (H)$, for the initial value problem \eqref{E:numerical-initial-value}, is a function $m (h)$.  The modified midpoint method is chosen for extrapolation using Bulirsch-Stoer because, like the function $d (h)$ in \eqref{E:numerical-centred-difference} employed by \code{derivative}, in a power series of $m (h)$ about $h = 0$, all odd powers of $h$ cancel out, and so the extrapolation can be performed in $h^2$.

        The modified midpoint method is represented in \grwb\ by the class \code{modified_midpoint_stepper}, whose code is listed in Section~\ref{S:grwb-code-modified-midpoint}.  It must be supplied with the derivatives function $f$ and the initial data $\mathbf{y}_0$.  The only routine, \code{step}, takes $H$ and $n$ as arguments, and returns the estimate $y (H)$.

        The difficult problem of choosing the optimal value for $H$, so that the Richardson extrapolation will not take too many steps, but so that a significant distance in $x$ will be covered, is discussed in \cite{R:numerical-recipes}, pages 726--728.

        The class \code{bulirsch_stoer}, whose code is listed in Section~\ref{S:grwb-code-bulirsch-stoer}, is adapted from the implementation of the Bulirsch-Stoer method in \cite{R:numerical-recipes}.  The class must be supplied with the same information as \code{modified_midpoint_stepper}, as well as: a characteristic length scale in $x$, over which $f$ in \eqref{E:numerical-initial-value} varies significantly; the maximum number of steps\footnote{A \emph{step} is a successful Richardson extrapolation of the results of as many calls to \code{modified_midpoint_stepper} as are necessary.} to try before giving up; and the desired relative accuracy of the solution.  The routine \code{step} takes an argument indicating the desired final value of $x$, after which the routines \code{x} and \code{y} return, respectively, the final values of $x$ and $\mathbf{y}$ obtained by the algorithm; if the result of a call to the routine \code{x} equals the argument given to \code{step}, then the integration was successful.

        The new implementation of numerical \textsc{ode} integration in \grwb\ is more general than the \slatec\ Runge-Kutta algorithm.  Previously, the \textsc{ode} integrator required the function $f$ to satisfy $f \colon \mathbb{R}^n \times \mathbb{R} \to \mathbb{R}^n$, and to be encoded using the built-in array notation of C++ (rather than in terms of \code{nvector}, or some other type).  Now, the function $f$ can satisfy $f \colon V \times \mathbb{R} \to V$, where $V$ is any vector space.

    \section{Minimisation of functions}
        \label{S:numerical-minimisation}
                
        Previously, the applications of \grwb\ did not necessitate a mechanism to find local minima of functions.  The development of tools for numerical experimentation, as described in Chapter~\ref{C:numerical-experiments}, highlighted the need for a general algorithm which, for a function $f \colon \mathbb{R}^n \to \mathbb{R}$, can locate a minimum of $f$ near a given initial `guess' point $\mathbf{x}$.

        If $f \colon \mathbb{R} \to \mathbb{R}$, then a local minimum of $f$ can be \emph{bracketed} by three numbers $a < b < c$ which satisfy $f (a) > f (b) < f (c)$.  More efficient algorithms exist for this special case; \grwb\ employs \emph{Brent's method}, from \cite{R:numerical-recipes}, pages 402--405, which repeatedly refines the bracket on a minimum by fitting the three smallest function values found so far (the smallest of which will be $f (b)$) to a parabola, and using the exact minimum of that parabola as the next trial point; it converges quadratically near the minimum.  Brent's method is represented in \grwb\ by the functor class \mbox{\code{brent_minimiser},} whose constructor must be supplied with the function $f$; it is then the function
        \begin{align}
            &\text{brent\_minimiser} \colon \mathbb{R} \times \mathbb{R} \times \mathbb{R} \to \mathbb{R} \times \mathbb{R}, \notag \\
            &\text{brent\_minimiser} (x_0, \mu, \epsilon) = (x_\text{min}, f (x_\text{min})),
        \end{align}
        where $x_\text{min}$ is within relative precision $\epsilon$ of a local minimum of $f$ near $x_0$, and $\mu$ is a characteristic length scale over which $f$ varies significantly.  The code of the \code{brent_minimiser} class is listed in Section~\ref{S:grwb-code-brent}.

        \subsection{Multi-dimensional minimisation}
            \label{S:numerical-powell}
        
            In the general case of multi-dimensional minimisation, minima cannot be bracketed, and minimisation consists, more or less, of `rolling' downhill from the initial guess $\mathbf{x}_0$.  \grwb\ employs \emph{Powell's method}, from \cite{R:numerical-recipes}, pages 412--418, which proceeds by using \code{brent_minimiser} to minimise the function one-dimensionally in each of $n$ linearly independent directions.  The $n$ basis directions are then updated, based on the overall distance moved from $\mathbf{x}_0$, and the process is repeated with the new directions.  The problem of how to choose the right basis directions is discussed in \cite{R:numerical-recipes}.
    
            Powell's method is represented in \grwb\ by the functor class \code{powell_minimiser}, whose constructor must be supplied with the function $f \colon \mathbb{R}^n \to \mathbb{R}$; it is then the function
            \begin{align}
                &\text{powell\_minimiser} \colon \mathbb{R}^n \times M_{n \times n} \times \mathbb{R} \to \mathbb{R}^n \times \mathbb{R}, \notag \\
                &\text{powell\_minimiser} (\mathbf{x}_0, B, \epsilon) = (\mathbf{x}_\text{min}, f (\mathbf{x}_\text{min})),
            \end{align}
            where $\mathbf{x}_\text{min}$ is within relative precision $\epsilon$ (in the Euclidean norm on $\mathbb{R}^n$) of a local minimum of $f$ near $\mathbf{x}_0$, $M_{n \times n}$ is the set of $n \times n$ matrices with real entries, and $B$ is the matrix whose columns are the initial directions to minimise over.  The minimisation is made over the subspace of $\mathbb{R}^n$ spanned by the columns of $B$; this will be all of $\mathbb{R}^n$ only if the columns of $B$ are linearly independent.
    
            The code of the \code{powell_minimiser} class is listed in Section~\ref{S:grwb-code-powell}.  The implementation of Powell's method in \cite{R:numerical-recipes} requires a separately coded implementation of Brent's method\footnote{See \cite{R:numerical-recipes}, pages 418--419.} to perform the minimisations over one-dimensional subspaces of $\mathbb{R}^n$; the quite general interface of the \code{brent_minimiser} class makes this inelegance unnecessary in the implementation of Powell's method in \grwb. 

    \section{Conclusion}

        The rewritten and extended numerical engine of \grwb\ is more efficient, robust, and general.  The implementation of sophisticated algorithms for key operations yields increased computation speed.  The \code{relative_difference} abstraction enables algorithms to be encoded with consistent notions of approximate equality, making them more robust and elegant.  Through the C++ template mechanism, numerical methods can be encoded such that they can be applied to any sets which have the required structure defined upon them. \clearemptydoublepage
    \chapter{Functional differential geometry}
    \label{C:differential-geometry}
    
    The differential geometric engine of \grwb\ has been rewritten within the framework of functional programming, using the functional numerical tools of Chapter~\ref{C:numerical}.  The definition of charts, and the components of the metric on charts, is discussed in Section~\ref{S:differential-geometry-charts}.  Collections of charts, and inter-chart maps, are introduced in Section~\ref{S:differential-geometry-inter-chart-maps}.  The representation of points and tangent vectors as C++ classes is described in Section~\ref{S:differential-geometry-points-and-vectors}.
   
    \begin{table}
        \begin{center}
            \begin{tabular}{| l | l | c |}
                \hline
                Concept & Representation in \grwb & Section \\
                \hline                        
                Space-time & \code{atlas} & \ref{S:differential-geometry-atlases} \\
                Coordinates & \code{nvector<double>} & \ref{S:differential-geometry-charts} \\
                Metric components & \code{nvector<nvector<double>>} & \ref{S:differential-geometry-charts} \\
                Inter-chart map & See \eqref{E:differential-geometry-inter-chart-map-grwb} & \ref{S:differential-geometry-inter-chart-maps} \\
                Point & \code{point} & \ref{S:differential-geometry-points} \\
                Tangent vector & \code{tangent_vector} & \ref{S:differential-geometry-vectors} \\
                Metric & {\footnotesize \code{function<double (tangent_vector, tangent_vector)>}} & \ref{S:differential-geometry-vectors-metric} \\
                World-line & \code{function<point (double)>} & \ref{S:differential-geometry-world-lines} \\
                \hline
            \end{tabular}
            \caption{Representation of important differential geometric concepts in \grwb}
            \label{T:differential-geometry-correspondence}
        \end{center}
    \end{table}
    
    Table~\ref{T:differential-geometry-correspondence} summarises the correspondence between important concepts in differential geometry and their representations in \grwb.  Each correspondence is described in detail in this chapter, but, as the concepts are interrelated, Table~\ref{T:differential-geometry-correspondence} will be useful when reading the earlier sections.
   
    \section{Charts and the metric components}
        \label{S:differential-geometry-charts}

        A \emph{chart} is a subset $C \subset \mathbb{R}^n$, representing a coordinate system on a subset $\mathcal{M}_C \subset \mathcal{M}$ of the space-time manifold $\mathcal{M}$.  We denote by $\phi_C \colon \mathcal{M}_C \to C$ the one-to-one and onto function which maps points in $\mathcal{M}_C$ into the chart $C$.
        
        A space-time in \grwb\ consists of the definition of the components of the metric tensor on one or more charts, and the definition of maps (coordinate transformations) between those charts.  In this section we describe the definition of the metric components on charts; discussion of the inter-chart maps is deferred until Section~\ref{S:differential-geometry-inter-chart-maps}.

        The coordinates of a point on a chart, $\{ x^i \}_{i = 1}^n \in \mathbb{R}^n$, or simply $x^i \in \mathbb{R}^n$, where $n$ is the dimensionality of the space-time, are represented by a variable of type \code{nvector<double>} (see Table~\ref{T:functional-correspondence}).  The components of the metric tensor $g_{ab}$ at a point on a chart are represented as an $n \times n$ matrix, by a variable of type \code{nvector<nvector<double>>}.  A function which defines the metric components $g_{ab}$, as a function of the chart coordinates $x^i$, might then be of the form
        \begin{align}
            \label{E:differental-geometry-naive-chart}
            &\text{chart} \colon \mathbb{R}^n \to M_{n \times n}, \notag \\
            &\text{chart} (x^i) = g_{ab} |_{x^i},
        \end{align}
        represented in \grwb\ by a function of signature \code{nvector<nvector<double>> (nvector<double>)}.  In general, however, the chart coordinates are an open subset of $\mathbb{R}^n$, and so \eqref{E:differental-geometry-naive-chart} will not be defined everywhere in $\mathbb{R}^n$.  A mechanism is required to represent functions which are only defined on a subset of some other, standard, set.\footnote{By `standard set' we mean a set which is represented by a type in C++, such as those in Tables~\ref{T:functional-correspondence} and \ref{T:differential-geometry-correspondence}.}

        \subsection{The \code{optional} mechanism}
            \label{S:differential-geometry-optional}
        
            \grwb\ employs the \emph{Boost Optional Library} \cite{R:boost-optional} to represent functions which are undefined for some values of their arguments.  The Optional Library provides a templatised type \code{optional<T>}, which represents the set $S \cup \{ \varnothing \}$, where $S$ is the set corresponding to the template parameter type \code{T}, and $\varnothing$ is a special value taken by functions at points where they are undefined.

            The \code{optional} template might be used in the following way:
            \begin{codeblock}
optional<double> square_root(double x)
{
    if (x < 0)
    {
        // undefined; return the special value `undefined'
        return optional<double>();
    }
    else
    {
        // defined; return the result of the standard C++ square root algorithm, `sqrt'
        return optional<double>(sqrt(x));
    }
}
            \end{codeblock}
            Thus, by returning a variable of type \code{optional<double>}, instead of a variable of type \code{double}, the \code{square_root} routine can return the special value $\varnothing$ (using the code \code{return optional<double>();}) to indicate points where the algorithm is undefined; in this case, $\varnothing$ is returned for negative values of the argument \code{x}.

            The \code{optional} mechanism is most useful when the caller of a function cannot know beforehand whether the function will be defined at the arguments to be given to it.  This would be the case for callers of the function \eqref{E:differental-geometry-naive-chart}; the differential geometric algorithms in \grwb\ must be coded in such a way that they can operate on any space-time definition, without prior knowledge of the particular coordinate systems (charts) they will be working in.

            We can now modify \eqref{E:differental-geometry-naive-chart} to support charts defined on subsets of $\mathbb{R}^n$, using the \code{optional} mechanism.  Thus, in \grwb, functions which return the metric components $g_{ab}$, as a function of the chart coordinates $x^i$, are of the form
            \begin{align}
                \label{E:differental-geometry-chart}
                &\text{chart} \colon \mathbb{R}^n \to M_{n \times n} \cup \{ \varnothing \}, \notag \\
                &\text{chart} (x^i) =
                \begin{cases}
                    g_{ab} |_{x^i}, &\text{if the $x^i$ are valid chart coordinates;} \\
                    \varnothing, &\text{otherwise.}
                \end{cases}
            \end{align}
            The corresponding C++ type is
            \begin{equation}
                \label{E:differental-geometry-chart-c++}
                \text{\small \code{function<optional<nvector<nvector<double>>> (nvector<double>)>}},
            \end{equation}
            for which \grwb\ declares a short synonym, \code{chart}, using the C++ \code{typedef} mechanism:
            \begin{codeblock}
typedef function<optional<nvector<nvector<double>>> (nvector<double>)> chart;
            \end{codeblock}
            References to \code{chart}s are stored in variables of type \code{shared_ptr<chart>}, using the \emph{Boost Smart Pointers Library} \cite{R:boost-shared-ptr}.
            
        \subsection{Example chart and metric components}
            \label{S:differential-geometry-example}
        
            In this section we demonstrate the encoding of the flat space metric of special relativity, in cylindrical coordinates $(t, r, \phi, z)$, using a C++ function of type \code{chart}.  The line element is
            \begin{equation}
                \label{E:differential-geometry-example-metric}
                ds^2 = - dt^2 + dr^2 + r^2 \, d\phi^2 + dz^2,
            \end{equation}
            so the metric components, as functions of the chart coordinates $(t, r, \phi, z)$, are $- g_{tt} = g_{rr} = g_{zz} = 1$, $g_{\phi\phi} = r^2$, and all other $g_{ab} = 0$.  The chart coordinates are valid in the open subset of $\mathbb{R}^n$ satisfying
            \begin{align}
                \label{E:differential-geometry-example-chart}            
                t &\in (- \infty, \infty), \notag \\
                r &\in (0, \infty), \notag \\
                \phi &\in (0, 2 \pi), \notag \\
                z &\in (- \infty, \infty).
            \end{align}
            
            The following routine encodes \eqref{E:differential-geometry-example-metric} and \eqref{E:differential-geometry-example-chart} in C++:
            \begin{codeblock}
optional<nvector<nvector<double>>> flat_metric_cylindrical(nvector<double> x)
{
    // t = x[0], r = x[1], phi = x[2], and z = x[3]
    if (x[1] <= 0 or x[2] <= 0 or x[2] >= 2 * pi)
    {
        // invalid chart coordinates; return `undefined'
        return optional<nvector<nvector<double>>>();
    }
    else
    {
        // valid chart coordinates; compute and return metric components
        nvector<nvector<double>> gab;
        
        gab[0][0] = -1;
        gab[1][1] = 1;
        gab[2][2] = x[1] * x[1];
        gab[3][3] = 1;
        // all other gab = 0
        
        return optional<nvector<nvector<double>>>(gab);
    }
}
            \end{codeblock}
            The operator \code{[i]}, applied to an \code{nvector} such as in \code{x[i]}, returns the \code{i}th component of the vector.

            The opening \code{if} statement determines whether the argument \code{x} represents valid chart coordinates; if so, the metric components are computed in the variable \code{gab}, and returned; if not, $\varnothing$ is returned.  All space-times in \grwb\ have the metric components defined on each of their charts by functions like \code{flat_metric_cylindrical}, above.

        \subsection{The connection}
        
            The components of the connection, or the \emph{Christoffel symbols}, are the useful quantities defined in terms of the metric components $g_{ab}$ by
            \begin{equation}
                \label{E:differential-geometry-connection}
                \Gamma^c_{ab} = \frac{1}{2} g^{dc} (g_{ad,b} + g_{bd,a} - g_{ab,d}),
            \end{equation}
            where $g_{ab,c}$ denotes partial differentiation of $g_{ab}$ with respect to the coordinate $x^c$, and $g^{ab}$ denotes the contravariant components of the metric tensor.  The Christoffel symbols are used by the numerical differential geometric functions of Chapter~\ref{C:numerical-experiments}.

            The \grwb\ routine \code{connection} accepts an argument of type \code{chart}, and returns a variable of type \code{function<optional<nvector<nvector<nvector<double>>>> (nvector<double>)>}, representing the function which returns the components \eqref{E:differential-geometry-connection} as a function of the chart coordinates.

            The differentiation of the metric components $g_{ab}$ is accomplished using the numerical tools of Chapter~\ref{C:numerical}.  A function which returns the components of $g_{ab,c}$, as a function of the chart coordinates, is given simply by \code{gradient(c)}, where \code{c} is the function, of type \code{chart}, which returns the metric components $g_{ab}$ as a function of the chart coordinates.

            The matrix of {contravariant components $g^{ab}$ of the the metric is simply the inverse of the matrix $g_{ab}$ of covariant components.  This matrix inversion is performed in \grwb\ using standard row reduction techniques (see for example \cite{R:lay}, pages 115-116).
        
    \section{Inter-chart maps}
        \label{S:differential-geometry-inter-chart-maps}
        
        As mentioned at the beginning of Section~\ref{S:differential-geometry-charts}, space-times are defined by specifying, together with the metric components on each chart, maps between the various charts.

        For two charts $A, B \subset \mathbb{R}^n$, the \emph{inter-chart map} from $A$ to $B$ is
        \begin{align}
            \label{E:differential-geometry-inter-chart-map}
            &\phi_{AB} \colon A \to B, \notag \\
            &\phi_{AB} (x^i) =  (\phi_B |_{\mathcal{M}_A} \circ \phi_A^{-1}) (x^i),
        \end{align}
        where $\phi_B |_{\mathcal{M}_A}$ is the function $\phi_B$ restricted to the set $\mathcal{M}_A$, and $\circ$ denotes function composition.  The inter-chart maps must be specified to complete the definition of a space-time.

        In the definition \eqref{E:differential-geometry-inter-chart-map}, the domain $A$ of $\phi_{AB}$ is, in general, a subset of $\mathbb{R}^n$.  Hence $\phi_{AB}$ cannot be represented by a variable of type \code{function<nvector<double> (nvector<double>)>}; instead, the \code{optional} mechanism of Section~\ref{S:differential-geometry-optional} is again employed.  Thus, in \grwb, an inter-chart map from a chart $A$ to a chart $B$ is represented by a function of the form
        \begin{align}
            \label{E:differential-geometry-inter-chart-map-final}
            &\text{map} \colon \mathbb{R}^n \to \mathbb{R}^n \cup \{ \varnothing \}, \notag \\
            &\text{map} (x^i) =
            \begin{cases}
                (\phi_B |_{\mathcal{M}_A} \circ \phi_A^{-1}) (x^i), &\text{if $(x^i) \in A$ and $\phi_A^{-1} (x^i) \in \mathcal{M}_B$;} \\
                \varnothing, &\text{otherwise.}
            \end{cases}
        \end{align}
        The corresponding C++ type is
        \begin{equation}
            \label{E:differential-geometry-inter-chart-map-grwb}
            \text{\code{function<optional<nvector<double>> (nvector<double>)>}}.
        \end{equation}
        As with \code{chart}s, the C++ \code{typedef} mechanism is used to define a synonym \code{map} for the type \eqref{E:differential-geometry-inter-chart-map-grwb}.  References to \code{map}s are stored in variables of type \code{shared_ptr<map>}.
        
        \subsection{Example inter-chart map}
            
            In this section we demonstrate the encoding in \grwb\ of an inter-chart map of the form \eqref{E:differential-geometry-inter-chart-map-grwb}, which transforms between two cylindrical coordinate systems like example \eqref{E:differential-geometry-example-chart} in Section~\ref{S:differential-geometry-example}, with the coordinate systems displaced from each other by $\pi$ in the $\phi$ coordinate.  Together, the two coordinate systems thus cover the entire flat-space manifold $\mathbb{R}^4$ of special relativity, except for the line $r = 0$.

            The coordinate transformation, of the form \eqref{E:differential-geometry-inter-chart-map-final}, is
            \begin{align}
                \label{E:differential-geometry-example-inter-chart-map}
                &\text{revolve} \colon \mathbb{R}^n \to \mathbb{R}^n \cup \{ \varnothing \}, \notag \\
                &\text{revolve} (t, r, \phi, z) =
                \begin{cases}
                    (t, r, \phi + \pi, z), &\text{if $\phi < \pi$;} \\
                    \varnothing, &\text{if $\phi = \pi$;} \\
                    (t, r, \phi - \pi, z), &\text{otherwise,}
                \end{cases}
            \end{align}
            and is encoded in C++ in the following way:
            \begin{codeblock}
optional<nvector<double>> revolve(nvector<double> x)
{
    // t = x[0], r = x[1], phi = x[2], and z = x[3]
    if (x[2] == pi)
    {
        // mapping not defined; return `undefined'
        return optional<nvector<double>>();
    }
    else
    {
        // mapping defined; perform transformation
        nvector<double> y;
        
        y[0] = x[0];
        y[1] = x[1];
        if (x[2] < pi)
            y[2] = x[2] + pi;
        else
            y[2] = x[2] - pi;
        y[3] = x[3];
        
        return optional<nvector<double>>(y);
    }
}                
            \end{codeblock}
            The operator \code{==}, used in the first \code{if} statement, is the test for equality in C++.
            
            By using a functor class (Section~\ref{S:functional-functors}), we could parameterise the transformation \code{revolve} on the angle of rotation, which is currently $\pi$.  All space-times in \grwb\ have their inter-chart maps specified by routines or functors like \code{revolve}, above.
        
    \section{Atlases}
        \label{S:differential-geometry-atlases}
            
        A collection of charts with the metric components defined on them, of the form \eqref{E:differental-geometry-chart}, and a collection of inter-chart maps, of the form \eqref{E:differential-geometry-inter-chart-map-final}, together comprising a \emph{space-time}, are represented in \grwb\ by the class \code{atlas}.  The \code{atlas} class uses C++ Standard Template Library (\textsc{stl}) \cite{R:stl} containers to maintain the collections of charts and maps.

        An \code{atlas} contains a \code{std::set} of \code{chart}s, and a \code{std::map} from \code{std::pair}s of \code{chart}s to inter-chart map definitions of type \code{map}.\footnote{\code{std::set}, \code{std::map}, and \code{std::pair} are \textsc{stl} templates; see \cite{R:stl}.}  An \code{atlas} also contains an \code{int} named \code{dimension} which stores the dimensionality of the space-time.

        The members \code{charts} and \code{maps} of class \code{atlas} are used by the differential geometric algorithms of \grwb\ to, respectively, enumerate the set of all charts, and retrieve the inter-chart map between any two charts.  If two charts do not overlap at all, there will be no inter-chart map between them; this is equivalent to there being an inter-chart map between them that always returns $\varnothing$.
    
    \section{Points and tangent vectors}
        \label{S:differential-geometry-points-and-vectors}
        
        For a point, a \emph{valid chart} is a chart containing the point; for a tangent vector, a valid chart is a chart containing the point whose tangent space contains the tangent vector.  While points and tangent vectors may be represented by their coordinates on a valid chart, it is useful to have a representation of these objects which is not linked to any particular chart.  The \grwb\ representation for points is described in Section~\ref{S:differential-geometry-points}, and the representation for tangent vectors is described in Section~\ref{S:differential-geometry-vectors}.

        \subsection{Points}
            \label{S:differential-geometry-points}
        
            The abstract notion of a point $p \in \mathcal{M}$, independent of any particular coordinate system, is represented in \grwb\ by the class \code{point}.  A \code{point} is constructed from three pieces of information: the \code{atlas} to which it belongs, a \code{chart} which contains it, and its coordinates on that chart.
    
            The \code{context} and \code{valid_chart} routines of class \code{point} return, respectively, the \code{atlas} and the \code{chart} from which the point was constructed.  Numerical operations involving points can only be performed in terms of a valid coordinate system, so the \code{valid_chart} routine is used whenever a variable of type \code{point} is an argument to a numerical differential geometric routine in \grwb.

            \subsubsection{Change of coordinates}
    
                The \code{operator[]} routine of class \code{point}, which takes one argument, a variable of type \code{chart}, returns a variable of type \code{optional<nvector<double>>}, representing the coordinates of the point on the given chart.  (The \code{optional} mechanism of Section~\ref{S:differential-geometry-optional} is used because a particular point may, or may not, have coordinates on the given chart.)  Thus, if \code{p} is a variable of type \code{point}, and \code{c} is a variable of type \code{chart}, then the coordinates of \code{p} on \code{c} are given by \code{p[c]}.

                Let \code{a} be the variable of type \code{chart} from which \code{p} was constructed.  If \code{c} and \code{a} represent the same chart, then \code{p[c]} will simply return the coordinates from which \code{p} was constructed.  If, on the other hand, \code{c} and \code{a} are different charts, then \grwb\ will use the \code{maps} member of the \code{atlas} class to determine if there is an inter-chart map from \code{a} to \code{c} defined; if so, then the inter-chart map is used to compute the coordinates of \code{p} on \code{c}, which are then returned; if not, then $\varnothing$ is returned, indicating that \code{p} is not contained in the chart \code{c}.
                
        \subsection{World-lines}
            \label{S:differential-geometry-world-lines}
            
            A curve in space-time, such as a world-line, is a function $\lambda \colon \mathbb{R} \to \mathcal{M}$; such functions are represented by variables of type \code{function<point (double)>}.  However, if the curve $\lambda$ is not defined for all values of its real parameter, then it will instead be represented by a variable of type \code{function<optional<point> (double)>}.  All curves in \grwb\ are in fact represented in this latter form, because they are often defined in terms of numerical processes which may not converge to a solution.  The computation of geodesics, discussed in Section~\ref{S:num-exp-geodesics}, exemplifies this.

            The C++ \code{typedef} mechanism is used to define the synonym \code{worldline} for the type \code{function<optional<point> (double)>}:
            \begin{codeblock}
typedef function<optional<point> (double)> worldline;
            \end{codeblock}             
            
        \subsection{Tangent vectors}
            \label{S:differential-geometry-vectors}            
            
            The abstract notion of a tangent vector $v \in T_p$, where $T_p$ is the tangent space of a point $p \in \mathcal{M}$, is represented in \grwb\ by the class \code{tangent_vector}.  Like a \code{point}, a \code{tangent_vector} is constructed from three pieces of information: the \code{point} to whose tangent space it belongs, a \code{chart} containing that point, and the contravariant components\footnote{Whenever we discuss the components of a tangent vector, we always mean its contravariant components.} of the tangent vector on that chart.

            The \code{context} routine of class \code{tangent_vector} returns the \code{point} from which the tangent vector was constructed; through the \code{valid_chart} routine of this point, a valid chart for the tangent vector can be obtained.  As with the \code{point} class, the \code{operator[]} routine of the \code{tangent_vector} class, taking one argument, a variable of type \code{chart}, returns the components of the tangent vector on the given chart, in a variable of type \code{optional<nvector<double>>}.

            \subsubsection{Change of coordinates}
            
                As with the \code{point} class, when the components of a tangent vector are requested on a chart other than that from which the tangent vector was constructed, \grwb\ uses the inter-chart map, if it exists, to compute the components.  If $v^i$ are the components of a tangent vector $v$ at a point $p$ on a chart with coordinates $x^i$, then the components on another chart, with coordinates $x^{i'}$, are
                \begin{equation}
                    \label{E:differential-geometry-vector-transform}
                    v^{i'} = \left. \frac{\partial x^{i'}}{\partial x^i} \right|_p v^i = A^{i'}_i v^i.
                \end{equation}
                The columns of the matrix $A^{i'}_i$ are the derivatives of the inter-chart map $\phi \colon \mathbb{R}^n \to \mathbb{R}^n$ with respect to the coordinates $x^i$ of its argument, evaluated at $p$.  \grwb\ computes $A^{i'}_i$, and thereby the components $v^{i'}$, by using the methods of Chapter~\ref{C:numerical} to numerically evaluate the derivatives.

        \subsection{Tangent vectors and the metric}
            \label{S:differential-geometry-vectors-metric}
            
            At a point $p$, the metric $g_{ab}$ is naturally considered as the inner product
            \begin{align}
                \label{E:differential-geometry-metric-vectors}
                &\text{metric} \colon T_p \times T_p \to \mathbb{R}, \notag \\
                &\text{metric} (u, v) = g_{ab} u^a v^b.
            \end{align}
            If $u = v$ in \eqref{E:differential-geometry-metric-vectors}, then the sign of $\text{metric} (u, u)$ determines whether $u$ is space-like, null, or time-like. If $\text{metric} (u, u) = -1$ then $u$ represents the time direction of a physical observer---this is discussed in Section~\ref{S:num-exp-observers}.

            The function \eqref{E:differential-geometry-metric-vectors} is encoded in \grwb\ in the routine \code{metric}, whose signature is \code{double (tangent_vector, tangent_vector)}.  Also, the \code{operator*} routine of the class \code{tangent_vector} is defined to call \code{metric}, so that if \code{u} and \code{v} are variables of type \code{tangent_vector}, then the expression \code{u * v} is equivalent to the expression \code{metric(u, v)}.  This notation is reminiscent of the two equivalent forms
            \begin{equation}
                g_{ab} u^a v^b = u_b v^b
            \end{equation}
            for the inner product of two vectors. 
    
    \section{Conclusion}
    
        The implementation of the differential geometric structure of \grwb\ within the framework of functional programming, using the numerical methods of Chapter~\ref{C:numerical}, is robust and elegant.  The representation of abstract objects such as points and tangent vectors, independent of any particular chart, will be useful in the construction of the numerical experiments of Chapter~\ref{C:numerical-experiments}. \clearemptydoublepage
    \chapter{Numerical experiments}
    \label{C:numerical-experiments}
    
    A \emph{numerical experiment} is a model of a physical situation in \grwb, from which a measurement of a physical quantity is obtained.  Tools for simulating physical situations in \grwb\ have been implemented using the methods of Chapters~\ref{C:numerical} and \ref{C:differential-geometry}.  Basic operations on points and tangent vectors are described in Section~\ref{S:num-exp-basic}.  Geodesic tracing and the parallel transport operation are the topics of Sections~\ref{S:num-exp-geodesics} and \ref{S:num-exp-parallel-transport}, respectively.  In Section~\ref{S:num-exp-implicit} we discuss methods for finding geodesics that are defined implicitly in terms of boundary conditions.

    In Chapter~\ref{C:karim-grwb}, the methods of this chapter are used to numerically investigate the claim to be discussed in Chapter~\ref{C:karim}.

    \section{Basic operations}
        \label{S:num-exp-basic}

        In this section we describe some operations on points, tangent vectors, and world-lines, which will be useful for constructing numerical experiments.
    
        \subsection{Tangent vectors and observers}
            \label{S:num-exp-observers}
        
            As was mentioned at the end of Section~\ref{S:differential-geometry-vectors}, a tangent vector $u$, such that $\text{metric} (u, u) = -1$, represents the proper time direction of a physical observer.  More precisely: physical observers are defined by their time-like world-lines, with parameter $t$; if the tangent vector $u$ to the world-line always satisfies $\text{metric} (u, u) = -1$, then the parameter $t$ is the (proper) time coordinate in the frame of reference of the observer.

            Normalisation of a tangent vector is defined by
            \begin{align}
                \label{E:num-exp-normalise}
                &\text{normalise} \colon T_p \to T_p, \notag \\
                &\text{normalise} (u) = \frac{u}{\sqrt{|\text{metric} (u, u)|}}.
            \end{align}
            Thus, the normalisation of a vector $u$ is a vector $v$ such that $\text{metric} (v, v) = \pm 1$, according as whether $u$ was space-like or time-like.  The definition \eqref{E:num-exp-normalise} is encoded in \grwb\ in the routine \code{normalise}, which has signature \code{tangent_vector (tangent_vector)}.

            Also useful is the operation of \emph{orthonormalisation}.  The orthonormalisation of a vector $u$ with respect to another vector $v$ is defined by
            \begin{align}
                \label{E:num-exp-orthonormalise}
                &\text{orthonormalise} \colon T_p \times T_p \to T_p, \notag \\
                &\text{orthonormalise} (u, v) = \text{normalise} (\text{metric} (u, v) v - \text{metric} (v, v) u),
            \end{align}
            which is encoded in \grwb\ in the routine \code{orthonormalise}, which has signature \code{tangent_vector (tangent_vector, tangent_vector)}.  Orthonormalisation has the property that, if $w = \text{orthonormalise} (u, v)$, then $\text{metric} (v, w) = 0$, and either $\text{metric} (w, w) = 1$ or $\text{metric} (w, w) = -1$.

        \subsection{Orthonormal tangent bases}
            \label{S:num-exp-orthonormal-basis}

            An \emph{orthonormal tangent basis} for $T_p$ at a point $p$ is a set of $n$ vectors in $T_p$ that are mutually orthonormal.  The metric components expressed in an orthonormal tangent basis form a diagonal matrix; this will be useful in Section~\ref{S:num-exp-implicit}.  The determination of an orthonormal tangent basis is also called \emph{diagonalising the metric}.

            \grwb\ constructs an orthonormal tangent basis by finding the eigenvectors of the matrix $g$ of metric components $g_{ab}$.  The eigenvectors are orthogonal, because the matrix $g$ is symmetric.  The process of determining the eigenvectors of a matrix is represented in \grwb\ by the class \code{eigen}, which is constructed from a variable of type \code{nvector<nvector<double>>}, representing the matrix whose eigenvectors are to be determined.  The routine \code{vectors} of class \code{eigen} then returns a variable of type \code{nvector<nvector<double>>}, representing the $n$ eigenvectors, and the routine \code{values} of class \code{eigen} returns a variable of type \code{nvector<double>}, a list of the corresponding eigenvalues.

            The \code{eigen} class uses an iterative method to find the eigenvectors of $g$ (see \cite{R:searle}, page 25).  Starting with a coordinate basis vector $e_1$, the sequence of vectors $g^n e_1$ converges, as $n \to \infty$, to an eigenvector $v_1$ of $g$.  A second eigenvector $v_2$ is obtained by seeding the process with $e_2$.  Because the sequence $g^n e_i$ will tend to converge to the eigenvector which has the largest eigenvalue, each successive estimate is orthogonalised with respect to the previously determined eigenvectors, before the next left-multiplication by $g$.  Once this process has been completed, starting with each coordinate basis vector, the full set of orthogonal eigenvectors are known.

            If the metric is Lorentzian, then one of the eigenvectors will have a negative eigenvalue, corresponding to a time-like direction, and all the others will have positive eigenvalues, corresponding to space-like directions.  The normalised eigenvectors constitute an orthonormal tangent basis.  The \grwb\ routine \code{orthonormal_tangent_basis} takes one argument of type \code{point}, and one argument of type \code{chart}, and uses the \code{eigen} class to return a variable of type \code{nvector<nvector<double>>}, representing a matrix whose columns are the components of an orthonormal basis of the tangent space of the given point in the given chart.
    
        \subsection{Coordinate lines}
            \label{S:num-exp-coordinate-lines}
        
            If a particular coordinate system on a space-time has known properties, such as the metric being independent of one of the coordinates, then it may be useful to specify space-time curves explicitly in terms of the coordinates.  Straight lines in a particular coordinate system are obtained in \grwb\ through the \code{coordinate_line} routine, which takes three arguments: a \code{point} on the curve; the \code{chart} on which the curve is to be a straight line; and an \code{nvector<double>} giving the components of the tangent vector to the coordinate line at the given \code{point}.

            The \code{coordinate_line} routine returns a variable of type \code{worldline}, as defined in Section~\ref{S:differential-geometry-world-lines}.  If the coordinate line intersects a chart boundary, then it is undefined beyond it; hence the use of the \code{optional} mechanism.

    \section{Geodesics}
        \label{S:num-exp-geodesics}
        
        \emph{Geodesics}, the straightest possible lines in a curved space-time, are physically important.  Geodesics whose tangent vectors are time-like are the world-lines of freely-falling observers; geodesics whose tangent vectors are space-like represent straight `rulers', for observers whose world-lines intersect them orthogonally; and geodesics whose tangent vectors are null represent the world-lines of photons.

        Geodesics are uniquely defined by a point $p$ on the geodesic and the tangent vector $v$ of the geodesic at $p$.  The coordinates $x^c$ of a geodesic on a chart $A$, as functions of an \emph{affine parameter} $t$, satisfy the \emph{geodesic equation},
        \begin{equation}
            \label{E:num-exp-geodesic-equation}
            \frac{d^2 x^c}{dt^2} + \Gamma^c_{ab} \frac{dx^a}{dt} \frac{dx^b}{dt} = 0,
        \end{equation}
        which involves the connection \eqref{E:differential-geometry-connection}.  Note that the components of $\Gamma^c_{ab}$ in \eqref{E:num-exp-geodesic-equation} are a function of the coordinates $x^c$.

        The equation \eqref{E:num-exp-geodesic-equation} is a system of $n$ second order \textsc{ode}s in the coordinates $x^c$; we may rewrite it as a system of $2 n$ first order \textsc{ode}s.  Together with the $n$ components of an initial point $p$ on $A$, and the $n$ components of an initial vector $v \in T_p$ on $A$, \eqref{E:num-exp-geodesic-equation} defines an initial value problem, which can be solved on the chart $A$ using the numerical \textsc{ode} integration techniques of Section~\ref{S:numerical-ode}.

        In general, no single chart will cover the entire space-time.  Equation \eqref{E:num-exp-geodesic-equation} can only be integrated up to a chart boundary; beyond that, the metric components $g_{ab}$, and hence the Christoffel symbols $\Gamma^c_{ab}$, are undefined on that chart.

        Let $y$ be a point near the boundary of a chart $A$, beyond which numerical integration of \eqref{E:num-exp-geodesic-equation} fails.  If there is another chart $B$ containing $y$, and an inter-chart map from $A$ to $B$, then integration of \eqref{E:num-exp-geodesic-equation} can be attempted on $B$:  Using the inter-chart map, the components $x^c$, in \eqref{E:num-exp-geodesic-equation}, can be computed on $B$ from those on $A$; using \eqref{E:differential-geometry-vector-transform}, the components $dx^i / dt$, in \eqref{E:num-exp-geodesic-equation}, can be computed on $B$ from those on $A$; and, using \eqref{E:differential-geometry-connection}, the components of $\Gamma^c_{ab}$ at $y$ can be computed on $B$.

        \subsection{Implementation in \grwb}
        
            A point on a geodesic, and the tangent vector to the geodesic at that point, are represented in \grwb\ by a variable of type \code{tangent_vector}.  (The \code{context} routine of class \code{tangent_vector} returns the \code{point} at which the tangent vector exists.)  To determine a new \code{tangent_vector} on the geodesic, at a desired value $t = t_\text{final}$ of the affine parameter, \grwb\ uses the \code{operator[]} routines of the classes \code{point} and \code{tangent_vector} to obtain the initial data for equation \eqref{E:num-exp-geodesic-equation} on each chart, one by one, until it finds a chart on which \eqref{E:num-exp-geodesic-equation} can be integrated.

            If no chart exists on which \eqref{E:num-exp-geodesic-equation} could be successfully integrated to the desired value $t_\text{final}$ of the affine parameter, then integration to affine parameter $t_\text{final} / 2$ is attempted, followed by integration to affine parameter $t_\text{final}$.  If either of these integrations fail, then the corresponding interval in $t$ is further subdivided, up to a maximum of 7 bisections.\footnote{The maximum number of bisections, 7, was empirically determined to be adequate for current applications of \grwb.}  If the maximum number of bisections is reached without successful integration to $t = t_\text{final}$, then $\varnothing$ is returned, indicating that the geodesic is undefined at the value $t_\text{final}$ of the affine parameter.

            This definition of a geodesic from its initial data is represented in \grwb\ by the functor class \code{geodesic}, which is constructed from a variable of type \code{tangent_vector}.  Upon construction, a \code{geodesic} is a function of type \code{worldline}, as defined in Section~\ref{S:differential-geometry-world-lines}.  The code of the \code{geodesic} class is listed in Section~\ref{A:grwb-code-geodesic}.

            The class \code{geodesic} maintains a list (\emph{cache}) of all \code{tangent_vector}s found so far on the geodesic.  The \code{operator()} routine of class \code{geodesic}, which takes $t_\text{final}$ as its only argument, uses the class \code{bulirsch_stoer} of Section~\ref{S:numerical-ode} to attempt to numerically integrate \eqref{E:num-exp-geodesic-equation} from initial data in the cache.  The particular initial data chosen is that whose affine parameter is nearest to $t_\text{final}$.

    \section{Parallel transport}
        \label{S:num-exp-parallel-transport}
        
        The operation of \emph{parallel transport} represents the notion of transporting a vector along a curve while changing its direction as little as possible.  It is defined in a similar way to a geodesic.\footnote{A geodesic is, by definition, a curve whose tangent vector is the parallel transport of itself along the curve.}

        A parallel transport is defined by a curve, and a tangent vector at a point on that curve.  It then defines a unique tangent vector at each other point on the curve.  On a chart, the components $v^c$ of the parallelly-transported tangent vector satisfy the equation
        \begin{equation}
            \label{E:num-exp-parallel-transport}
            \frac{dv^c}{dt} + \Gamma^c_{ab} v^a \frac{dx^b}{dt} = 0,
        \end{equation}
        where $x^b (t)$ are the coordinates of the curve as a function of the curve parameter $t$.
        
        Just as for geodesics, \eqref{E:num-exp-parallel-transport} must in general be integrated on multiple charts to determine the tangent vector at a desired value $t = t_\text{final}$ of the curve parameter.  The operation of parallel transport is represented in \grwb\ by the functor class \code{parallel_transport}, which is constructed from a \code{tangent_vector} and a \code{worldline}.  It is then a function with signature \code{optional<tangent_vector> (double)}, representing the tangent vector as a function of the curve parameter $t$.  The \code{parallel_transport} class uses a similar algorithm to the \code{geodesic} class to integrate \eqref{E:num-exp-parallel-transport} on any chart for which it is possible, bisecting the interval of integration if integration cannot proceed on any chart.

        A parallelly-transported vector has a physical interpretation which makes it potentially useful in constructing numerical experiments: it is a fixed coordinate direction for a locally non-rotating physical observer who is moving on a geodesic.  For locally non-rotating physical observers moving on non-geodesic world-lines, the operation with the corresponding physical interpretation is Fermi-Walker transport (see \cite{R:stephani}, pages 47--49), which has not yet been implemented in \grwb.
               
    \section{Implicitly-defined geodesics}
        \label{S:num-exp-implicit}
        
        The methods of Section~\ref{S:num-exp-geodesics} allow the computation of the unique geodesic solving the initial value problem comprising \eqref{E:num-exp-geodesic-equation} together with the initial coordinates $x^i$ and the initial components of the tangent vector $d x^i / dt$.  However, there are ways other than the initial value problem to define a geodesic.  Two physically important examples are discussed in this section.

        \subsection{Unique connecting geodesics}
            \label{S:num-exp-connecting}

            Around every point there is a neighbourhood such that, given two points within it, there will be a unique geodesic that intersects both points.  The way to find this connecting geodesic is the topic of this section.  The problem can be formulated in the following way: given two points $a$ and $b$, find a tangent vector $v \in T_a$ such that the unique geodesic passing through $a$ with tangent $v$ also passes through $b$.  If $v$ is a solution to this problem, then so is $\alpha v$ for any $\alpha \neq 0$; changing the value of $\alpha$ simply changes the affine parameter value at which the geodesic intersects $b$.

            The problem of finding the tangent vector $v$, up to scaling by a real number, can be thought of as determining which direction, in space and time, to launch a geodesic from $a$ such that it `hits' $b$.  We solve this problem by minimising, over all possible directions at $a$, the amount by which the launched geodesic `misses' $b$.  To do this, we need a definition for the amount by which the geodesic misses---a real-valued function to minimise.

            The function $f \colon T_a \to \mathbb{R}$, which gives the amount by which the geodesic, launched from $a$ with the given tangent vector, misses $b$, must satisfy certain properties.  It must be zero for a geodesic which exactly intersects the point $a$, and strictly greater than zero otherwise; and it must be continuous, in the sense that, whenever a sequence of vectors $v_n$ satisfy $\lim_{n \to \infty} f (v_n) = 0$, then we must have $v_n \to v$, where $v$ is an exact solution to the problem.

            \subsubsection{\code{min_euclidean_separation}}
            
                A simple definition for the function $f$, satisfying the requirements listed above, is as follows:
                \begin{equation}
                    \label{E:num-exp-simple-separation}
                    f (v) = \min_{t \in \mathbb{R}} \min_{\text{charts $C$}} \| \delta x^i \|, \quad \delta x^i = \text{geodesic} (v) (t) |_C - b |_C,
                \end{equation}
                where $\text{geodesic} (v)$ denotes the geodesic with tangent vector $v \in T_a$ at $a$, $\| \cdot \|$ denotes the standard Euclidean norm on $\mathbb{R}^n$, and we have used the notation that, for any point $q$ and chart $C$, $q_C$ denotes the coordinates of $q$ on $C$.  That is, the distance between the curve $\text{geodesic} (v)$ and the point $b$ is defined as the closest they ever get, in the Euclidean norm, in the coordinates of any chart.  The quantity $\delta x^i$ is intended to be a small displacement in the coordinates of the chart $C$; in any case, it will certainly be zero if $\text{geodesic} (v)$ intersects $b$ at affine parameter value $t$.

                The definition \eqref{E:num-exp-simple-separation} is adequate, and was briefly employed in \grwb, but it has a practical disadvantage: By using the Euclidean norm on $\mathbb{R}^n$, it effectively assigns equal importance to each of the coordinates.  This is not ideal for some common coordinate systems.  For example, consider, in the cylindrical coordinate system  $(t, r, \phi, z)$ of \eqref{E:differential-geometry-example-chart}, the point $p = (0, 10^4, 0, 0)$.  Then the two points $p_{+r} = (0, 10^4 + 1, 0, 0)$ and $p_{+\phi} = (0, 10^4, 1, 0)$ are equidistant from $p$ in the sense of \eqref{E:num-exp-simple-separation}, but $p_{+r}$ is much closer than $p_{+\phi}$ to $p$ in the sense of the standard flat metric \eqref{E:differential-geometry-example-metric}, essentially because the coefficient of the $dr^2$ term in \eqref{E:differential-geometry-example-metric} is 1, whereas the coefficient of the $d\phi^2$ term is $r^2 = 10^8$.

                If the metric is diagonal, as above, then we can assign to each coordinate direction $x^i$ an approximate `importance' equal to the coefficient of $d{x^i}^2$ in the line element.  If the metric is not diagonal then we diagonalise it at $b$, using the methods of Section~\ref{S:num-exp-orthonormal-basis}, and express the coordinate displacement $\delta x^i$ in terms of the resulting orthonormal basis $B$ of $T_b$:\footnote{In \eqref{E:num-exp-displacement}, $B$ is the matrix whose columns are the components of the orthonormal basis of $T_b$ on the chart $C$.}
                \begin{equation}
                    \label{E:num-exp-displacement}
                    \delta x^i |_B = B^{-1} \delta x^i.
                \end{equation}
                Like $\delta x^i$, the coordinate displacement $\delta x^i |_B \in \mathbb{R}^n$ will depend on the chart $C$.  The value $\| \delta x^i |_B \|$ is, in general, a better definition than $\| \delta x^i \|$ for the amount by which $\text{geodesic} (v) (t)$ `misses' $b |_C$, because it accounts for the difference in importance of the various coordinate directions at $b$.

                We rewrite \eqref{E:num-exp-simple-separation}, using \eqref{E:num-exp-displacement}, as
                \begin{equation}
                    \label{E:num-exp-separation-final}
                    f (v) = \min_{t \in \mathbb{R}} \min_{\text{charts $C$}} \| \delta x^i |_B \|.
                \end{equation}
                Definition \eqref{E:num-exp-separation-final} is implemented in the routine \code{min_euclidean_separation}, which takes one argument of type \code{worldline}, and one argument of type \code{point}; it performs the minimisation of $f$ over the curve parameter $t$ using the one-dimensional minimisation routine \code{brent_minimiser}, of Section~\ref{S:numerical-minimisation}.

            \subsubsection{Parameterising the search space}

                We want to minimise the function $f (v)$, \eqref{E:num-exp-separation-final}, over the variable $v \in T_a$.  The tangent space $T_a$ has dimension $n$, but, as already noted, $f (\alpha v) = f (v)$ whenever $\alpha \neq 0$, and so the space to minimised over has dimension $n - 1$.

                In \grwb, the minimisation is performed in the following way:  The vector $v \in T_a$ is expressed in terms of its components $v^i \in \mathbb{R}^n$ in an orthonormal tangent basis $B$ (Section~\ref{S:num-exp-orthonormal-basis}).  Then, we minimise $f (v)$ with $v^i$ ranging over the unit sphere in $\mathbb{R}^n$, by parameterising the unit sphere by the $n - 1$ coordinates $(\theta_1, \ldots, \theta_{n - 1})$ using the generalised spherical polar coordinate transformation,
                \begin{align}
                    \label{E:num-exp-generalised-spherical}
                    v^1 &= \sin \theta_1, \notag \\
                    v^m &= \sin \theta_m \prod_{i = 1}^{m - 1} \cos \theta_i, \quad (1 < m < n), \notag \\
                    v^n &= \prod_{i = 1}^{n - 1} \cos \theta_i.
                \end{align}
                The multi-dimensional minimisation of $f (v)$ is thus performed over the $n - 1$ variables $(\theta_1, \ldots, \theta_{n - 1})$.

                If the determined minimum value of $f (v)$ is approximately equal to zero (in the sense of Section~\ref{S:numerical-relative-difference}), then the solution values $(\theta_1, \ldots, \theta_{n - 1})$ of the minimisation problem define, via \eqref{E:num-exp-generalised-spherical}, the components $v^i$ of $v$ in the orthonormal tangent basis $B$, which in turn defines the solution vector $v \in T_a$, which finally defines, with $a$, the initial data for a geodesic intersecting $a$ and $b$, as required.

            \subsubsection{Implementation in \grwb}
            
                The generalised spherical polar transformation \eqref{E:num-exp-generalised-spherical} is encoded in \grwb\ in the routines \code{from_polar} and \code{to_polar}, both of which have signature \code{nvector<double> (nvector<double>)}.  The routine \code{from_polar} encodes \eqref{E:num-exp-generalised-spherical}, and \code{to_polar} encodes the inverse transformation to \eqref{E:num-exp-generalised-spherical}.  The code for these routines is listed in Section~\ref{A:grwb-code-generalised-spherical}.

                The entire process of first solving the minimisation problem,
                \begin{equation}
                    \label{E:num-exp-min}
                    \min_{v \in T_a} f (v),
                \end{equation}
                by parameterising the space $T_a$ and minimising over the generalised spherical polar coordinates, and then constructing and returning the geodesic defined by the solution to \eqref{E:num-exp-min}, is encapsulated in the routine \code{connecting_geodesic} of \grwb, which has signature \code{optional<geodesic> (point, point)}.  The \code{optional} mechanism is employed because it may not be possible to find the connecting geodesic; for example, the numerical minimisation of \eqref{E:num-exp-min} may converge to a local, rather than a global, minimum, where $f (v) \neq 0$.  The code of \code{connecting_geodesic} is listed in Section~\ref{A:grwb-code-connecting-geodesic}.

                The \code{connecting_geodesic} routine uses the routines \code{to_polar} and \code{from_polar} to perform the generalised spherical polar coordinate transformation, and the functor class \code{powell_minimiser} of Section~\ref{S:numerical-powell} to perform the multi-dimensional minimisation.

                The minimisation class \code{powell_minimiser} requires an initial guess for the location of the minimum, around which it looks for an exact minimum; the guess supplied to \code{powell_minimiser} by \code{connecting_geodesic} is simply the coordinate difference between the two points $a$ and $b$ on some chart, transformed to the generalised spherical polar coordinates by the routine \code{to_polar}.  This guess is good if the space-time curvature between $a$ and $b$ is small.

        \subsection{Connecting null geodesics}
            \label{S:num-exp-connecting-null}
        
            Given any world-line $\lambda (s)$ and a nearby point $p$, there will be two null geodesics which connect $p$ with a point on $\lambda$, corresponding to the intersections of $\lambda$ with the the past and future null cones of $p$.  These null geodesics are important because, if $\lambda$ is the world-line of a physical observer, they represent the world-lines of photons travelling to the event $p$ from the observer, and from the event $p$ to the observer.  The determination of these null geodesics is the topic of this section.

            We solve the problem in a very similar way to the solution of the problem of Section~\ref{S:num-exp-connecting}, above:  We minimise, over all null vectors $v \in T_p$, the amount by which a geodesic launched from $p$ with tangent vector $v$ `misses' the world-line $\lambda$.  There are two important differences between the two problems:  We require a definition for the amount by which a curve misses another curve, analagous to the function $f$ of \eqref{E:num-exp-separation-final} which gives the amount by which a curve misses a point; and we only wish to minimise over null vectors in $T_p$, rather than all vectors in $T_p$.
        
            \subsubsection{\code{min_euclidean_separation} of two curves}
            
                We require a function $g \colon T_p \to \mathbb{R}$, analagous to $f$ of \eqref{E:num-exp-separation-final}, which we can minimise to find the tangent vector at $p$ of a null geodesic intersecting the point $p$ and the world-line $\lambda$.  We define $g$ in a similar way to $f$, as
                \begin{equation}
                    \label{E:num-exp-curve-curve-separation}
                    g (v) = \min_{s \in \mathbb{R}} \min_{t \in \mathbb{R}} \min_{\text{charts $C$}} \| \delta x^i |_B \|,
                \end{equation}
                where the quantity $\delta x^i |_B$ is defined in terms of the quantity $\delta x^i$ as in \eqref{E:num-exp-displacement}, using an orthonormal tangent basis $B$ at $\lambda (s)$, and $\delta x^i$ is redefined as
                \begin{equation}
                    \delta x^i = \text{geodesic} (v) (t) |_C - \lambda (s) |_C,
                \end{equation}
                so that it is now a function of $s$, as well as $t$ and $C$.
                
                We can summarise \eqref{E:num-exp-curve-curve-separation} as follows: the distance between two curves is defined as the closest they ever get, in the Euclidean norm, in the coordinates of any chart.  The definition \eqref{E:num-exp-curve-curve-separation} is encoded in a specialisation of the \grwb\ routine \code{min_euclidean_separation}, which takes two arguments of type \code{worldline}, representing the space-time curves; it performs the minimisation \eqref{E:num-exp-curve-curve-separation}, over the two real parameters $s$ and $t$, using the multi-dimensional minimisation routine \code{powell_minimiser} of Section~\ref{S:numerical-powell}.
            
            \subsubsection{Parameterisation of the null cone}
            
                We want to minimise the function $g (v)$, \eqref{E:num-exp-curve-curve-separation}, over the variable $v \in T_p$.  The null subspace of $T_p$ has dimension $n - 1$, and, as in Section~\ref{S:num-exp-connecting}, $g (\alpha v) = g (v)$ whenever $\alpha \neq 0$, and so the space to be minimised over has dimension $n - 2$.

                The search space is parameterised in a similar way to that of Section~\ref{S:num-exp-connecting}:  The vector $v \in T_p$ is expressed in terms of its components $v^i \in \mathbb{R}^n$ in an orthonormal tangent basis $B$ at $p$.  Let the first vector in the basis $B$ be the time-like eigenvector, and thus let the remaining eigenvectors be space-like.\footnote{In doing this, we implicitly assume that the metric is Lorentzian, which is the usual case for physical applications of \grwb.}  The component $v^1$ thus represents the `time-like part' of $v$, and the remaining $n - 1$ components $v^\beta$, $\beta = 2, \ldots, n$, represent the `space-like part' of $v$.  Now, given any values for the $v^\beta$, if we set
                \begin{equation}
                    \label{E:num-exp-null-space}
                    v^1 = \sqrt{\sum_{\beta = 2}^n |v^\beta|^2},
                \end{equation}
                then the vector $v$ defined by the components $v^i$ is null, since the tangent basis $B$ is orthonormal.  Thus, to restrict our minimisation to the null space of $T_p$, we minimise $g$ over the components $v^\beta$, and fix the remaining component $v^1$ using \eqref{E:num-exp-null-space}.

                We minimise $g (v)$ over the components $v^\beta \in \mathbb{R}^{n - 1}$ by using the generalised spherical polar coordinate transformation \eqref{E:num-exp-generalised-spherical} to obtain from the $v^\beta$ the coordinates $(\theta_1, \ldots, \theta_{n - 2})$, which parameterise the unit sphere in $\mathbb{R}^{n - 1}$, and then minimise $g (v)$ over the $n - 2$ variables $(\theta_1, \ldots, \theta_{n - 2})$.

                As in Section~\ref{S:num-exp-connecting}, if the minimum located value of $g (v)$ is approximately equal to zero, then the solution values $(\theta_1, \ldots, \theta_{n - 2})$ define the components $v^\beta$ via the generalised spherical polar coordinate transformation, and the $v^\beta$, together with \eqref{E:num-exp-null-space}, define the components $v^i$ of $v$ in the tangent basis $B$, which in turn define the solution vector $v \in T_p$, which finally, together with the point $p$, defines initial data for a solution geodesic intersecting both $p$ and $\lambda$.  The geodesic is guaranteed to be null, due to \eqref{E:num-exp-null-space}.

            \subsubsection{Implementation in \grwb}         
            
                The \grwb\ routine \code{connecting_null_geodesic} implements the process described above for minimising the function $g (v)$ over all $v$ in the null space of $T_p$, and constructing the resulting geodesic, using the \code{to_polar} and \code{from_polar} routines, and the \code{powell_minimiser} class.  The signature of \mbox{\code{connecting_null_geodesic}} is \code{optional<std::pair<double, geodesic>> (functional<optional<point> (double)>, point, double)}.  The code of \code{connecting_null_geodesic} is listed in Section~\ref{A:grwb-code-connecting-null-geodesic}.

                The first and second arguments represent $\lambda$ and $p$, respectively.  The third argument is an initial guess for the value of the parameter $s$ of the world-line $\lambda$, such that the connecting null geodesic will intersect $\lambda (s)$.  This third argument is necessary for two reasons:  There is otherwise no natural way for the \code{connecting_null_geodesic} routine to choose an initial guess for the values of the generalised spherical polar coordinates $(\theta_1, \ldots, \theta_{n - 2})$ to pass to \code{powell_minimiser}; and it permits a degree of control over which of the two possible connecting null geodesics (corresponding to either the backward or forward null cone of $T_p$) the \code{connecting_null_geodesic} routine will converge to.

                The return type of \code{connecting_null_geodesic}, \code{optional<std::pair<double, geodesic>>}, represents, in the first element of the \code{std::pair}, the parameter value $s$ of the curve $\lambda$ at which the null geodesic intersects $\lambda$; and in the second element of the \code{std::pair}, the null geodesic itself.  By convention, the null geodesic returned by the routine \code{connecting_null_geodesic} intersects the curve $\lambda$ at the parameter value 1.        

    \section{Conclusion}
    
        Various tools useful for the simulation of physical situations have been implemented in \grwb.  The tools are written within the functional framework of \grwb, allowing them to be easily interfaced with one-another to construct potentially complex physical models.  Algorithms for the determination of implicitly-defined geodesics, in particular, demonstrate the numerical solution of an important physical problem using the numerical methods of Chapter~\ref{C:numerical} and the differential geometric framework of Chapter~\ref{C:differential-geometry}. \clearemptydoublepage
    \chapter{Analysis of a recent claim}
    \label{C:karim}

    In this chapter we introduce and investigate a recent claim by Karim \etal\ \cite{R:karim} that the mass of the Milky Way can be determined using a small Michelson interferometer located on the surface of the Earth.  After summarising their calculation in Section~\ref{S:karim-summary}, we analyse consequences of the physical model employed by Karim \etal\ in Section~\ref{S:karim-analysis}.  An alternative model, argued to be the correct one on physical grounds, is proposed and investigated in Section~\ref{S:karim-geodesic-interferometer}.

    In Chapter~\ref{C:karim-grwb} we describe numerical experiments performed in \grwb\ using both models, and compare the results.  

    \section{Summary of the claim}
        \label{S:karim-summary}
        Employing a model metric of our galaxy, Karim \etal\ approximate the world-lines of the beam-splitter, end-mirrors, and connecting photons of an idealised Michelson interferometer located on the surface of the orbiting Earth.  The proper time elapsed at the beam-splitter between the departure and return of photons along each interferometer arm is computed. 

        The galaxy is modelled using a Kerr black hole metric.  In Boyer-Lindquist coordinates\footnote{See for example \cite{R:hawking-ellis}, page 161.} $(t, r, \theta, \phi)$, the Kerr metric takes the form
        \begin{equation}
            \label{E:karim-kerr}
            ds^2 = g_{tt} \, dt^2 + 2 g_{t\phi} \, dt \, d\phi + g_{rr} \, dr^2 + g_{\theta\theta} \, d\theta^2 + g_{\phi\phi} \, d\phi^2.
        \end{equation}
        The metric components $g_{ab}$ depend on two parameters, $m$ and $a$, which represent, respectively, the mass and specific angular momentum,\footnote{(angular momentum per unit mass)} as measured from infinity, of the field source.  Using the approximation employed by Karim \etal,  that $a$ is small compared to $m$, and that $m$ is small compared to the radius of the orbit of the Earth about the centre of the galaxy, the metric components are\footnote{Throughout, we use geometric units in which times are scaled by a factor $c$, and masses by a factor $G / c^2$, so that physical quantities are measured in powers of metres.  For example, angular momentum ($\text{kg} \, \text{m}^2 \, \text{s}^{-1}$) is measured in square metres.}
        \begin{align}
            \label{E:karim-low-a-kerr}
            g_{tt} &= \frac{a^2 \sin^2 \theta - \xi}{\rho^2} \simeq - (1 - 2 m / r), \notag \\
            g_{t\phi} &= - \frac{2 m a r \sin^2 \theta}{\rho^2} \simeq - \frac{2 m}{r} a \sin^2 \theta, \notag \\
            g_{rr} &= \frac{\rho^2}{\xi} \simeq \frac{1}{1 - 2 m / r}, \notag \\
            g_{\theta\theta} &= \rho^2 \simeq r^2, \notag \\
            g_{\phi\phi} &= \frac{(r^2 + a^2)^2 - \xi a \sin^2 \theta}{\rho^2} \simeq r^2 \sin^2 \theta,
        \end{align}
        where
        \[
            \xi = r^2 - 2 m r + a^2, \quad \rho^2 = r^2 + a^2 \cos^2 \theta.
        \]
        
        The world-line of the beam-splitter is modelled as a circular equatorial orbit about the centre of the galaxy: $r = R$, $\theta = \pi / 2$, and $\phi = \phi_0 + (v / R) t$, where $R$ is the coordinate distance of the beam-splitter from the field centre, $v$ is the coordinate speed of the beam-splitter, and $v / R$ is the corresponding angular coordinate speed.  The constant $\phi_0$ is chosen to be zero.

        Karim \etal\ compute light travel times, to go up and back an interferometer arm, for three possible orientations of the interferometer arm: inward-radially directed, positive-$\phi$ directed, and positive-$\theta$ directed.  Each arm is intended to have the same length $L$.

        The world-line of the end-mirror of the inward-radially directed arm (henceforth `radial arm') is approximated as a circular orbit inside that of the beam-splitter: $r = R - L$, $\theta = \pi / 2$, and $\phi = (v / R) t$.  The world-line of the end-mirror of the positive-$\phi$ directed arm (henceforth `$\phi$ arm') is approximated as a circular equatorial orbit which leads the beam-splitter in the $\phi$ direction by the angle $\Phi = L / R$: $r = R$, $\theta = \pi / 2$, and $\phi = \Phi + (v / R) t$.  The world-line of the end-mirror of the positive-$\theta$ directed arm (henceforth `$\theta$ arm') is approximated as differing from that of the beam-splitter only in the $\theta$ direction, again by the angle $\Phi$: $r = R$, $\theta = \pi / 2 + \Phi$, $\phi = (v / R) t$.

        The world-line of a photon travelling along an interferometer arm will in reality be a null geodesic which intersects the beam-splitter world-line, then intersects an end-mirror world-line, and finally intersects the beam-splitter world-line once again.  To simplify the analytic calculation, Karim \etal\ make the approximation that the coordinates $(r, \theta, \phi)$ are linearly related along each photon world-line.  The values of the remaining coordinate $t$ for each world-line are fixed by requiring the tangent vector to the world-line to be null ($ds = 0$ in \eqref{E:karim-kerr}).\footnote{The photon world-lines so defined, while null, will not, in general, be null geodesics.}

        Explicitly, for photons travelling along the radial arm (where $\theta = \pi / 2$ is constant by symmetry), $d\phi / dr$ is assumed to be constant; for photons travelling along the $\phi$ arm, $r$ and $\theta$ are assumed to be constant; and for photons travelling along the $\theta$ arm, $r$ and $d\phi / d\theta$ are assumed to be constant.

        To summarise, Karim \etal\ make the following assumptions and approximations:
        \begin{enumerate}
            \item Our galaxy is modelled by the Kerr black hole metric \eqref{E:karim-kerr} in the low angular-momentum approximation \eqref{E:karim-low-a-kerr}.
            \item The Michelson interferometer is modelled in terms of the Boyer-Lindquist coordinates as described above.
            \item Photon world-lines are approximated as null curves in which the coordinates $(r, \theta, \phi)$ are linearly related to one-another.
        \end{enumerate}

        \subsection{Main results of the claim}
            \label{S:karim-main-results}

            With the assumptions described above, Karim \etal\ solve for the coordinates of the arrival of a photon at the end-mirror, and for the  return of the reflected photon to the beam-splitter.  The $t$ coordinate of the return event, scaled by the factor $\sqrt{-g_{tt}}$, gives the proper time elapsed at the beam-splitter.  In terms of the dimensionless parameter $\mu \equiv 2 m / R$ and the coordinate speed $v$, Karim \etal\ find that the elapsed proper times for the radial, $\phi$, and $\theta$ arms are, respectively,
            \begin{align}
                \label{E:karim-travel-times}
                \tau_r &= 2 L \left[ 1 + \frac{1}{2} \mu - \frac{5}{8} \mu^2 - \frac{1}{2} v^2 + \cdots \right], \notag \\
                \tau_\phi &= 2 L \left[ 1 - \frac{1}{2} \mu^2 + \frac{1}{2} v^2 + \cdots \right], \notag \\
                \tau_\theta &= 2 L \left[ 1 - \frac{1}{2} v^2 + \cdots \right].
            \end{align}
            The $\cdots$ denote terms of higher order in $\mu$, $v$, and the parameter $\kappa \equiv a / R$.
            
            For interferometry, the measurable quantity is the light travel time difference between two arms.  Karim \etal\ find that
            \begin{align}
                \label{E:karim-time-differences}
                \delta \tau_{r\theta} &= \tau_r - \tau_\theta \simeq L \mu \left[ 1 - \frac{5}{4} \mu \right], \notag \\
                \delta \tau_{\phi\theta} &= \tau_\phi - \tau_\theta \simeq 2 L v^2,
            \end{align}
            and propose to determine $\mu$ (and hence the galactic mass $m$) by measuring the time differences \eqref{E:karim-time-differences}.
            
            Karim \etal\ estimate the order of magnitude of the effect \eqref{E:karim-time-differences} due to the Earth, Sun, and Milky Way, for an interferometer of length 10 cm.  The calculation is summarised in Table~\ref{T:karim-order-of-magnitude-estimate}.  The effect due to the Milky Way is found to be largest, with
            \begin{equation}
                \label{E:karim-time-difference-seconds}
                \delta \tau_{r\theta} \sim 6 \times 10^{-15} \text{ s}.
            \end{equation}
            Karim \etal\ conclude that the galactic mass can be determined by measuring $\delta \tau_{r\theta}$ with a small interferometer.
    
        \begin{table}
            \begin{center}
                \begin{tabular}{| l | c | c | c | c |}
                    \hline
                    field source & $2 m$ (m) & $R$ (m) & $\mu = 2 m / R$ & $\delta \tau_{r\theta}$ (s) \\
                    \hline
                    Milky Way & $\sim 10^{14}$ & $\sim 2.8 \times 10^{20}$ & $\sim 10^{-6}$ & $\sim 6 \times 10^{-15}$ \\
                    Sun & $\sim 10^3$ & $\sim 10^{11}$ & $\sim 10^{-8}$ & smaller \\
                    Earth & $\sim 10^{-2}$ & $\sim 6 \times 10^6$ & $\sim 10^{-8}$ & smaller \\
                    \hline
                \end{tabular}
                \caption{Order of magnitude estimates of \eqref{E:karim-time-differences} for various bodies with $L = 10$ cm, from \cite{R:karim}.}
                \label{T:karim-order-of-magnitude-estimate}
            \end{center}
        \end{table}

    \section{Theoretical analysis of the claim}
        \label{S:karim-analysis}
        
        We now investigate properties of the physical model employed by Karim \etalnodot.  In Section~\ref{S:karim-geodesic-interferometer} we propose an alternative interferometer model, and investigate its properties.

        The main result in \cite{R:karim}, upon which the proposed experiment depends, is the approximate light travel time difference \eqref{E:karim-time-differences}.  It is independent of $\kappa$ and hence independent of $a$, the specific angular momentum of the gravitational field source.  Thus, the result will be unchanged if the galaxy is instead modelled using a Schwarzschild black hole metric (setting $a = 0$ in \eqref{E:karim-low-a-kerr}).  In this case $g_{t\phi}$ vanishes, and the algebra is simplified.  We adopt this simpler model for the analytical calculations in Sections~\ref{SS:karim-coordinate-interferometer} and \ref{S:karim-geodesic-interferometer} and the numerical investigation of Chapter~\ref{C:karim-grwb}.

        In discussing why the predicted time difference $\delta \tau_{r\theta}$ is proportional to $\mu \propto 1 / R$, Karim \etal\ note that the proposed effect depends on the variation of the gravitational potential\footnote{Karim \etal\ in fact describe $2 m / R$ as the gravitational potential.  In any case, since $m / R \propto 2 m / R$, the line of reasoning is unchanged.} $m / R$ over the volume of the interferometer, and suggest that it is therefore reasonable to expect an effect proportional to this potential.  However, the variation of the potential over the volume of the interferometer will be approximately
        \begin{equation}
            \label{E:karim-potential-variation}
            L \frac{\partial}{\partial R} \frac{m}{R} = - L \frac{m}{R^2} \propto \frac{1}{R^2} \propto \mu^2.
        \end{equation}
        Thus, it would seem that we should instead expect $\delta \tau_{r\theta} \propto \mu^2$.

        \subsection{Properties of the coordinate-defined interferometer}
            \label{SS:karim-coordinate-interferometer}

            The interferometer of \cite{R:karim} is defined in terms of the Boyer-Lindquist coordinates $(t, r, \theta, \phi)$: The radial arm has coordinate length $L$ in the $r$ direction, and the $\theta$ and $\phi$ arms have coordinate length $\Phi = L / R$ in the positive $\theta$ and positive $\phi$ directions, respectively.  The justification for such a model is that, as $R / 2m \to \infty$, the metric components \eqref{E:karim-low-a-kerr} asymptote to those of the flat metric in spherical polar coordinates,
            \begin{align}
                \label{E:karim-flat-spherical-polar}
                - g_{tt} &= g_{rr} = 1, \notag \\
                g_{t\phi} &= 0, \notag \\
                g_{\theta\theta} &= r^2, \notag \\
                g_{\phi\phi} &= r^2 \sin^2 \theta,
            \end{align}
            and in that metric all of the arms of the coordinate-defined interferometer would have proper length $L$.

            Since $g_{\theta\theta}$ and $g_{\phi\phi}$ in \eqref{E:karim-low-a-kerr} are equal to those in \eqref{E:karim-flat-spherical-polar}, the $\theta$ and $\phi$ arms of the coordinate-defined interferometer have proper length $L$.  On the other hand, since $g_{rr}$ in \eqref{E:karim-low-a-kerr} differs from that in \eqref{E:karim-flat-spherical-polar}, the radial arm of the coordinate-defined interferometer does not have proper length $L$.  In fact, the proper length $s$ of the radial arm is
            \begin{align}
                s &= \int_{R - L}^R \sqrt{g_{rr}} \, dr = \int_{R - L}^R \frac{1}{\sqrt{1 - 2 m / r}} \, dr \notag \\
                &\simeq \int_{R - L}^R \left( 1 + \frac{1}{2} \frac{2 m}{r} \right) \, dr \notag \\
                &= L - \frac{1}{2} 2 m \ln \frac{R - L}{R} \notag \\
                &\simeq L + \frac{1}{2} 2m \frac{L}{R} \notag \\
                &= L + \frac{1}{2} L \mu.
            \end{align}
            
            \subsubsection{Consequences of model}
            
                The proper length of the radial arm differs from $L$ by an amount proportional to $\mu$.  The estimated time difference $\delta \tau_{r\theta}$ is also proportional to $\mu$.  This raises the possibility that the calculated value for $\delta \tau_{r\theta}$ is due, at least in part, to the proper length difference between the $r$ and $\theta$ arms of the coordinate-defined interferometer.
    
                The total difference in proper length along and back each arm is $2 (s - L) \simeq L \mu$.  From \eqref{E:karim-time-differences}, the lowest order term in $\delta \tau_{r\theta}$ is also $L \mu$.  This is exactly the time difference expected for an interferometer in flat space, with arms of differing proper lengths $s$ and $L$.  We therefore conclude that the largest term in $\delta \tau_{r\theta}$, proportional to $\mu$, is entirely due to the difference in proper lengths between the $r$ and $\theta$ arms of the coordinate-defined interferometer, and not to space-time curvature.
    
                Note that it does not follow from the above argument that there is no term proportional to $\mu$ in the true physical value of $\delta \tau_{r\theta}$; it merely shows that, in the analysis of \cite{R:karim}, the term proportional to $\mu$ is an artifact of the coordinate-dependent manner in which the interferometer is defined.  Due to \eqref{E:karim-potential-variation}, however, we have good reason to believe that the lowest-order term in $\delta \tau_{r\theta}$ is proportional to $\mu^2$, and not to $\mu$.
        
    \section{Geodesic-defined interferometer}
        \label{S:karim-geodesic-interferometer}

        The problems resulting from the coordinate-dependent interferometer model of Karim \etal\ suggest that we should look for a coordinate-\emph{independent} model; we develop such a model in this section.  Its properties are explored in Sections~\ref{SS:karim-geodesic-interferometer-comparison} and \ref{SS:karim-geodesic-interferometer-estimate}.  Along with the original model of Karim \etal, this alternative model is investigated numerically using \grwb\ in Chapter~\ref{C:karim-grwb}.

        \subsection{Definition}
            \label{S:karim-geodesic-definition}

            We begin by specifying the world-line of the beam-splitter in the same way as Karim \etal: $r = R$, $\theta = \pi / 2$, and $\phi = (v / R) t$.  Since this world-line will not, in general, be a geodesic,\footnote{For each value of $v$ there will be one value of $R$ such that the world-line of the beam-splitter is, in fact, a circular equatorial geodesic.} it models an accelerating interferometer, rather than a freely-falling one.
    
            When deciding how to model the world-lines of the end-mirrors of each interferometer arm, the most obvious requirement is that the arms have length $L$.  While the proper distance between two nearby \emph{points} in a space-time may be defined as the proper length of the unique geodesic connecting them, it is a consequence of special relativity that there is no such \emph{observer-independent} definition of the distance between two nearby \emph{world-lines}.  There is, however, a natural choice for a preferred observer: the beam-splitter, since proper time along the world-line of the beam-splitter is the physical quantity to be measured.
    
            With respect to a preferred observer, we can define the property of simultaneity of two events.\footnote{For discussion regarding this definition of simultaneity see \cite{R:de-felice-clark}, pages 274--280.}  Let $\mathbf{b} (\tau)$ be the world-line of the beam-splitter, where $\tau$ is the proper time on $\mathbf{b}$, let $p$ be a point on $\mathbf{b}$, let $T_p$ be the tangent space of $p$, and let $\lambda_0 \in T_p$ be the tangent vector to $\mathbf{b} (\tau)$ at $p$.  The vector $\lambda_0$ is the time direction of the beam-splitter at $p$.  Let $S_p$ be the space-like subspace of $T_p$ orthogonal to $\lambda_0$.  The vectors in $S_p$ are the space directions of the beam-splitter at $p$.  An event $q$ not on $\mathbf{b}$ is \emph{simultaneous} with $p$ if the unique geodesic connecting $p$ and $q$ has tangent $v \in S_p$ at $p$.  That is, $q$ is simultaneous to $p$ if it is reachable from $p$ by a (space-like) geodesic orthogonal to $\mathbf{b}$.
    
            If $q$ is simultaneous to $p \in \mathbf{b}$ then the distance between $p$ and $q$ is defined as the proper length of the space-like geodesic connecting them.
    
            Using the above definitions, we can construct an end-mirror world-line which is always a distance $L$ from the beam-splitter.  At each point $p \in \mathbf{b}$ take a geodesic through $p$ whose tangent vector $\lambda_1$ is orthogonal to $\lambda_0$, and trace it out to proper length $L$.  The end-point of this geodesic segment defines a point on the world-line of the end-mirror.  To construct a second interferometer arm, take a second geodesic through $p$ whose tangent vector $\lambda_2$ is orthogonal to both $\lambda_0$ and $\lambda_1$, and trace it out to proper length $L$, defining the end-point as a point on the world-line of the second end-mirror.  A third interferometer arm can be similarly constructed by taking a third vector $\lambda_3$ which is orthogonal to $\lambda_0$, $\lambda_1$, and $\lambda_2$.
    
            Since the vectors $(\lambda_1, \lambda_2, \lambda_3)$ must be chosen for each $p = \mathbf{b} (\tau)$, they are functions of $\tau$.  We require $(\lambda_1, \lambda_2, \lambda_3)$ to satisfy the following condition: each vector must vary continuously\footnote{The components of each vector must be continuous in any coordinate system.} with $\tau$.  This ensures that the orientation of the interferometer does not vary discontinuously.
    
            The vector $\lambda_0$ is fixed by our choice for the world-line $\mathbf{b} (\tau)$ of the beam-splitter, and $g (\lambda_0, \lambda_0) = - 1$.  We then choose $\lambda_1$, $\lambda_2$, and $\lambda_3$ to model as closely as possible the same physical situation as Karim \etal:\footnote{The shorthand notation $\partial_{x^i}$ represents the coordinate basis vector $\partial / \partial x^i$.}
            \begin{align}
                \label{E:karim-geodesic-arm-directions}
                \lambda_0 &\propto \partial_t + \frac{v}{R} \partial_\phi, \notag \\
                \lambda_1 &= - \partial_r, \notag \\
                \lambda_2 &= \partial_\theta, \notag \\
                \lambda_3 &= \partial_\phi + g (\lambda_0, \partial_\phi) \lambda_0.
            \end{align}
            To see that this set is orthogonal, observe that in the Kerr space-time $(\partial_t, \partial_r, \partial_\theta)$ are mutually orthogonal, as are $(\partial_\phi, \partial_r, \partial_\theta)$, while
            \begin{align}
                g (\lambda_0, \lambda_3) &= g (\lambda_0, \partial_\phi + g (\lambda_0, \partial_\phi) \lambda_0) \notag \\
                &= g (\lambda_0, \partial_\phi) + g (\lambda_0, \partial_\phi) g (\lambda_0, \lambda_0) \notag \\
                &= g (\lambda_0, \partial_\phi) - g (\lambda_0, \partial_\phi) \notag \\
                &= 0.
            \end{align}
            Note that in the Schwarzschild space-time $\partial_t$ is orthogonal to $\partial_\phi$, and so if $v = 0$ then $\lambda_0$ is orthogonal to $\partial_\phi$ and thus $\lambda_3 = \partial_\phi$.
    
            It remains to specify the world-lines of the photons connecting the beam-splitter to the end-mirrors.  Now, from Section~\ref{S:num-exp-connecting-null}, given any world-line $\mathbf{c}$ and a nearby point $p$, there will be two null geodesics which connect $p$ with a point on $\mathbf{c}$, corresponding to the intersections of $\mathbf{c}$ with the the past and future null cones of $p$.  Thus, for each interferometer arm, we let the world-line of the outgoing photon be the (locally unique) future directed null geodesic joining the origin event $O$ to some point $q$ on the world-line of the end-mirror, and we let the world-line of the returning photon be the future directed null geodesic joining $q$ to some point $r$ on $\mathbf{b}$.
    
            For each arm, the proper length of $\mathbf{b}$ between the origin event $O$ and the return event $r$ is the time experienced by the beam-splitter between the departure and return of a photon travelling along that arm.  The point $r$ will in general be different for each interferometer arm, and the proper length along $\mathbf{b}$ between two such points gives the measurable light travel time difference between the corresponding interferometer arms: $\delta \tau_{r\theta}$, $\delta \tau_{r\phi}$, or $\delta \tau_{\theta\phi}$.

        \subsection{Comparison with the coordinate-defined interferometer}
            \label{SS:karim-geodesic-interferometer-comparison}

            \noindent The geodesic-defined interferometer has the following properties:
            \begin{enumerate}
                \item \label{I:karim-length} The arms are of proper length $L$.
                \item \label{I:karim-straight} The arms are straight, in the sense of a geodesic being the straightest possible line in a curved space.
                    \item \label{I:karim-orthogonal} At their point of intersection, the arms are orthogonal to:
                        \begin{enumerate}
                            \item \label{I:karim-orthogonal-arms} one-another;
                            \item \label{I:karim-orthogonal-world-line} the world-line of the beam-splitter.
                        \end{enumerate}
            \end{enumerate}
            We have seen in Sections~\ref{S:karim-summary} and \ref{S:karim-analysis} that properties \ref{I:karim-length} and \ref{I:karim-straight} are not shared by the coordinate-defined interferometer of Karim \etal.

            Property \ref{I:karim-orthogonal-arms} \emph{is} shared by the coordinate-defined interferometer, because the tangent vectors to the arms are $\partial_r$, $\partial_\theta$, and $\partial_\phi$, which are mutually orthogonal.  Property \ref{I:karim-orthogonal-world-line} does not hold in general because, when $v \neq 0$, the tangent vector to the world-line of the beam-splitter (equal to $\lambda_0$, above) is not orthogonal to $\partial_\phi$; and because in the Kerr space-time $\partial_t$ is not orthogonal to $\partial_\phi$.  In the special case of the Schwarzschild space-time with $v = 0$, property \ref{I:karim-orthogonal-world-line} does hold for the coordinate-defined interferometer.

        \subsection{Estimate of light travel time}
            \label{SS:karim-geodesic-interferometer-estimate}

            In this section we estimate $\tau_r$ for the geodesic-defined interferometer, for the simplest case of $v = 0$ in the Schwarzschild space-time.  We will find that the result differs from $2 L$ by an amount proportional to $\mu^2$, in contrast to the corresponding result \eqref{E:karim-travel-times} for the coordinate-defined interferometer.

            From symmetry it follows that the world-line of the outgoing radial light ray has constant $\theta = \pi / 2$ and constant $\phi$.  Since the world-line is null, $ds = 0$ along it.  Thus, from \eqref{E:karim-kerr} and \eqref{E:karim-low-a-kerr}, with $a = 0$,
            \begin{equation}
                \label{E:karim-estimate-metric}
                0 = g_{tt} \, dt^2 + g_{rr} \, dr^2,
            \end{equation}
            where
            \[
                g_{tt} = - (1 - 2 m / r), \quad g_{rr} = \frac{1}{1 - 2 m / r}.
            \]

            Let the time coordinate of the photon leaving the beam-splitter be $t = 0$, and let the time coordinate of the photon reflecting at the mirror be $t = t_\text{r}$.  Then, since the space-time is static and time-reversible,\footnote{The Schwarzschild space-time is \emph{static} because the metric is independent of $t$, and \emph{time-reversible} because it is invariant under the exchange $t \to -t$, $dt \to - dt$.  The Kerr space-time is thus static but not time-reversible.} the time coordinate of the return of the photon to the beam-splitter is
            \begin{equation}
                t = 2 t_\text{r}, 
            \end{equation}
            in terms of which
            \begin{equation}
                \label{E:karim-proper-time-estimate}
                \tau_r = 2 t_\text{r} \sqrt{- g_{tt}}.
            \end{equation}
            From \eqref{E:karim-estimate-metric} we have
            \begin{equation}
                \label{E:karim-time-integral}
                t_\text{r} = \int_{R - \Delta}^R \sqrt{- \frac{g_{rr}}{g_{tt}}} \, dr,
            \end{equation}
            where $\Delta$ is the coordinate distance on the $r$ axis corresponding to a proper length $L$.
            
            \subsubsection{Relation between coordinate length and proper length}

                To find an expression for $\Delta$ in terms of $L$, we first find $L$ in terms of $\Delta$:
                \begin{equation}
                    L = \int_{R - \Delta}^R \sqrt{g_{rr}} \, dr = \int_{R - \Delta}^R \frac{1}{\sqrt{1 - 2 m / r}} \, dr.
                \end{equation}
                The solution to this integral can be expressed in closed form, but we only require the first few terms in $\Delta_* = \Delta / 2 m$.  Using \mathematica\ we obtain\footnote{\mathematica\ input: \code[Mathematica]{Simplify[Series[Integrate[1 / Sqrt[1 - 1 / r], \{r, R - Delta, R\}], \{Delta, 0, 3\}]]}}
                \begin{align}
                    \label{E:karim-coordinate-length-series}
                    L_* = \sqrt{\frac{R_*}{R_* - 1}} \Delta_* &+ \frac{1}{4 \sqrt{R_*} (R_* - 1)^{3/2}} \Delta_*^2 \notag \\
                    &{} + \frac{4 R_* - 1}{24 R_*^{3/2} (R_* - 1)^{5/2}} \Delta_*^3 + \cdots,
                \end{align}
                where $R_* = R / 2 m = 1 / \mu$ and $L_* = L / 2 m$.  We can invert\footnote{The general process of finding a series which is the inverse function of another series is called \emph{series inversion} or \emph{series reversion}.} this series to obtain a series for $\Delta_*$ in terms of $L_*$.  We begin by rewriting \eqref{E:karim-coordinate-length-series} as
                \begin{equation}
                    \label{E:karim-coordinate-length-series-simpler}
                    L_* = a_1 \Delta_* + a_2 \Delta_*^2 + a_3 \Delta_*^3 + \cdots,
                \end{equation}
                and then writing a general series for $\Delta_*$ in terms of $L_*$:
                \begin{equation}
                    \label{E:karim-proper-length-series}
                    \Delta_* = b_1 L_* + b_2 L_*^2 + b_3 L_*^3 + \cdots.
                \end{equation}
                Substituting \eqref{E:karim-coordinate-length-series-simpler} into \eqref{E:karim-proper-length-series}, equating powers of $\Delta_*$, and solving for the $b_i$ yields\footnote{For a formula for the general coefficient $b_n$, see, for example, \cite{R:morse-feshbach}, page 412.}
                \begin{align}
                    b_1 &= \frac{1}{a_1}, \notag \\
                    b_2 &= - \frac{a_2}{a_1^3}, \notag \\
                    b_3 &= \frac{2 a_2^2 - a_1 a_3}{a_1^5}.
                \end{align}
                The series for $\Delta_*$ in terms of $L_*$ is thus
                \begin{equation}
                    \label{E:karim-proper-length-series-finished}
                    \Delta_* = \sqrt{\frac{R_* - 1}{R_*}} L_* - \frac{1}{4 R_*^2} L_*^2 - \frac{1}{6 R_*^3} \sqrt{\frac{R_* - 1}{R_*}} L_*^3 + \cdots.
                \end{equation}
            
            \subsubsection{Solution}
            
                We evaluate the integral \eqref{E:karim-time-integral} for $t_{\text{r}*} = t_\text{r} / 2 m$ using the reduced variable $r_* = r / 2 m$:
                \begin{equation}
                    \label{E:karim-time-integral-evaluated}
                    t_{\text{r}*} = \int_{R_* - \Delta_*}^{R_*} \frac{1}{1 - 1 / r_*} \, dr_* = \Delta_* + \ln{\frac{R_* - 1}{R_* - 1 - \Delta_*}}.
                \end{equation}
                Substituting \eqref{E:karim-proper-length-series-finished} and \eqref{E:karim-time-integral-evaluated} into \eqref{E:karim-proper-time-estimate} and expanding in powers of $L_*$ and $1 / R_*$  yields, after some simplification,
                \begin{align}
                    \label{E:karim-proper-time-estimate-dimensionless}
                    \frac{\tau_r}{2 m} &= 2 L_* - \frac{L_*^2}{2 R_*^2} + \frac{L_*^3 / 3 - L_*^2 / 4}{R_*^3} + \cdots \notag \\
                    {} &= 2 L_* - \frac{\mu^2}{2} L_*^2 + \cdots,
                \end{align}
                or
                \begin{equation}
                    \label{E:karim-proper-time-estimate-finished}
                    \tau_r = 2 L - \frac{m L^2}{R^2} + \cdots.
                \end{equation}
                
                Thus, $\tau_r$ differs from $2 L$ by an amount proportional to $\mu^2$, in agreement with the argument of \eqref{E:karim-potential-variation}.  \eqref{E:karim-proper-time-estimate-finished}, along with \eqref{E:karim-travel-times}, will also be useful in validating the numerical analysis of Chapter~\ref{C:karim-grwb}.

                The quantity $m L^2 / R^2$ in \eqref{E:karim-proper-time-estimate-finished} differs from the corresponding quantity $2 m L / R$ from the analysis of Karim \etal\ \eqref{E:karim-travel-times} by a factor of $L / 2 R$.  For $L = 1$\,m and $R = 8$\,kpc,\footnote{The estimate for $R$ is taken from \cite{R:carroll-ostlie}, page 917.} we have $L / 2 R \simeq 2 \times 10^{-21}$.  Thus we might expect a change in the time difference estimate \eqref{E:karim-time-difference-seconds} of roughly a factor of $10^{-21}$, so that $\delta \tau_{r\theta} \sim 10^{-35}$ s, which is too small to detect with current methods.  An accurate estimate of the time difference $\delta \tau_{r\theta}$ for the geodesic-defined interferometer is obtained numerically in Chapter~\ref{C:karim-grwb}.

    \section{Intermission}

        Because it is defined explicitly and simply in terms of the Boyer-Lindquist coordinates $(t, r, \theta, \phi)$, the coordinate-defined interferometer of Karim \etal\ is more susceptible to analytic methods than the geodesic-defined interferometer of Section~\ref{S:karim-geodesic-interferometer}.  Nonetheless, to keep the algebra manageable, various approximations were necessarily employed in \cite{R:karim}.  In particular, by approximating null geodesics as null curves in which the coordinates $(r, \theta, \phi)$ are linearly related, Karim \etal\ completely avoid the geodesic equation in their analysis.\footnote{Similarly, the analysis of Section~\ref{SS:karim-geodesic-interferometer-estimate} was relatively simple because the radial geodesics were easily found via the symmetries present in the special case $v = a = 0$.}

        The geodesic-defined interferometer, on the other hand, is defined explicitly terms of space-like geodesics, and so an analysis of it akin to that of \cite{R:karim} would be even more complicated.  We do, however, have the methods of Chapter~\ref{C:numerical-experiments} at our disposal.  In Chapter~\ref{C:karim-grwb} we directly simulate both interferometers, bypassing the algebraic complexities of the metric and the geodesic equation.  By performing a range of numerical experiments, we can characterise the behaviour of both interferometers in terms of the parameters $R$, $L$, and $v$. 
\clearemptydoublepage
    \chapter{Numerical investigation of the claim}
    \label{C:karim-grwb}

     Using the methods of Chapter~\ref{C:numerical-experiments}, the coordinate-defined interferometer of Karim \etal\ and the geodesic-defined interferometer of Section~\ref{S:karim-geodesic-interferometer} were simulated in \grwb, in the Schwarzschild space-time.  In Section~\ref{S:karim-grwb-method} the modelling of the interferometers in \grwb\ is described.  The results of the numerical experiments are presented in Section~\ref{S:karim-grwb-results}.  In Section~\ref{S:karim-grwb-estimate-physical} the results for the geodesic-defined interferometer are used to obtain a new estimate for the size of the predicted effect on Earth due to the Milky Way.  Conclusions are drawn in Section~\ref{S:karim-grwb-conclusion}.

     The motivation for the experiments was twofold:  Under the assumption that the geodesic-defined interferometer is more physically realistic than the coordinate-defined interferometer of \cite{R:karim}, we aimed to obtain a new estimate on the size of the effect $\delta \tau_{r\theta}$, in order to determine whether the Milky Way can in fact be weighed with a small interferometer on Earth; and we aimed to verify the analysis of the coordinate-defined interferometer made in \cite{R:karim}.  By directly simulating the coordinate-defined interferometer, we can bypass the approximations necessary in an analytic argument, including the approximation of light rays as certain (non-geodesic) null curves, and thus determine the extent to which those approximations affected the final result of Karim \etalnodot.
    
    \section{Modelling the interferometers}
        \label{S:karim-grwb-method}

        In this section we describe how to simulate the two interferometer models defined in Chapter~\ref{C:karim}, using the tools for numerical experimentation described in Chapter~\ref{C:numerical-experiments}.  The coordinate-defined interferometer of Karim \etal\ is constructed in terms of straight lines in coordinate space, using the \code{coordinate_line} tool of Section~\ref{S:num-exp-coordinate-lines}, while the geodesic-defined interferometer also makes use of the \code{geodesic} functor class of Section~\ref{S:num-exp-geodesics}.  For both interferometers, null geodesics, representing photon world-lines, are determined using the implicit methods of Section~\ref{S:num-exp-implicit}.

        Each interferometer model depends on the three parameters $R$, $L$, and $v$, corresponding, respectively, to the coordinate distance of the beam-splitter from the field centre, the interferometer arm length, and the coordinate speed of the beam-splitter (Section~\ref{S:karim-summary}).  The important physical quantities obtained from each simulation are the light travel time differences $\delta \tau_{r\theta}$, $\delta \tau_{r\phi}$, and $\delta \tau_{\theta\phi}$, which are arc lengths along the world-line of the beam-splitter.  By simulating each interferometer model for a wide range of values of $R$, $L$, and $v$, the effect of each parameter on the light travel time differences can be characterised.

        \subsection{Beam-splitter world-line}
            \label{S:karim-grwb-beam-splitter-world-line}

            For both interferometer models, the world-line of the beam-splitter is modelled as a circular equatorial orbit, which is a straight line in the Boyer-Lindquist coordinates $(t, r, \theta, \phi)$.  The world-line satisfies (Section~\ref{S:karim-summary})
            \begin{equation}
                \label{E:karim-grwb-beam-splitter-world-line}
                t = s, \quad r = R, \quad \theta = \pi / 2, \quad \phi = \phi_0 + (v / R) s,
            \end{equation}
            where $s$ is a curve parameter; the tangent vector to the curve \eqref{E:karim-grwb-beam-splitter-world-line} everywhere has the components $(1, 0, 0, v / R)$.  However, the parameter $s$ does not correspond to the proper time $\tau$ of the beam-splitter, because the vector $\lambda_0$ with components $\lambda_0^i = (1, 0, 0, v / R)$ does not satisfy $\text{metric} (\lambda_0, \lambda_0) = -1$.  We normalise $\lambda_0$ using the routine \code{normalise} of Section~\ref{S:num-exp-observers}, and use the resulting vector $u$ to construct a \code{coordinate_line} whose parameter is the proper time $\tau$.  The arbitrary constant $\phi_0$ is chosen to be $\pi / 2$.  Note that $\lambda_0$ as defined here is simply the $\lambda_0$ of \eqref{E:karim-geodesic-arm-directions}.

            The following code fragment demonstrates the construction of the beam-splitter world-line in \grwb:
            \begin{codeblock}
// construct the point representing the origin event
nvector<double> origin_coordinates = make_vector(0, R, half_pi, half_pi);
point origin(a, c, origin_coordinates);

// construct the world-line of the beam-splitter
nvector<double> coordinate_direction = make_vector(1, 0, 0, v / R);
tangent_vector beam_splitter_tangent = normalise(tangent_vector(origin, c, coordinate_direction));
worldline beam_splitter_worldline = coordinate_line(beam_splitter_tangent, c);
            \end{codeblock}
            The variable \code{c} is assumed to be of type \code{chart}, representing a chart which uses the Boyer-Lindquist coordinates, and the variable \code{a} is assumed to be of type \code{atlas}, representing the Schwarzschild space-time encoded in \grwb.  After execution of the code fragment, above, the beam-splitter world-line, represented by a function of type \code{worldline} (Section~\ref{S:differential-geometry-world-lines}), is stored in the variable \code{beam_splitter_worldline}, and the argument to the function \code{beam_splitter_worldline}, of type \code{double}, corresponds to the proper time of the beam-splitter.

            Note that the \code{coordinate_line} on the last line of the code fragment, above, is constructed from a \code{tangent_vector} and a \code{chart}; the information regarding the origin point is contained in the \code{context} routine of the \code{tangent_vector} class; see Section~\ref{S:differential-geometry-vectors}.

        \subsection{End-mirror world-lines}
            \label{S:karim-grwb-end-mirror-world-lines}

            Both interferometer models have all parts of the interferometer orbiting the field centre at a constant value of the $r$ coordinate.  Hence, the end-mirror world-lines, like the beam-splitter world-line, have tangent vectors whose components are proportional to $(1, 0, 0, v / R)$.  The only difference between the construction of an end-mirror world-line in \grwb, and the construction of the beam-splitter world-line in the code listing, above, will be the definition of the variable \code{origin} of type \code{point}.

            \subsubsection{Coordinate-defined interferometer}

                For the coordinate-defined interferometer, the origin events of the end-mirrors are defined simply in terms of the Boyer-Lindquist coordinates.  For the inward-radial arm, the origin event has coordinates $(0, R - L, \pi / 2, \pi / 2)$; for the positive-$\phi$ arm, the origin event has coordinates $(0, R, \pi / 2, \pi / 2 + L / R)$; and for the positive-$\theta$ arm, the origin event has coordinates $(0, R, \pi / 2 + L / R, \pi / 2)$.  The following code fragment demonstrates the construction of the end-mirror world-lines in \grwb:
                \begin{codeblock}
// (choose one of the following three lines)
nvector<double> mirror_origin_coordinates = make_vector(0, R - L, half_pi, half_pi); // inward-radial arm
nvector<double> mirror_origin_coordinates = make_vector(0, R, half_pi + L / R, half_pi); // positive-theta arm
nvector<double> mirror_origin_coordinates = make_vector(0, R, half_pi, half_pi + L / R); // positive-phi arm

// construct the point representing the origin event
point mirror_origin(a, c, mirror_origin_coordinates);

// construct the world-line of the end-mirror
nvector<double> coordinate_direction = make_vector(1, 0, 0, v / R);
tangent_vector mirror_tangent = normalise(tangent_vector(mirror_origin, c, coordinate_direction));
worldline mirror_worldline = coordinate_line(mirror_tangent, c);
                \end{codeblock}
                The only significant difference between this code fragment, and the code fragment demonstrating the construction of the beam-splitter world-line, above, is in the definition of the coordinates of the origin point.

            \subsubsection{Geodesic-defined interferometer}

                As described in Section~\ref{S:karim-geodesic-definition}, the origin events for the end-mirrors of the geodesic-defined interferometer are the end-points of space-like geodesics of length $L$ emanating from the origin event of the beam-splitter, and orthogonal to the world-line of the beam-splitter.  The tangent vectors of the space-like geodesics at the origin event are the mutually orthogonal vectors $\lambda_1$, $\lambda_2$, and $\lambda_3$, of \eqref{E:karim-geodesic-arm-directions}.

                The vectors $\lambda_1$, $\lambda_2$, and $\lambda_3$ are obtained from the coordinate basis vectors $\partial_r$, $\partial_\theta$, and $\partial_\phi$ by using the \code{orthonormalise} routine of Section~\ref{S:num-exp-observers}.  Specifically, $\lambda_1$ is defined as the orthonormalisation of $\partial_r$ with respect to the tangent $\lambda_0$ to the world-line of the beam-splitter; $\lambda_2$ is defined as the orthonormalisation of $\partial_\theta$ with respect to both $\lambda_0$ and $\lambda_1$ (obtained by two applications of \code{orthonormalise}); and $\lambda_3$ is defined as the orthonormalisation of $\partial_\phi$ with respect to $\lambda_0$, $\lambda_1$, and $\lambda_2$.  This process is equivalent to applying the Gram-Schmidt process (see for example \cite{R:lay}, page 399) to the vectors $\lambda_0$, $\partial_r$, $\partial_\theta$, and $\partial_\phi$, to obtain an orthonormal basis for the tangent space at the origin.

                The following code fragment demonstrates the construction of the end-mirror world-lines of the geodesic-defined interferometer in \grwb:
                \begin{codeblock}
// coordinate basis vectors
tangent_vector r (mirror_origin, c, make_vector(0., -1., 0., 0.));
tangent_vector theta (mirror_origin, c, make_vector(0., 0., 1., 0.));
tangent_vector phi (mirror_origin, c, make_vector(0., 0., 0., 1.));

// gram-schmidt process
tangent_vector radial_mirror_direction = orthonormalise(
    r, beam_splitter_tangent);
tangent_vector theta_mirror_direction = orthonormalise(orthonormalise(
    theta, beam_splitter_tangent), radial_mirror_direction);
tangent_vector phi_mirror_direction = orthonormalise(orthonormalise(orthonormalise(
    phi, beam_splitter_tangent),
    radial_mirror_direction), theta_mirror_direction);

// construct the space-like geodesic representing the interferometer arm
// (choose one of the following three lines)
worldline interferometer_arm = geodesic(r_mirror_direction);
worldline interferometer_arm = geodesic(theta_mirror_direction);
worldline interferometer_arm = geodesic(phi_mirror_direction);

// determine the point representing the origin event of the end-mirror
point mirror_origin = interferometer_arm(L);

// construct the world-line of the end-mirror
nvector<double> coordinate_direction = make_vector(1, 0, 0, v / R);
tangent_vector mirror_tangent = normalise(tangent_vector(mirror_origin, c, coordinate_direction));
worldline mirror_worldline = coordinate_line(mirror_tangent, c);
                \end{codeblock}
                The difference between this code fragment, and the corresponding code fragment for the construction of the coordinate-defined interferometer, is in the definition of the origin event for the end-mirror---the variable \code{mirror_origin}.  For the geodesic-defined interferometer, above, it is constructed in terms of a space-like geodesic from the \code{mirror_origin} event, whereas, for the coordinate-defined interfermeter, it was constructed explicitly in terms of the Boyer-Lindquist coordinates.                

        \subsection{Photon world-lines}
                  
            In Sections~\ref{S:karim-grwb-beam-splitter-world-line} and \ref{S:karim-grwb-end-mirror-world-lines}, the origin event \code{mirror_origin}, from which photons are emitted, and the end-mirror world-lines (\code{mirror_worldline} in the code fragment above), with which the photons must intersect, were defined.  This is sufficient information for the application of the method of Section~\ref{S:num-exp-connecting-null} to obtain null geodesics representing the world-lines of outgoing photons.

            Once the outgoing geodesics have been obtained, their points of intersection with the end-mirror world-lines define \code{reflection} events.  The \mbox{\code{reflection}} events, together with the beam-splitter world-line, \code{beam_splitter_worldline}, constitute sufficient information to again apply the method of Section~\ref{S:num-exp-connecting-null}, to obtain null geodesics representing the world-lines of ingoing photons.

            The points of intersection of the ingoing geodesics with the world-line of the beam-splitter will occur at various values of the world-line parameter $\tau$, the proper time of the beam-splitter.  The difference between these values of $\tau$ define the light travel time differences $\delta \tau_{r\theta}$, $\delta \tau_{r\phi}$, and $\delta \tau_{\theta\phi}$, which are the quantities to be obtained.

            The following code fragment demonstrates the application of the routine \code{connecting_null_geodesic} of Section~\ref{S:num-exp-connecting-null} to determine the light travel time for one interferometer arm:
            \begin{codeblock}
geodesic outward_ray = connecting_null_geodesic(beam_splitter_origin, mirror_worldline, L)->second;
point reflection = outward_ray(1);       
double light_travel_time = connecting_null_geodesic(reflection, beam_splitter_worldline, 2 * L)->first;
            \end{codeblock}
            In the first line, the routine \code{second} obtains the second element of the \code{std::pair<double, geodesic>} returned by the routine \code{connecting_null_geodesic} (see the end of Section~\ref{S:num-exp-connecting-null}).  In the second line, we make use of the convention that the null geodesic returned by \code{connecting_null_geodesic} intersects \code{mirror_worldline} at parameter value 1.  In the third line, the routine \mbox{\code{first}} obtains the first element of the \code{std::pair<double, geodesic>} returned by \code{connecting_null_geodesic}, which corresponds to the parameter $\tau$ of the world-line of the beam-splitter at which the ingoing photon arrives.

            Note that the third argument to \code{connecting_null_geodesic}, an initial guess for the parameter value of the curve at which the null geodesic will intersect, is chosen to be $L$ for the outgoing ray intersecting with the end-mirror world-line, and $2 L$ for the ingoing ray intersecting with the beam-splitter world-line.  These guesses correspond to the exact points of intersection for an interferometer in flat space, where the light travel time will be $L$ to reach the mirror, and $2 L$ to return to the beam-splitter; they are good guesses if the space-time curvature is small in the region of interest.
            
            \begin{figure}
                \begin{center}
                    \includegraphics[width = 12cm]{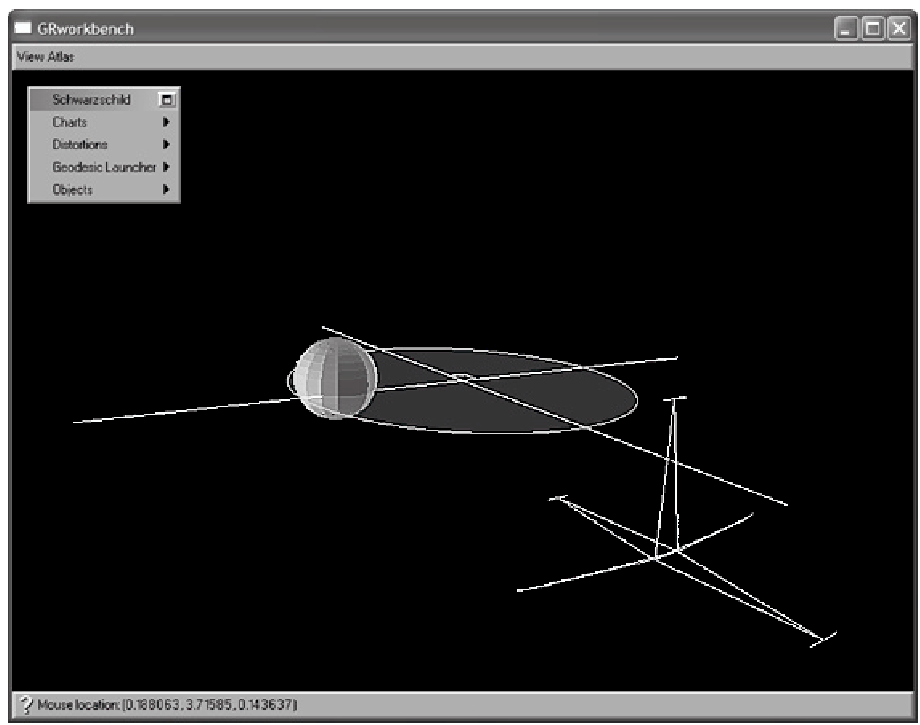}
                \end{center}
                \caption{The coordinate-defined interferometer with 5 orthogonal arms, simulated in \grwb.}
                \label{F:karim-grwb-interferometer}
            \end{figure}
            
            Figure~\ref{F:karim-grwb-interferometer} shows the coordinate-defined interferometer modelled in \grwb, as described in this section.  There are 5 interferometer arms: inward-radial, outward-radial, positive-$\phi$, negative-$\phi$, and positive-$\theta$.  (By symmetry, the negative $\theta$ arm has the same light travel time as the positive $\theta$ arm.)  The photon world-lines, determined by \code{connecting_null_geodesic}, are visible for both of the radial arms and the positive-$\theta$ arm.

    \section{Experiment}
        \label{S:karim-grwb-experiment}
        
        Using the methods of Section~\ref{S:karim-grwb-method}, we can simulate either the coordinate-defined interferometer of Karim \etal, or the geodesic-defined interferometer of Section~\ref{S:karim-geodesic-interferometer}, for any values of the parameters $R$, $L$, and $v$.  Because physical values of $L / R$ are smaller than $10^{-15}$, the precision of the \code{double} type in C++, it is not possible to directly simulate an interferometer on Earth under the influence of the galactic gravitational field.  However, by simulating the interferometer for a wide range of values of $R$, $L$, and $v$, the dependence of the light travel time difference on each parameter can be discovered, and the effect at Earth due to the galactic gravitational field can be predicted.

        Appendix~\ref{A:karim-code} lists the code of the numerical experiment performed in \grwb\ to characterise each of the interferometer models.  The simulation of the coordinate-defined interferometer is represented by the class \code{karim_interferometer}, and the simulation of the geodesic-defined interferometer is represented by the class \code{geodesic_interferometer}.  The \code{reflect} routine of each class performs the simulation of the corresponding interferometer; it takes three arguments of type \code{double}, representing the values of the dimensionless parameters $R_* = R / 2m$, $L_* = L / 2m$, and $v$, where $2 m$ is the Schwarzschild radius for a black hole of mass $m$.

        The \code{reflect} routine computes the light travel times $\tau_r$, $\tau_\theta$, and $\tau_\phi$, as described in Section~\ref{S:karim-grwb-method}, and takes their difference to form the travel time differences $\delta \tau_{r\theta}$, $\delta \tau_{r\phi}$, and $\delta \tau_{\theta\phi}$.  The computed travel time differences are in units of $2 m$.

        For each interferometer, 5 experiments were performed, with each experiment comprising many calls to \code{reflect}, that is, many simulations of the interferometer.  The 5 experiments were
        \begin{enumerate}
            \item \label{I:vary-r-zero-v} $v = 0$, $L_* = 1$, $3 \le R_* \le 50$, and
            \item \label{I:vary-r-nonzero-v} $v = 10^{-2}$, $L_* = 1$, $3 \le R_* \le 50$ (varying $R_*$);
            \item \label{I:vary-l-zero-v} $v = 0$, $R_* = 10$, $10^{-2} \le L_* \le 6$, and
            \item \label{I:vary-l-nonzero-v} $v = 10^{-2}$, $R_* = 10$, $10^{-2} \le L_* \le 6$ (varying $L_*$); and
            \item \label{I:vary-v} $R_* = 10$, $L_* = 1$, $10^{-3} \le v \le 0.5$ (varying $v$).            
        \end{enumerate}
        In the experiments, $R_*$ was varied over 17 values in a geometric progression starting with $R_* = 3$; $L_*$ was varied over 37 values in a geometric progression starting with $L_* = 10^{-2}$; and $v$ was varied over 37 values in a geometric progression starting with $v = 10^{-3}$.  Thus, each interferometer model was simulated for a total of 145 different sets of values for the parameters $R_*$, $L_*$, and $v$.

    \section{Results}
        \label{S:karim-grwb-results}
    
        In this section we present the results of the numerical experiments described in Section~\ref{S:karim-grwb-experiment}.
    
        \subsection{Validation}
            \label{S:karim-grwb-validation}
        
            An analytic calculation for the light travel time along the radial arm of the geodesic-defined interferometer, for the special case $v = 0$, was made in Section~\ref{SS:karim-geodesic-interferometer-estimate}, resulting in a power series expansion in $L_*$ and $R_*$ for the travel time $\tau_r$, \eqref{E:karim-proper-time-estimate-dimensionless}.  This travel time was compared with the values for $\tau_r$ obtained in the numerical experiments of Section~\ref{S:karim-grwb-experiment}, for various values of $R_*$ and $L_*$.  In all cases the numerical experiment results were found to agree with the analytic calculation in the first 8 or 9 significant figures.  The relative precision used by the \code{approx_equal} mechanism of Section~\ref{S:numerical-relative-difference} was $10^{-9}$ for the numerical experiments described in this chapter.

            The case $v = 0$ is not special from the point of view of the numerical differential geometric engine of \grwb.  It can thus be extrapolated that the light travel times determined by the numerical methods of this chapter when $v \neq 0$ are also as accurate as permitted by the relative precision of the numerical methods.
    
        \subsection{Varying orbital radius}
                    
            \begin{figure}
                \begin{center}
                    \includegraphics[width = 12 cm]{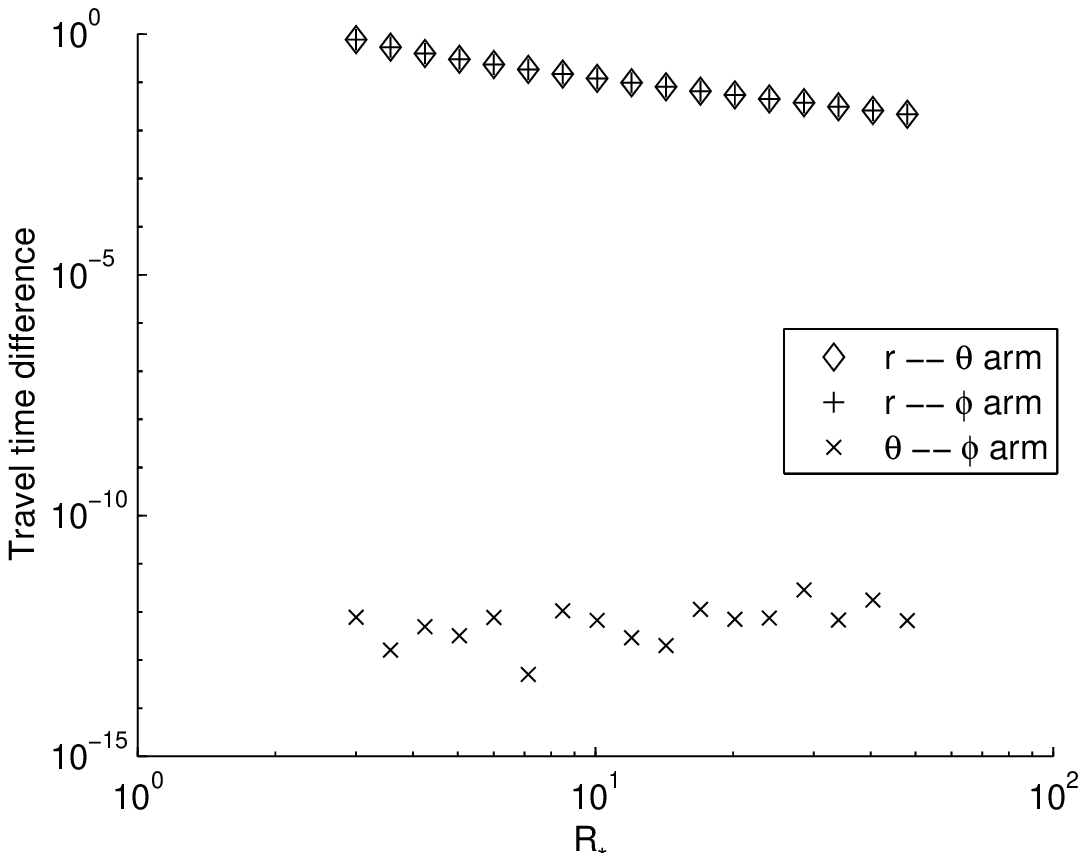}
                \end{center}
                \caption{Light travel time difference for the coordinate-defined interferometer, for various values of $R_*$, with fixed $L_* = 1$ and $v = 0$.}
                \label{F:karim-varying-r-zero-v}
            \end{figure}
               
            \begin{figure}
                \begin{center}
                    \includegraphics[width = 12 cm]{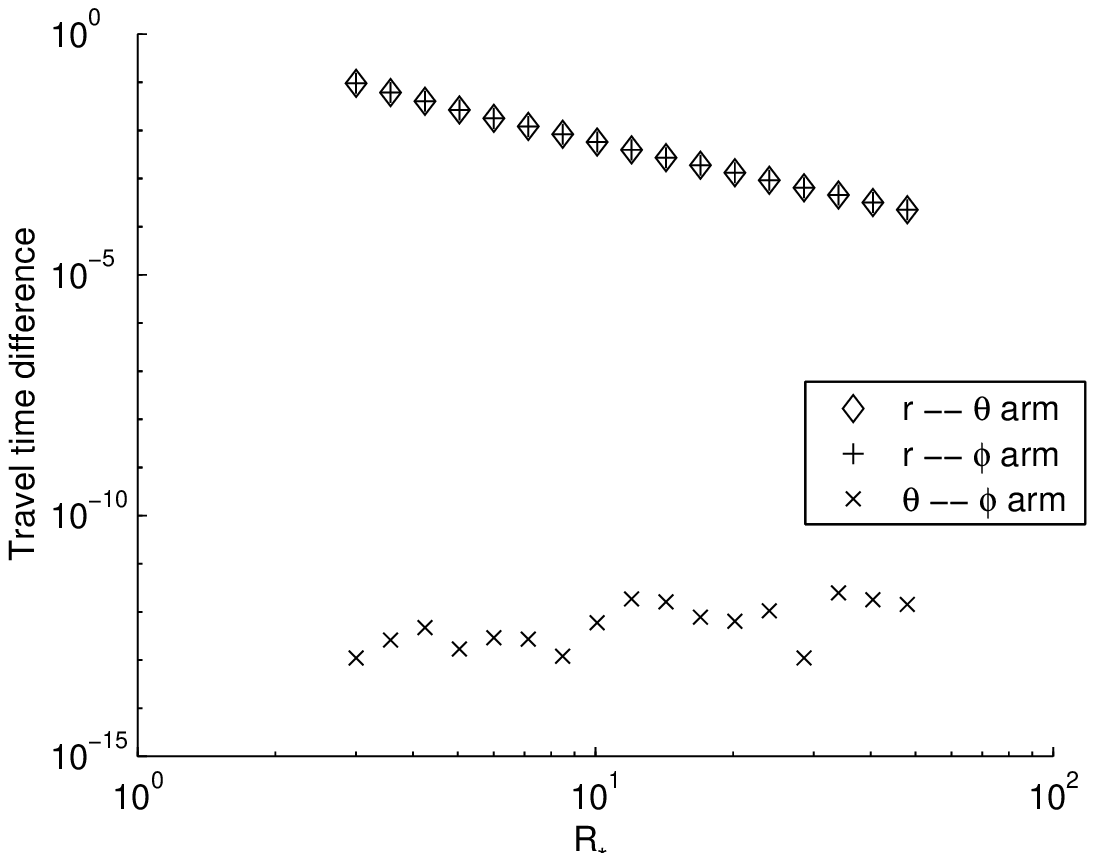}
                \end{center}
                \caption{Light travel time difference for the geodesic-defined interferometer, for various values of $R_*$, with fixed $L_* = 1$ and $v = 0$.}
                \label{F:geodesic-varying-r-zero-v}
            \end{figure}
            
            Figures~\ref{F:karim-varying-r-zero-v} and \ref{F:geodesic-varying-r-zero-v} show the light travel time differences for Experiment~\ref{I:vary-r-zero-v}, of Section~\ref{S:karim-grwb-experiment}, for the coordinate-defined interferometer and the geodesic-defined interferometer, respectively.  Note the logarithmic axes on these plots, and all plots in this section.

            In all figures in this section, three sets of data are plotted, corresponding to the light travel time differences between the three pairs of interferometer arms: $r$--$\theta$, $r$--$\phi$, and $\theta$--$\phi$.

            The data for the $r$--$\theta$ time difference coincides with the data for the $r$--$\phi$ time difference on Figures~\ref{F:karim-varying-r-zero-v} and \ref{F:geodesic-varying-r-zero-v} because, when $v = 0$, the $\phi$ and $\theta$ arms are equivalent, owing the spherical symmetry of the Schwarzschild space-time.

            The relative precision of the numerically determined light travel time differences is at best $10^{-9}$; we see from Figures~\ref{F:karim-varying-r-zero-v} and \ref{F:geodesic-varying-r-zero-v} that the $\theta$--$\phi$ time differences are well below the numerical precision limit---they are effectively zero.  This is to be expected because, since the $\theta$ and $\phi$ arms are equivalent when $v = 0$, the light travel time along them should be exactly the same (within the numerical precision).

            From Figure~\ref{F:karim-varying-r-zero-v}, for large values of $R_*$, the slope of the $r$--$\theta$ time difference data is very close to $-1$ on the logarithmic scale, corresponding to the travel time difference $\delta \tau_{r\theta}$ being proportional to $1 / R_*$ for the coordinate-defined interferometer.  This $1 / R_*$ scaling is in agreement with the calculation \eqref{E:karim-time-differences} of Karim \etal\ and, comparing the values of the $r$--$\theta$ data in Figure~\ref{F:karim-varying-r-zero-v} with the predicted travel time differences, \eqref{E:karim-time-differences} is found to be accurate to several significant figures.  Thus, the analysis of the coordinate-defined interferometer by Karim \etal\ is validated.

            For large values of $R_*$, the slope of the $r$--$\theta$ time difference data for the geodesic-defined interferometer (Figure~\ref{F:geodesic-varying-r-zero-v}) is found to be very close to $-2$ on the logarithmic scale, corresponding to the travel time difference $\delta \tau_{r\theta}$ being proportional to $1 / R_*^2$.  This is in agreement with the argument \eqref{E:karim-potential-variation} of Section~\ref{S:karim-analysis}.
    
            \begin{figure}
                \begin{center}
                    \includegraphics[width = 12 cm]{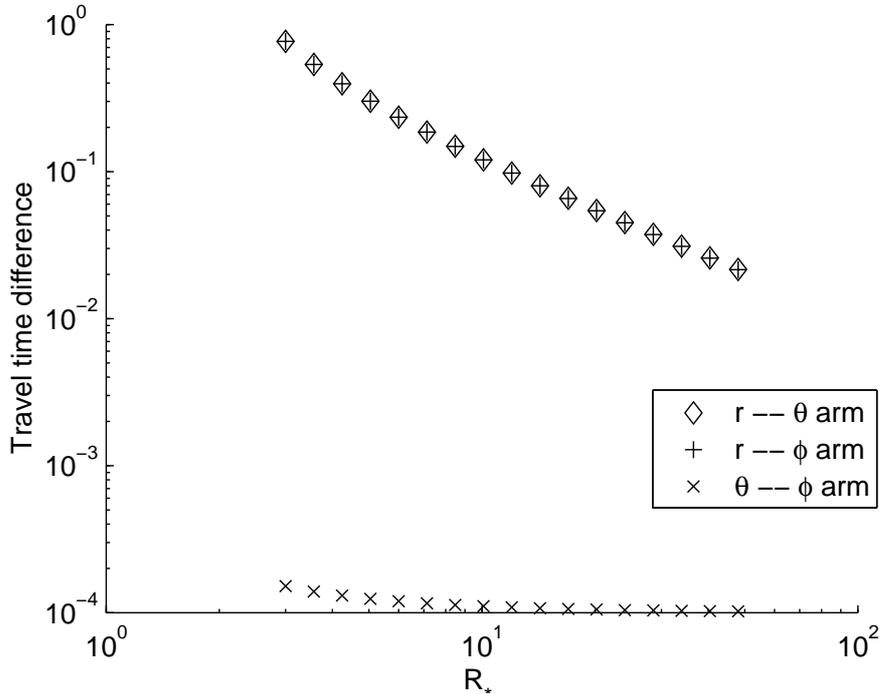}
                \end{center}
                \caption{Light travel time difference for the coordinate-defined interferometer, for various values of $R_*$, with fixed $L_* = 1$ and $v = 10^{-2}$.}
                \label{F:karim-varying-r-nonzero-v}
            \end{figure}
    
            \begin{figure}
                \begin{center}
                    \includegraphics[width = 12 cm]{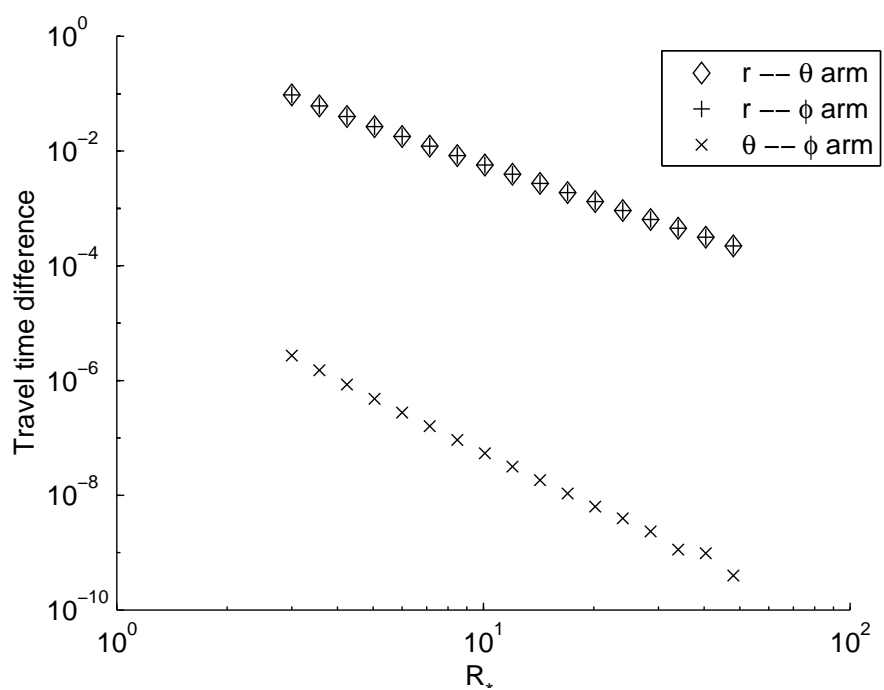}
                \end{center}
                \caption{Light travel time difference for the geodesic-defined interferometer, for various values of $R_*$, with fixed $L_* = 1$ and $v = 10^{-2}$.}
                \label{F:geodesic-varying-r-nonzero-v}
            \end{figure}
            
            Figures~\ref{F:karim-varying-r-nonzero-v} and \ref{F:geodesic-varying-r-nonzero-v} show the light travel time differences for Experiment~\ref{I:vary-r-nonzero-v}, for the coordinate-defined interferometer and the geodesic-defined interferometer, respectively.  The physical situation modelled in producing these plots differs from that of Figures~\ref{F:karim-varying-r-zero-v} and \ref{F:geodesic-varying-r-zero-v} only in the interferometer coordinate speed $v$ being non-zero for these plots.

            The $r$--$\theta$ data and the $r$--$\phi$ data of Figures~\ref{F:karim-varying-r-nonzero-v} and \ref{F:geodesic-varying-r-nonzero-v} do not differ significantly from the corresponding data for Figures~\ref{F:karim-varying-r-zero-v} and \ref{F:geodesic-varying-r-zero-v}, despite the non-zero interferometer coordinate speed.  In particular, the data for the $r$--$\theta$ time differences still coincides with data for the $r$--$\phi$ time differences, despite the two arms $\theta$ and $\phi$ being no longer equivalent.  The coincidence of these two data sets is, in fact, a feature of all the plots in this section.

            The $\theta$--$\phi$ time difference data for the coordinate-defined interferometer (Figure~\ref{F:karim-varying-r-nonzero-v}) can be seen to be roughly independent of $R_*$, for large values of $R_*$.  This is in agreement with the estimate \eqref{E:karim-time-differences} of the time difference $\delta \tau_{\theta\phi}$ of Karim \etal.  Once again, examining the data comprising Figure~\ref{F:karim-varying-r-nonzero-v}, it is found to be in agreement with the estimate \eqref{E:karim-time-differences} in the first several significant figures, validating the analysis of Karim \etal.
    
            Interestingly, for large values of $R_*$, the $\theta$--$\phi$ time difference data for the geodesic-defined interferometer has a slope very close to $-3$ on the logarithmic scale, corresponding to the time difference $\delta \tau_{\theta\phi}$ being proportional to $1 / R_*^3$.  Thus, while the $\theta$--$\phi$ time difference is already smaller than the $r$--$\theta$ time difference on Figure~\ref{F:geodesic-varying-r-nonzero-v} by several orders of magnitude, at physical values of $R_*$ ($R_* > 10^5$), it will be comparatively even smaller.  This is in contrast with the situation for the coordinate-defined interferometer: On Figure~\ref{F:karim-varying-r-nonzero-v} it would appear that, if we extrapolate the data to physical values of $R_*$, we might enter a regime where the $\theta$--$\phi$ time difference is larger than the $r$--$\theta$ time difference.
            
        \subsection{Varying interferometer length}
    
            \begin{figure}
                \begin{center}
                    \includegraphics[width = 12 cm]{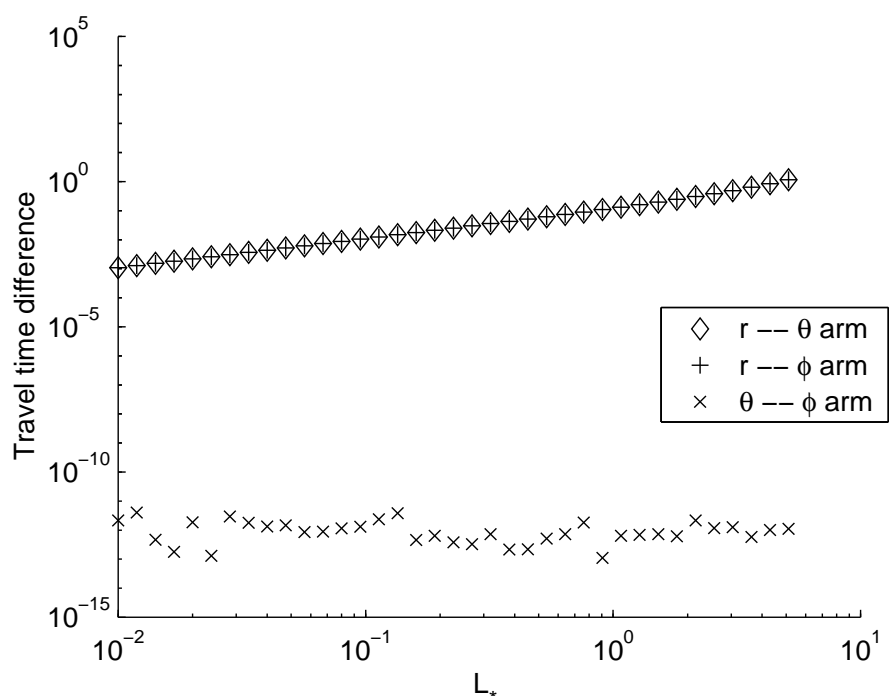}
                \end{center}
                \caption{Light travel time difference for the coordinate-defined interferometer, for various values of $L_*$, with fixed $R_* = 10$ and $v = 0$.}
                \label{F:karim-varying-length-zero-v}
            \end{figure}
    
            \begin{figure}
                \begin{center}
                    \includegraphics[width = 12 cm]{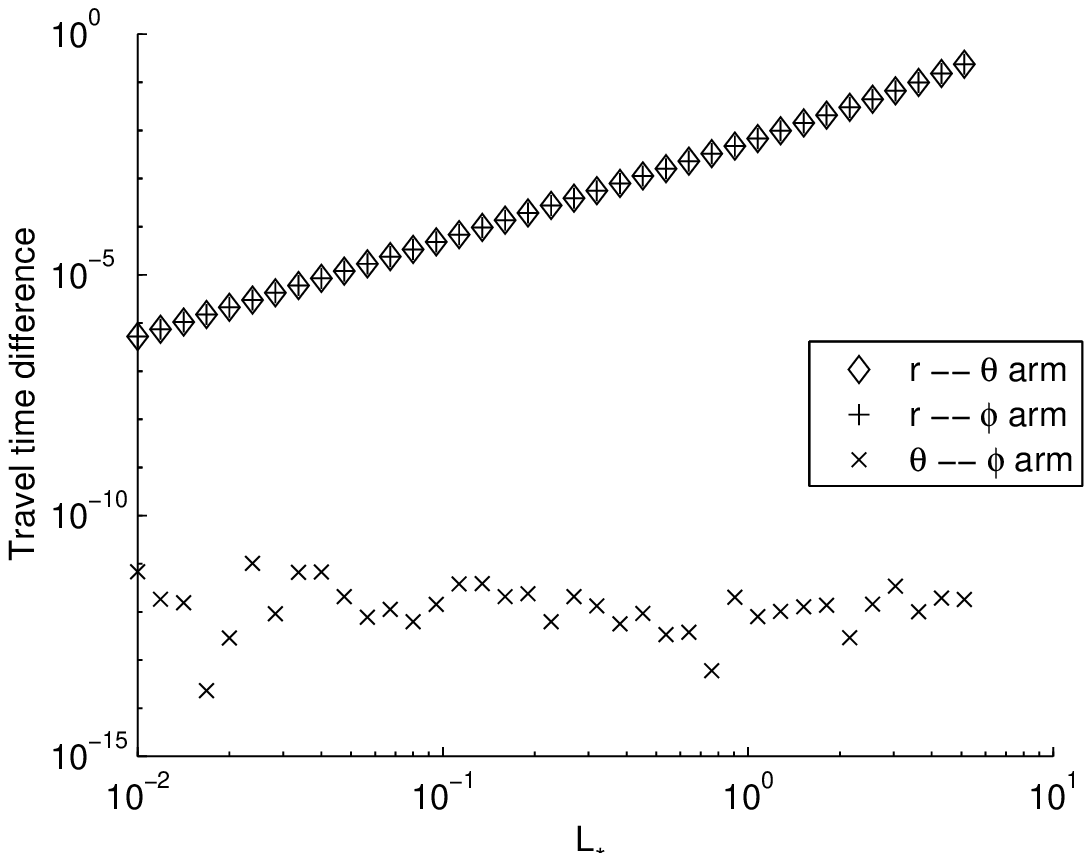}
                \end{center}
                \caption{Light travel time difference for the geodesic-defined interferometer, for various values of $L_*$, with fixed $R_* = 10$ and $v = 0$.}
                \label{F:geodesic-varying-length-zero-v}
            \end{figure}
            
            Figures~\ref{F:karim-varying-length-zero-v} and \ref{F:geodesic-varying-length-zero-v} show the light travel time differences for Experiment~\ref{I:vary-l-zero-v}, for the coordinate-defined interferometer and the geodesic-defined interferometer, respectively.

            As with the other experiment with $v = 0$ (Experiment~\ref{I:vary-r-zero-v}), and as expected, the $\theta$--$\phi$ time difference data is everywhere zero, within the numerical precision.

            For small values of $L_*$, the $r$--$\theta$ data for the coordinate-defined interferometer has slope very close to $1$ on the logarithmic scale of Figure~\ref{F:karim-varying-length-zero-v}, corresponding to the travel time difference $\delta \tau_{r\theta}$ being proportional to $L_*$.  Again, the scaling is in agreement with the estimate \eqref{E:karim-time-differences} of Karim \etal.

            For the geodesic-defined interferometer, for small values of $L_*$, the $r$--$\theta$ data has slope very close to $2$ on the logarithmic scale of Figure~\ref{F:karim-varying-length-zero-v}, corresponding to the travel time difference $\delta \tau_{r\theta}$ being proportional to $L_*^2$.
    
            \begin{figure}
                \begin{center}
                    \includegraphics[width = 12 cm]{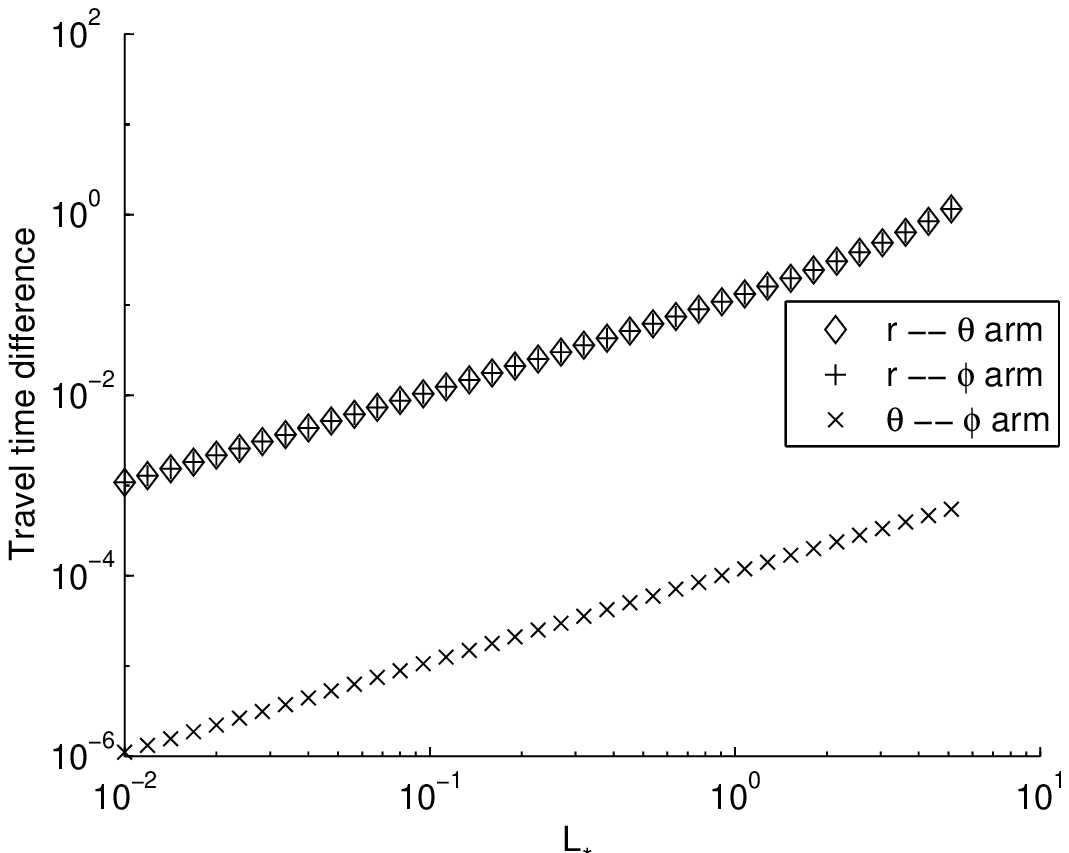}
                \end{center}
                \caption{Light travel time difference for the coordinate-defined interferometer, for various values of $L_*$, with fixed $R_* = 10$ and $v = 10^{-2}$.}
                \label{F:karim-varying-length-nonzero-v}
            \end{figure}
    
            \begin{figure}
                \begin{center}
                    \includegraphics[width = 12 cm]{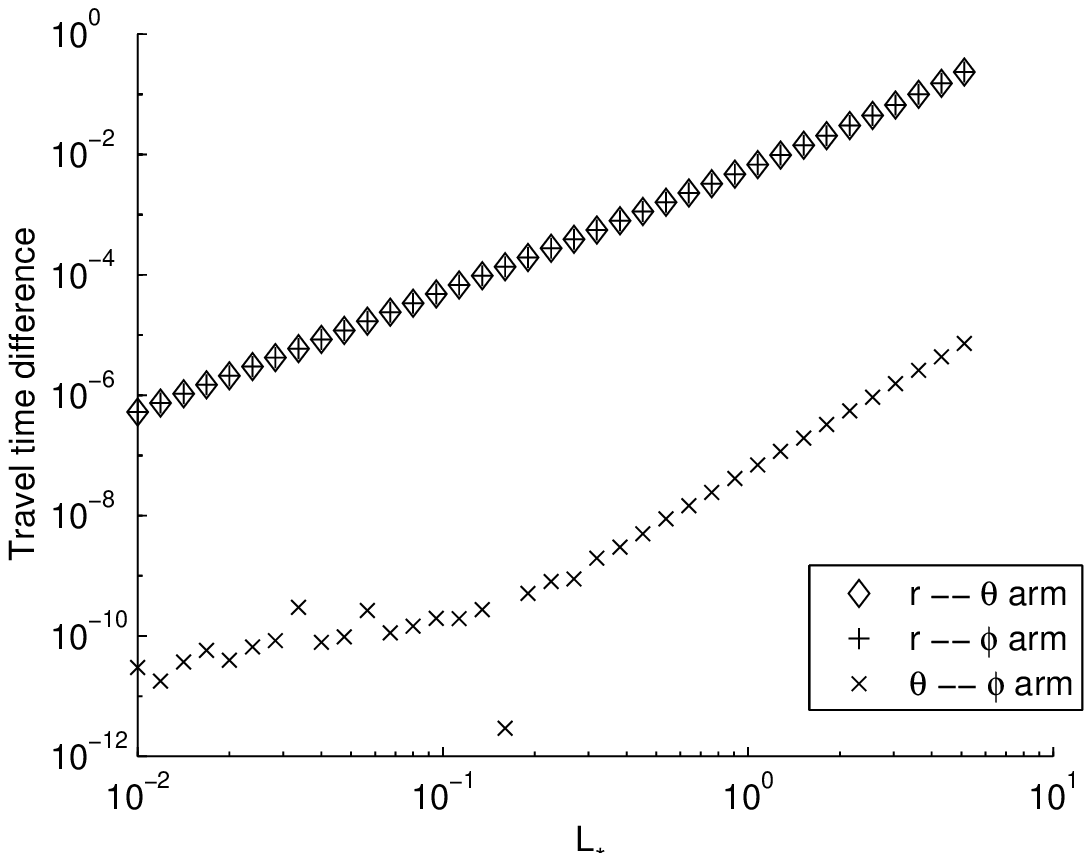}
                \end{center}
                \caption{Light travel time difference for the geodesic-defined interferometer, for various values of $L_*$, with fixed $R_* = 10$ and $v = 10^{-2}$.}
                \label{F:geodesic-varying-length-nonzero-v}
            \end{figure}
            
            Figures~\ref{F:karim-varying-length-nonzero-v} and \ref{F:geodesic-varying-length-nonzero-v} show the light travel time differences for Experiment~\ref{I:vary-l-nonzero-v}, for the coordinate-defined interferometer and the geodesic-defined interferometer, respectively.

            As with Experiments~\ref{I:vary-r-zero-v} and \ref{I:vary-r-nonzero-v}, there is no significant difference between the $r$--$\theta$ data of Figures~\ref{F:karim-varying-length-nonzero-v} and \ref{F:geodesic-varying-length-nonzero-v} and the corresponding data from Figures~\ref{F:karim-varying-length-zero-v} and \ref{F:geodesic-varying-length-zero-v}.

            For small values of $L_*$, the slope of the $\theta$--$\phi$ data on the logarithmic scale of Figure~\ref{F:karim-varying-length-nonzero-v} is very close to $1$, corresponding to the travel time difference $\delta \tau_{\theta\phi}$ being proportional to $L_*$ for the coordinate-defined interferometer.  This scaling is in agreement with the calculation \eqref{E:karim-time-differences} of Karim \etal.

            Almost all of the $\theta$--$\phi$ data for the geodesic-defined interferometer (Figure~\ref{F:geodesic-varying-length-nonzero-v}) are near or below the relative precision of the numerical methods, $10^{-9}$, and so no reliable conclusions can be drawn about it.  Based on the few reliable data points, which are unfortunately at large (non-physical) values of $L_*$, we might conjecture an $L_*^2$ dependence of $\delta \tau_{\theta\phi}$ on $L_*$, consistent with the scaling of $\delta \tau_{r\theta}$, since the slope of the valid $\theta$--$\phi$ data points is roughly 2.
            
        \subsection{Varying interferometer coordinate speed}

            \begin{figure}
                \begin{center}
                    \includegraphics[width = 12 cm]{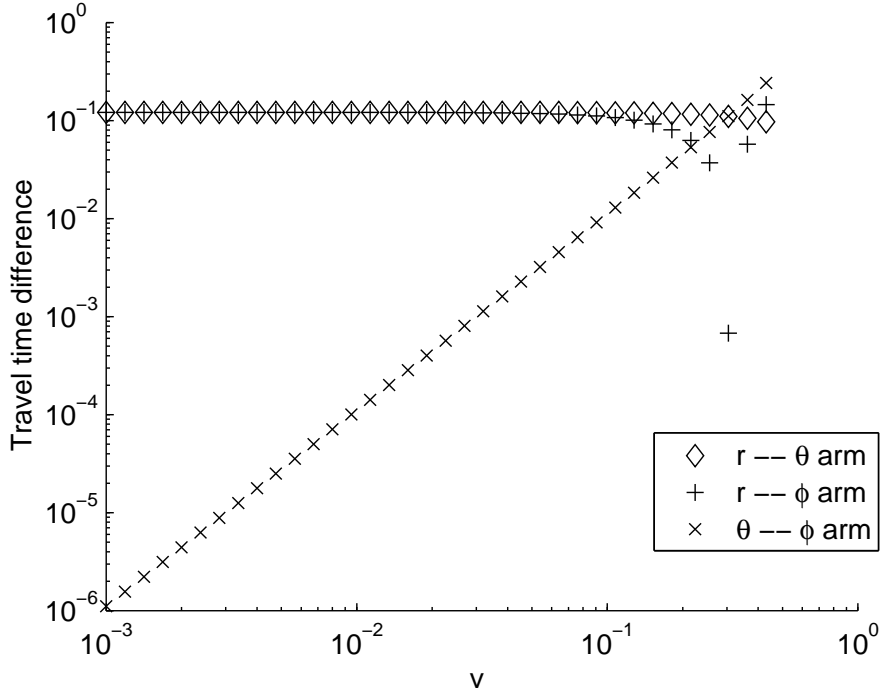}
                \end{center}
                \caption{Light travel time difference for the coordinate-defined interferometer, for various values of $v$, with fixed $R_* = 10$ and $L_* = 1$.}
                \label{F:karim-varying-v}
            \end{figure}
    
            \begin{figure}
                \begin{center}
                    \includegraphics[width = 12 cm]{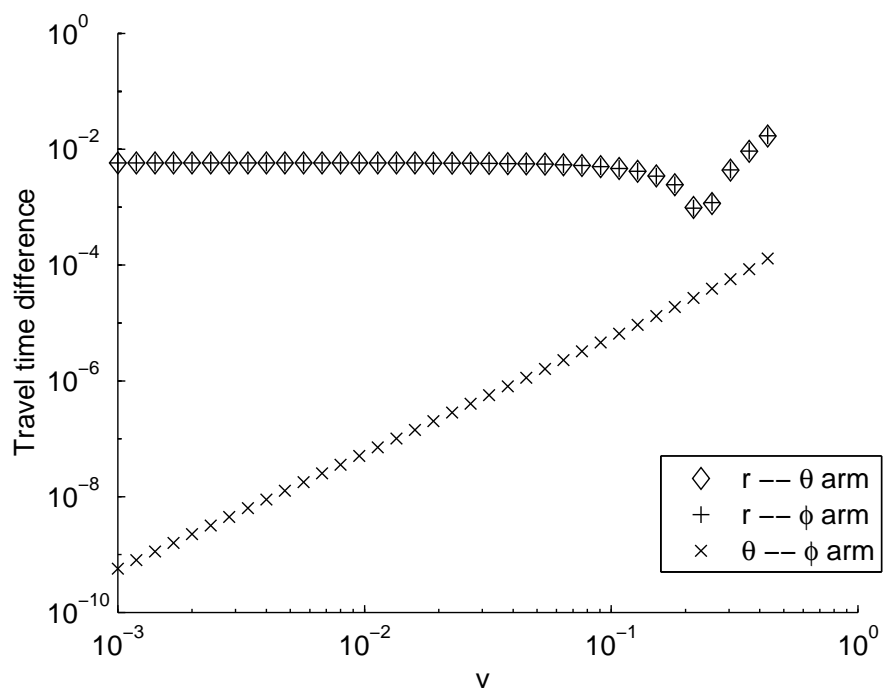}
                \end{center}
                \caption{Light travel time difference for the geodesic-defined interferometer, for various values of $v$, with fixed $R_* = 10$ and $L_* = 1$.}
                \label{F:geodesic-varying-v}
            \end{figure}
                        
            Figures~\ref{F:karim-varying-v} and \ref{F:geodesic-varying-v} show the light travel time differences for Experiment~\ref{I:vary-v}, for the coordinate-defined interferometer and the geodesic-defined interferometer, respectively.

            The most important property of these plots is that, for both interferometer models, for small values of $v$, the $r$--$\theta$ time difference data are independent of $v$.  For the coordinate-defined interferometer, this result is in agreement with the estimate \eqref{E:karim-time-differences} of Karim \etal.  For the geodesic-defined interferometer we conclude that, for physical values of $v$ ($v \sim 10^{-3}$), the travel time difference $\delta \tau_{r\theta}$ is independent of $v$.

            For small values of $v$, the slope of the $\theta$--$\phi$ data on the logarithmic scale of Figure~\ref{F:karim-varying-v} is very close to $2$, corresponding to the travel time difference $\delta \tau_{\theta\phi}$ being proportional to $v^2$ for the coordinate-defined interferometer.  This scaling is in agreement with the calculation \eqref{E:karim-time-differences} of Karim \etal.

            The slope of the $\theta$--$\phi$ data for the geodesic-defined interferometer (Figure~\ref{F:geodesic-varying-v}) is also very close to $2$ for small values of $v$, although it should be noted that the first few data points are near or below the relative precision $10^{-9}$ of the numerical methods employed.

            The unusual behaviour of the $r$--$\theta$ data on Figures~\ref{F:karim-varying-v} and \ref{F:geodesic-varying-v} for values of $v$ approaching unity is simply due to the light travel time difference passing through zero on the logarithmic axes.  Because $v$ is a coordinate speed, if it is increased beyond approximately unity, then the world-lines of the various parts of the interferometer will become space-like, which is certainly not physical.

        \subsection{Summary}
    
            The results of the all the numerical experiments simulating the coordinate-defined interferometer were in agreement with the estimated light travel time differences \eqref{E:karim-time-differences} of Karim \etalnodot.  Thus, the analysis of the coordinate-defined interferometer in \cite{R:karim} was validated.

            The light travel time differences for the geodesic-defined interferometer were investigated as a function of the dimensionless parameters $R_*$, $L_*$, and $v$.  The largest travel time difference was $\delta \tau_{r\theta}$ (or $\delta \tau_{r\phi}$), which was found to be proportional to $L_*^2 / R_*^2$, independent of $v$, for small values of $L_*$, large values of $R_*$, and small values of $v$.
        
    \section{Estimate of physical effect}
        \label{S:karim-grwb-estimate-physical}
    
        In this section we employ the relation
        \begin{equation}
            \label{E:karim-grwb-relation}
            \delta \tau_{r\theta} \propto L_*^2 / R_*^2
        \end{equation}
        for the geodesic-defined interferometer, which was discovered by numerical experimentation in Section~\ref{S:karim-grwb-results}, to estimate the size of the light travel time difference $\delta \tau_{r\theta}$ for a 1\,m interferometer on Earth.  Analagous to Table~\ref{T:karim-order-of-magnitude-estimate}, we estimate the effect due to three nearby gravitational fields: The Earth, the Sun, and the Milky Way.

        To use \eqref{E:karim-grwb-relation} we first need a data point to fix the constant of proportionality.  The data point selected is that with the largest value of $R_*$.  Noting that the light travel time differences computed by the \code{reflect} routine are in units of $2 m$, where $m$ is the geometric mass of the gravitational field source, the data point is
        \begin{equation}
            \label{E:karim-grwb-data-point}
            R_* = 48, \quad L_* = 1, \quad \frac{\delta \tau_{r\theta}}{2 m} = 2.06 \times 10^{-4}.
        \end{equation}
        
        From \eqref{E:karim-grwb-relation} and \eqref{E:karim-grwb-data-point} we have
        \begin{equation}
            \delta \tau_{r\theta} = (2.06 \times 10^{-4}) 2 m \frac{L_*^2}{(R_* / 48)^2},
        \end{equation}
        or, since $R_* = R / 2m$, $L_* = L / 2m$, and $m = G M / c^2$ where $M$ is the mass in \textsc{si} units,
        \begin{equation}
            \label{E:karim-grwb-extrapolate}
            \delta \tau_{\textsc{si}} = 48^2 \times (2.06 \times 10^{-4}) \times \frac{2 G M}{c^3} \frac{L^2}{R^2},
        \end{equation}
        where we have also divided by $c$ to obtain the time difference in seconds, rather than metres.
        
        \begin{table}
            \begin{center}
                \begin{tabular}{| l | c | c | c |}
                    \hline
                    field source & $M$ (kg) & $R$ (m) & $\delta \tau_{\textsc{si}}$ (s) \\
                    \hline
                    Earth & $5.97 \times 10^{24}$ & $6.38 \times 10^6$ & $3.5 \times 10^{-25}$ \\
                    Sun & $1.99 \times 10^{30}$ & $1.50 \times 10^11$ & $2.09 \times 10^{-28}$ \\
                    Milky Way & $2 \times 10^{41}$ & $2.5 \times 10^20$ & $8 \times 10^{-36}$ \\
                    \hline
                \end{tabular}
                \caption{Estimates of $\delta \tau_{\textsc{si}}$ for various bodies with $L = 1$\,m, for the geodesic-defined interferometer model.}
                \label{T:karim-grwb-extrapolate}
            \end{center}
        \end{table}
        
        Using \eqref{E:karim-grwb-extrapolate} we can estimate the effect due to the Earth, the Sun, and the Milky Way.  The calculation is summarised in Table~\ref{T:karim-grwb-extrapolate}.  Compare Table~\ref{T:karim-grwb-extrapolate} with Table~\ref{T:karim-order-of-magnitude-estimate} of Section~\ref{S:karim-main-results}.

        In Table~\ref{T:karim-grwb-extrapolate}, the effect due to Milky Way is $\sim 10^{-35}$\,s.  The smallest time-scale currently detectable with gravitational wave detectors is on the order of $10^{-20}$\,s.  We conclude that the Milky Way cannot be weighed by measuring $\delta \tau_{\textsc{si}}$.

        The ordering of the effects ($\text{Earth} > \text{Sun} > \text{Milky Way}$) is in opposition to that of Table~\ref{T:karim-order-of-magnitude-estimate}.  This may be thought of as due to the extra factor of $L / R$ in \eqref{E:karim-grwb-relation} compared with \eqref{E:karim-time-differences}.  (For the Milky Way, $L / R \sim 10^{-20}$, and for the Earth, $L / R \sim 10^{-7}$.)

        The estimate for the effect due to the Earth in Table~\ref{T:karim-grwb-extrapolate} cannot be assumed to be very accurate, because the Schwarzschild radius of the Earth is about 9\,mm and so, for a 1\,m interferometer, $L_* \simeq 113$, which is significantly larger than any value of $L_*$ tested in a numerical experiment in this chapter---the relation \eqref{E:karim-grwb-relation} may not hold in that regime, although we have no reason to think it will not.

    \section{Conclusions}
        \label{S:karim-grwb-conclusion}

        By simulating the coordinate-defined interferometer of Karim \etal\ in \grwb, we were able to validate the theoretical analysis of that interferometer, made in \cite{R:karim}.

        By simulating a physically realistic geodesic-defined interferometer, a more accurate estimate of the light travel time difference $\delta \tau_{r\theta}$ on Earth due to the Milky Way was obtained, and was found to be too small to be detected.  It was also found that, in contrast to the case for the coordinate-defined interferometer of Karim \etal, the light travel time difference due to the gravitational field of the Earth is the most important for an interferometer located on Earth, and that due to the gravitational field of the Milky Way is the least important of the major gravitational fields in the vicinity of Earth.

        We conclude that the experiment proposed by Karim \etal, to weigh the galaxy using a small interferometer on Earth, is not feasible, and that their conclusion is false because of the approximations implied in their coordinate-dependent interferometer model. 
\clearemptydoublepage
    \chapter{Conclusion}
    \label{C:conclusion}

    \grwb\ has been successfully and substantially extended to facilitate numerical experimentation in General Relativity.

    A functional programming framework has been crucial to the development of tools for numerical experimentation within \grwb.  The functional framework is more expressive, permitting important concepts in numerical programming and differential geometry to be directly represented in the C++ code of \grwb.

    New algorithms for key numerical operations have replaced pre-existing simpler methods.  The numerical engine is now expressed in the paradigm of functional programming, enabling algorithms to easily interface with one-another.  The sophisticated new algorithms are faster and more accurate, and an abstraction of the notion of approximate equality enables them to be encoded in a robust and elegant way.  Through the C++ template mechanism, numerical methods can be encoded such that they can be applied to any sets with the required structure defined upon them.

    The differential geometric engine of \grwb\ has been rewritten within the functional programming framework.  Abstract notions, such as points and tangent vectors, are represented by C++ classes.  Functions used in differential geometry, such as curves in space-time, are now represented and manipulated directly as functions.

    Using the new numerical and differential geometric core of \grwb, tools for numerical experimentation have been developed.  Geodesics and the parallel transport operation, both implemented in terms of the new \textsc{ode} integration algorithm, represent fundamental physical concepts in General Relativity.  Methods for determining unique geodesics, defined implicitly in terms of boundary conditions, have been developed using the new algorithm for function minimisation; these methods enable the construction of photon world-lines joining observers to particular events, representing an important physical situation.

    The utility of numerical experimentation in \grwb\ was demonstrated.  A traditional analysis of a physical problem in General Relativity, involving various simplifying approximations in the mathematical model, was investigated and found to yield an inaccurate estimate of the desired physical quantity.  A more physically motivated model was devised, and an accurate estimate of the quantity was obtained by simulating the new model in \grwb.  A physically meaningful result was thereby produced by a numerical experiment in \grwb, where analytic methods had proven to be inadequate. \clearemptydoublepage
        
     \clearemptydoublepage

    \appendix

    \chapter{\grwb\ code listings}
    \label{A:grwb-code}

    This appendix contains code listings from important parts of the rewritten numerical engine of \grwb, described in Chapter~\ref{C:numerical}, and some of the tools for numerical experimentation described in Chapter~\ref{C:numerical-experiments}.

    Whenever a conflict arises, the algorithms in \grwb\ are generally coded with execution speed taking priority over code brevity or simplicity.  As such, the code in this appendix may appear significantly different to the code in Chapters~\ref{C:functional}, \ref{C:numerical}, \ref{C:differential-geometry}, and \ref{C:numerical-experiments}.  In many cases, however, the algorithms may be more easily read by completely disregarding the symbols \code{const} and \code{&}, and by interpreting variable declarations of the form
    \begin{codeblock}
const Type variable(expression);
    \end{codeblock}
    as the more familiar
    \begin{codeblock}
Type variable = expression;
    \end{codeblock}

    \section{Relative difference}
        \label{S:grwb-code-relative-difference}

        \begin{codeblock}
template <typename T> struct relative_difference_implementation
{
    static double apply(const T& a, const T& b)
    {
        const double abs_a_abs_b(abs(a) * abs(b));
        const double abs_a_minus_b(abs(a - b));
        return abs_a_abs_b <= 1 ? abs_a_minus_b : abs_a_minus_b / sqrt(abs_a_abs_b);
    }
 
    static double apply_squared(const T& a, const T& b)
    {
        const double abs_a_abs_b(abs(a) * abs(b));
        const double abs_a_minus_b_squared(square(abs(a - b)));
        return abs_a_abs_b <= 1 ? abs_a_minus_b_squared : abs_a_minus_b_squared / abs_a_abs_b;
    }
};
 
template <typename T> double relative_difference(const T& a, const T& b)
{
    return relative_difference_implementation<T>::apply(a, b);
}
 
template <typename T> double relative_difference_squared(const T& a, const T& b)
{
    return relative_difference_implementation<T>::apply_squared(a, b);
}
 
template <typename T> struct relative_difference_implementation<nvector<T> >
{
    static double apply(const nvector<T>& a, const nvector<T>& b)
    {
        return sqrt(apply_squared(a, b));
    }
 
    static double apply_squared(const nvector<T>& a, const nvector<T>& b)
    {
        if (a.size() != b.size())
            throw nvector<T>::incompatible();
       
        typename nvector<T>::const_iterator i, j;
        double r(0.);
        for (i = a.begin(), j = b.begin(); i != a.end(); ++i, ++j)
            r += relative_difference_squared(*i, *j);
 
        return r;
    }
};
 
template <typename T, size_t N> struct relative_difference_implementation<grwb::vector<N, T> >
{
    static double apply(const grwb::vector<N, T>& a, const grwb::vector<N, T>& b)
    {
        return sqrt(apply_squared(a, b));
    }
 
    static double apply_squared(const grwb::vector<N, T>& a, const grwb::vector<N, T>& b)
    {
        typename grwb::vector<N, T>::const_iterator i, j;
        double r(0.);
        for (i = a.begin(), j = b.begin(); i != a.end(); ++i, ++j)
            r += relative_difference_squared(*i, *j);
 
        return r;
    }
};
        \end{codeblock}
    
    \section{Richardson extrapolation}
        \label{S:grwb-code-richardson}
        
        \begin{codeblock}
template <typename T> class richardson_extrapolation
{
public:
    richardson_extrapolation(const double& x, const T& y)
      : limit_(y),
        error_(y)
    {
        refine(x, y);
    }

    const T& limit() const
    {
        return limit_;
    }   

    const T& error() const
    {
        return error_;
    }
   
    void refine(const double& x, const T& y)
    {
        // adapted from Numerical Recipes in C (2nd Edition), p. 731
       
        data_.resize(data_.size() + 1, make_pair(x, y));
        error_ = limit_ = y;
       
        const size_t n(data_.size());
        if (n == 1)
            return;
       
        T c(y);
        for (size_t i(1); i < n; ++i)
        {
            const double x_i(data_[n - i - 1].first);
            const double delta(1. / (x_i - x));
            const double f1(x * delta);
            const double f2(x_i * delta);
            const T q(data_[i - 1].second);
            data_[i - 1].second = error_;
            const T d2(c - q);
            error_ = f1 * d2;
            c = f2 * d2;
            limit_ += error_;
        }
       
        data_[n - 1].second = error_;
    }

private:
    std::vector<pair<double, T> > data_;
    T limit_;
    T error_;
};
        \end{codeblock}
        
    \section{Differentiation}
        \label{S:grwb-code-derivative}
        
        \begin{codeblock}
template <typename T> class derivative_functor
{
public:
   derivative_functor(const function<optional<T> (const double&)>& f, const double& scale, const double& tolerance)
     : f_(f),
      scale_(scale),
      tolerance_(tolerance)
   {
   }

   optional<T> operator()(const double& x) const
   {
      if (!f_(x))
         return optional<T>();
      
      double h(scale_);
   
      optional<richardson_extrapolation<T> > extrapolator;
   
      for (size_t i(0); i < max_steps_; ++i)
      {
         const optional<T> right(f_(x + h));
         const optional<T> left(f_(x - h));
         if (left && right)
         {    
            const T diff((*right - *left) / (2. * h));
         
            if (!extrapolator)
               extrapolator.reset(richardson_extrapolation<T>(h * h, diff));
            else
            {
               extrapolator->refine(h * h, diff);
               if (tolerance_ > relative_difference(extrapolator->limit(), extrapolator->limit() + extrapolator->error()))
               return optional<T>(extrapolator->limit());
            }
         }
   
         h /= step_scale_;
      }

      return optional<T>();
   }

private:
   const function<optional<T> (const double&)> f_;
   const double scale_;
   const double tolerance_;

   static const size_t max_steps_ = 13;
   static const double step_scale_ = 1.7;
};

template <typename T> function<optional<T> (const double&)> derivative(const function<optional<T> (const double&)>& f, const double& scale = 1., const double& tolerance = default_approx_equal_tolerance)
{
   return derivative_functor<T>(f, scale, tolerance);
}        
        \end{codeblock}
        
        \subsection{Gradient}
            \label{S:grwb-code-gradient}
            
            \begin{codeblock}
namespace gradient_detail
{
   template <typename T> class single_coordinate_function
   {
   public:
      single_coordinate_function(const function<optional<T> (const nvector<double>&)>& f, const nvector<double>& x, const size_t& i)
        : f_(f),
         x_(x),
         i_(i)
      {
      }

      optional<T> operator()(const double& delta_x_i)
      {
         nvector<double> _x(x_);
         _x[i_] += delta_x_i;
         return f_(_x);
      }
   
   private:
      const function<optional<T> (const nvector<double>&)>& f_;
      const nvector<double>& x_;
      const size_t& i_;
   };
}

template <typename T> class gradient_functor
{
public:
   gradient_functor(function<optional<T> (const nvector<double>&)>f)
      : f_(f)
   {
   }

   optional<nvector<T> > operator()(const nvector<double>& x)
   {
      const optional<T> default_value(f_(x));
   
      if (!default_value)
         return optional<nvector<T> >();
   
      nvector<T> result(x.size(), unchanging(*default_value));
   
      for (size_t i = 0; i != x.size(); ++i)
      {
         optional<T> d(derivative<T>(gradient_detail::single_coordinate_function<T>(f_, x, i))(0.));
         if (!d)
            return optional<nvector<T> >();
         result[i] = *d;
      }

      return optional<nvector<T> >(result);
   }

private:
   const function<optional<T> (const nvector<double>&)> f_;
};

template<typename T> function<optional<nvector<T> >(const nvector<double>&)> gradient(const function<optional<T>(const nvector<double>&)>& f)
{
   return gradient_functor<T>(f);
}            
            \end{codeblock} 

    \section{Bulirsch-Stoer method}
        \label{S:grwb-code-bulirsch-stoer}

        \begin{codeblock}
template <class T, template <class> class U> class bulirsch_stoer
// adapted from Numerical Recipes in C (2nd Edition), p. 728
{
public:
  bulirsch_stoer(const function<optional<T> (const double&, const T&)>& f, const double& x_0, const T& y_0, const double& default_stepsize = 1., const size_t& maximum_steps = 100, const double& relative_error = default_approx_equal_tolerance)
    : f_(f),
      maximum_steps_(maximum_steps),
      relative_error_(relative_error),
      default_h_(default_stepsize),
      x_(x_0),
      y_(y_0)
  {
    const double safe_relative_error(relative_error * safe1_);

    typename std::map<double, vector<bulirsch_stoer_parameters<U>::k_total, vector<bulirsch_stoer_parameters<U>::k_total, double> > >::const_iterator i(alpha_cache_().find(safe_relative_error));
    if (i != alpha_cache_().end())
      alpha_ = i->second;
    else
    {
        for (size_t i = 1; i < bulirsch_stoer_parameters<U>::k_total; ++i)
            for (size_t j = 0; j < i; ++j)
                alpha_[j][i] = pow(safe_relative_error, (a_()[j] - a_()[i]) / ((a_()[i] - a_()[0] + 1.) * (2 * j + 3)));
                alpha_cache_()[safe_relative_error] = alpha_;
    }

    for (optimal_k_ = 1; optimal_k_ < bulirsch_stoer_parameters<U>::k_total - 1; ++optimal_k_)
        if (a_()[optimal_k_ + 1] > a_()[optimal_k_] * alpha_[optimal_k_ - 1][optimal_k_])
            break;
    max_k_ = optimal_k_;
  }

  const double& x() const
  {
    return x_;
  }

  const T& y() const
  {
    return y_;
  }

  bool step(const double& to_x)
  {
    double h(default_h_);
    if (to_x < x_)
      h *= -1;
      
    for (size_t i = 0; i < maximum_steps_; ++i)
    {
      bool reduced_step_size(false);
      bool success(false);
      size_t k(0), km(0);
      double stepsize_reduction_factor(0.);
      double err[bulirsch_stoer_parameters<U>::k_total];
      U<T> stepper(f_, x_, y_);

      if (to_x == x_)
       return true;

      if ((to_x - x_) * (to_x - x_ - h)  < 0.)
       h = to_x - x_;

      while (true)
      {
        optional<richardson_extrapolation<T> > extrapolator;

        for (k = 0; k < max_k_; ++k)
        {
            optional<T> y_est(stepper.step(h, bulirsch_stoer_parameters<U>::k_values[k]));
            if (!y_est)
                return false;
        
            const double little_h_squared(square(h / bulirsch_stoer_parameters<U>::k_values[k]));
            if (!extrapolator)
                extrapolator.reset(richardson_extrapolation<T>(little_h_squared, *y_est));
            else
            {
                extrapolator->refine(little_h_squared, *y_est);
                y_ = extrapolator->limit();
        
                const double error(relative_difference(y_, y_ + extrapolator->error()) / relative_error_);
                km = k - 1;
                err[km] = pow(error / safe1_, 1. / (2 * km + 3));
                
                if (k >= optimal_k_ - 1 || i == 0)
                {
                    if (error < 1.)
                    {
                        success = true;
                        break;
                    }
                    if (k == max_k_ || k == optimal_k_ + 1)
                    {
                        stepsize_reduction_factor = safe2_ / err[km];
                        break;
                    }
                    if (k == optimal_k_ && alpha_[optimal_k_ - 1][optimal_k_] < err[km])
                    {
                        stepsize_reduction_factor = 1. / err[km];
                        break;
                    }
                    if (optimal_k_ == max_k_ && alpha_[km][max_k_ - 1] < err[km])
                    {
                        stepsize_reduction_factor = alpha_[km][max_k_ - 1] * safe2_ / err[km];
                        break;
                    }
                    if (alpha_[km][optimal_k_] < err[km])
                    {
                        stepsize_reduction_factor = alpha_[km][optimal_k_ - 1] / err[km];
                        break;
                    }
                }
            }
        }
        
        if (success)
        break;
    
        if (stepsize_reduction_factor > min_stepsize_reduction_)
        stepsize_reduction_factor = min_stepsize_reduction_;
        if (stepsize_reduction_factor < max_stepsize_reduction_)
        stepsize_reduction_factor = max_stepsize_reduction_;
        h *= stepsize_reduction_factor;
        reduced_step_size = true;
      }
    
      x_ += h;
    
      double work_min(1.e300);
      double scale_factor(0.);
      for (size_t j = 0; j <= km; ++j)
      {
        const double s(err[j] < max_stepsize_increase_ ? max_stepsize_increase_ : err[j]);
        const double work(s * a_()[j + 1]);
        if (work < work_min)
        {
            scale_factor = s;
            work_min = work;
            optimal_k_ = j + 1;
        }
      }
      if (optimal_k_ >= k && optimal_k_ != max_k_ && !reduced_step_size)
      {
        double s(scale_factor / alpha_[optimal_k_ - 1][optimal_k_]);
        if (s < max_stepsize_increase_)
        s = max_stepsize_increase_;
        
        if (a_()[optimal_k_ + 1] * s <= work_min)
        {
            scale_factor = s;
            ++optimal_k_;
        }
      }
      h /= scale_factor;
    }

    cout << "Bulirsch-Stoer: Too many steps required." << endl;
    return false;
  }

private:
  const static double safe1_ = 0.25;
  const static double safe2_ = 0.7;
  const static double max_stepsize_reduction_ = 1.e-5;
  const static double min_stepsize_reduction_ = 0.7;
  const static double max_stepsize_increase_ = 0.1;

  const function<optional<T> (const double&, const T&)> f_;
  const size_t maximum_steps_;
  const double relative_error_;
  const double default_h_;
    
  vector<bulirsch_stoer_parameters<U>::k_total, vector<bulirsch_stoer_parameters<U>::k_total, double> > alpha_;

  static std::map<double, vector<bulirsch_stoer_parameters<U>::k_total, vector<bulirsch_stoer_parameters<U>::k_total, double> > >& alpha_cache_()
  {
    static std::map<double, vector<bulirsch_stoer_parameters<U>::k_total, vector<bulirsch_stoer_parameters<U>::k_total, double> > > _;
    return _;
  };

  static vector<bulirsch_stoer_parameters<U>::k_total + 1, double>& a_()
  {
    static optional<vector<bulirsch_stoer_parameters<U>::k_total + 1, double> > _;
    if (!_)
    {
      _.reset(vector<bulirsch_stoer_parameters<U>::k_total + 1, double>());
      (*_)[0] = bulirsch_stoer_parameters<U>::k_values[0] + 1;
      for (size_t i = 0; i < bulirsch_stoer_parameters<U>::k_total; ++i)
          (*_)[i + 1] = (*_)[i] + bulirsch_stoer_parameters<U>::k_values[i + 1];
    }
    return *_;
  }

  double x_;
  T y_;
  size_t optimal_k_;
  size_t max_k_;
};
        
        \end{codeblock}

        \subsection{Modified midpoint method}
            \label{S:grwb-code-modified-midpoint}

            \begin{codeblock}
template <class T> class modified_midpoint_stepper
{
public:
  modified_midpoint_stepper(const function<optional<T> (const double&, const T&)>& f, const double& x_0, const T& y_0)
    : f_(f),
      x_0_(x_0),
      y_0_(y_0),
      f_y_0_(f(x_0, y_0))
  {
  }

  optional<T> step(const double& total_h, const size_t& steps) const
  {
    // adapted from Numerical Recipes in C (2nd Edition), p. 724
      
    optional<T> ret;

    if (!f_y_0_)
      return ret;

    const double h(total_h / double(steps));
    const double two_h(2. * h);

    T ym(y_0_), yn(y_0_ + h * *f_y_0_);
    double x(x_0_ + h);

    optional<T> dydx(f_(x, yn));
    if (!dydx)
      return ret;
      
    for (size_t i(1); i < steps; ++i)
    {
      T y_next(ym + two_h * *dydx);
      ym = yn;
    
      dydx = f_(x += h, yn = y_next);
      if (!dydx)
       return ret;
    }

    ret.reset(0.5 * (ym + yn + h * *dydx));
    return ret;
    }

private:
  const function<optional<T> (const double&, const T&)> f_;
  const double x_0_;
  const T y_0_;
  const optional<T> f_y_0_;
};

template <> class bulirsch_stoer_parameters<modified_midpoint_stepper>
{
public:
  const static size_t k_total = 10;
  const static size_t k_values[k_total + 1];
};

template <> const size_t bulirsch_stoer_parameters<modified_midpoint_stepper>::k_values[] = {2, 4, 6, 8, 10, 12, 14, 16, 18, 20, 22};            
            \end{codeblock}
            
    \section{Powell's method}
        \label{S:grwb-code-powell}

        \begin{codeblock}
template <typename T, typename U> class powell_minimiser
{
public:
  powell_minimiser(const function<optional<T> (const U&)>& f)
    : f_(f)
  {
  }

  optional<pair<U, T> > operator()(const U& x, const nvector<U>& basis, const double& tolerance = default_approx_equal_tolerance) const
  {
    U minimum(x);
    optional<T> op(f_(minimum));
    if (!op)
      return optional<pair<U, T> >();
    T f_min(*op);
    nvector<U> basis_(basis);

    for (size_t i(0); i < max_steps; ++i)
    {
      const U prev_min(minimum);
      const T prev_f_min(f_min);
      T largest_decrease(zero(f_min));
      size_t largest_decrease_index(0);

      for (size_t j(0); j < basis.size(); ++j)
      {
        optional<pair<double, T> > line_minimum(brent_minimiser(linear_subspace(f_, minimum, basis_[j]))(0., 0., tolerance));
        if (!line_minimum)
            return optional<pair<U, T> >();

        if (f_min - line_minimum->second > largest_decrease)
        {
            largest_decrease = f_min - line_minimum->second;
            largest_decrease_index = j;
        }

        if (zero(line_minimum->first) != line_minimum->first && f_min > line_minimum->second)
        {
            basis_[j] *= line_minimum->first;
            minimum += basis_[j];
            f_min = line_minimum->second;
        }
      }

      if (approx_equal(prev_f_min, f_min, tolerance))
          return optional<pair<U, T> >(make_pair(minimum, f_min));

      const U new_direction(minimum - prev_min);
      const U extrapolated_min(minimum + new_direction);
      op = f_(extrapolated_min);
      if (!op)
          return optional<pair<U, T> >();
      const T f_extrapolated_min(*op);
        
      if (f_extrapolated_min < prev_f_min)
      {
        if (2. * (prev_f_min - 2. * f_min + f_extrapolated_min) * square(prev_f_min - f_min - f_extrapolated_min) <= largest_decrease * square(prev_f_min - f_extrapolated_min))
        {
            optional<pair<double, T> > line_minimum(brent_minimiser(linear_subspace(f_, minimum, new_direction))(0., 0., tolerance));
            if (!line_minimum)
                return optional<pair<U, T> >();
        
            if (zero(line_minimum->first) != line_minimum->first && f_min > line_minimum->second)
            {
                basis_[largest_decrease_index] = basis_[basis.size() - 1];
                minimum += (basis_[basis.size() - 1] = new_direction * line_minimum->first);
                f_min = line_minimum->second;
            }
        }
      }
    }

    return optional<pair<U, T> >();
  }

  optional<pair<U, T> > operator()(const U& x, const double& tolerance = default_approx_equal_tolerance) const
  {
    return operator()(x, default_basis(x), tolerance);
  }

private:
  const function<optional<T> (const U&)> f_;

  const static size_t max_steps = 100;
  const static double auto_scale = 1.e-2;

  class linear_subspace_functor
  {
  public:
    linear_subspace_functor(const function<optional<T> (const U&)>& f, const U& origin, const U& direction)
      : f_(f),
    origin_(origin),
    direction_(direction)
    {
    }

    optional<T> operator()(const double& t) const
    {
      return f_(origin_ + t * direction_);
    }
      
  private:
    const function<optional<T> (const U&)>& f_;
    const U& origin_;
    const U& direction_;
  };

  function<optional<T> (const double&)> linear_subspace(const function<optional<T> (const U&)>& f, const U& origin, const U& direction) const
  {
    return linear_subspace_functor(f, origin, direction);
  }

  nvector<U> default_basis(const U& x) const
  {
    nvector<U> r(unity(nvector<U>(x.size(), unchanging(x))));
    for (size_t i(0); i < x.size(); ++i)
    {
      const double scale(auto_scale * abs(x[i]));
      if (scale > 0)
          r[i] *= scale;
    }
    return r;
  }
};        
        \end{codeblock}

        \subsection{Brent minimiser}
            \label{S:grwb-code-brent}

            \begin{codeblock}
template <typename T, typename U> class bracketer
{
public:
  bracketer(const function<optional<T> (const U&)>& f)
    : _(f)
  {
  }

  optional<vector<3, pair<U, T> > > operator()(const U& x, const U& step_size = U()) const
  {
    optional<vector<3, pair<U, T> > > result;

    optional<T> op(_(x));
    if (!op)
      return result;
    vector<3, pair<U, T> > r(unchanging(make_pair(x, *op)));
    U step = step_size == U() ? (x == zero(x) ? unity(x) : auto_scale * abs(x)) : step_size;
      
    r[1].first += step;
    op = _(r[1].first);
    if (!op)
      return result;
    r[1].second = *op;

    if (r[1].second > r[0].second)
    {
      swap(r[0], r[1]);
      step *= -1;
    }

    for (size_t i(0); i < max_steps; ++i)
    {
      r[2].first = r[1].first + step;
      op = _(r[2].first);
      if (!op)
          return result;
      r[2].second = *op;

      if (r[2].second >= r[1].second)
      {
        if (step < 0)
            swap(r[0], r[2]);
        result.reset(r);
        return result;
      }

      r[0] = r[1];
      r[1] = r[2];
      step *= ratio;
    }

    return result;
  }

private:
  const function<optional<T> (const U&)> _;

  const static size_t max_steps = 100;
  const static double ratio = 1.6;
  const static double auto_scale = 1.e-2;
};

template <typename T, typename U> class brent_minimiser_functor
// adapted from Numeric Recipes in C (2nd Edition), p. 404
{
public:
    brent_minimiser_functor(const function<optional<T> (const U&)>& f)
    : f_(f)
    , bracketer_(f)
  {
  }

  optional<pair<U, T> > operator()(const U& x, const U& scale = U(), const double& tolerance = default_approx_equal_tolerance) const
  {
    const optional<vector<3, pair<U, T> > > bracket(bracketer_(x, scale));
    if (!bracket)
      return optional<pair<U, T> >();

    U left((*bracket)[0].first), best((*bracket)[1].first), right((*bracket)[2].first);
    U third_best(best), second_best(best);
    T f_best((*bracket)[1].second);
    T f_trial(f_best), f_third_best(f_best), f_second_best(f_best);

    U d(0.), prev_d(0.);

    const double two_tolerance(2. * tolerance);

    for (size_t i(0); i < max_steps; ++i)
    {
      const U mid(0.5 * (left + right));

      if (approx_equal(left, right, two_tolerance))
        return optional<pair<U, T> >(make_pair(best, f_best));

      if (abs(prev_d) > tolerance * abs(best))
      {
        const U r((best - second_best) * (f_best - f_third_best));
        U q((best - third_best) * (f_best - f_second_best));
        U p((best - third_best) * q - (best - second_best) * r);
        q = 2. * (q - r);
        if (q > 0.)
        p = -p;
        q = abs(q);
    
        U prev_prev_d(prev_d);
        prev_d = d;
    
        if (abs(p) >= abs(0.5 * q * prev_prev_d) || p <= q * (left - best) || p >= q * (right - best))
            d = cgold * (prev_d = (best >= mid ? left - best : right - best));
        else
            d = p / q;
      }
      else
        d = cgold * (prev_d = best >= mid ? left - best : right - best);

      const U trial(best + d);
      const optional<T> op(f_(trial));
      if (!op)
        return optional<pair<U, T> >();
      f_trial = *op;

      if (f_trial <= f_best)
      {
        if (trial >= best)
        left = best;
        else
        right = best;
        third_best = second_best; second_best = best; best = trial;
        f_third_best = f_second_best; f_second_best = f_best; f_best = f_trial;
      }
      else
      {
        if (trial < best)
            left = trial;
        else
            right = trial;
        if (f_trial <= f_second_best || second_best == best)
        {
            third_best = second_best; second_best = trial;
            f_third_best = f_second_best; f_second_best = f_trial;
        }
        else if (f_trial <= f_third_best || third_best == best || third_best == second_best)
        {
            third_best = trial;
            f_third_best = f_trial;
        }
      }
    }

    return optional<pair<U, T> >();
  }

private:
  const function<optional<T> (const U&)> f_;
  const bracketer<T, U> bracketer_;

  const static size_t max_steps = 100;
  const static double cgold = 0.3819660;
};

template <class T, class U> brent_minimiser_functor<T, U> brent_minimiser(const function<optional<T> (const U&)>& f)
{
  return brent_minimiser_functor<T, U>(f);
}            
            \end{codeblock}
            
    \section{Geodesic}
        \label{A:grwb-code-geodesic}
        
        \begin{codeblock}
class geodesic
{
public:
  geodesic(const tangent_vector& t)
    : atlas_(t.context().context()),
      least_upper_bound_(positive_infinity),
      greatest_lower_bound_(negative_infinity),
      cache_(new std::map<double, tangent_vector>)
  {
    cache_->insert(cache_value_type(0, t));
  }
    
  optional<point> operator()(const double& t) const
  {
    return operator()(t, 0.).second;
  }
    
  pair<double, optional<point> > operator()(const double& t, const double& epsilon) const
  {
    typedef pair<double, optional<point> > return_type;
      
    if (t >= least_upper_bound_)
      return return_type(least_upper_bound_, optional<point>());
    else if (t <= greatest_lower_bound_)
      return return_type(greatest_lower_bound_, optional<point>());
      
    cache_iterator_type initial_data(get_initial_data(t));
      
    if (abs(initial_data->first - t) <= epsilon)
      return return_type(initial_data->first, optional<point>(initial_data->second.context()));
      
    optional<cache_iterator_type> result(advance(initial_data->second, initial_data->first, t));
      
    if (result)
      return return_type(t, optional<point>((*result)->second.context()));
      
    if (t > 0)
      least_upper_bound_ = t;
    else
      greatest_lower_bound_ = t;
      
    return return_type(t, optional<point>());
  }
    
  optional<tangent_vector> tangent(const double& t) const
  {
    if (!operator()(t))
      return optional<tangent_vector>();
      
    return optional<tangent_vector>(cache_->find(t)->second);
  }

private:
  const weak_ptr<atlas> atlas_;    
    
  mutable double least_upper_bound_;
  mutable double greatest_lower_bound_;

  shared_ptr<std::map<double, tangent_vector> > cache_;

  typedef std::map<double, tangent_vector>::const_iterator cache_iterator_type;
  typedef std::map<double, tangent_vector>::value_type cache_value_type;

  class geodesic_callback
  {
  public:
    geodesic_callback(const shared_ptr<atlas::chart>& c)
      : _(c)
    {
    }
      
    optional<vector<2, nvector<double> > > operator()(const double& t, const vector<2, nvector<double> >& y) const
    {
      const int dim(y[0].size());
        
      vector<2, nvector<double> > ret(unchanging(y[1]));
        
      optional<ntensor_components<3>::type> con(connection(*_)(y[0]));
      if (!con)
         return optional<vector<2, nvector<double> > >();
        
      ret[1] *= 0.;
      for (int a = 0; a < dim; ++a)
        for (int b = 0; b < dim; ++b)
            for (int c = 0; c < dim; ++c)
                ret[1][a] -= (*con)[a][b][c] * y[1][b] * y[1][c];
            
      return optional<vector<2, nvector<double> > >(ret);
    }
      
  private:
    const shared_ptr<atlas::chart> _;
  };

  cache_iterator_type get_initial_data(const double& t) const
  {
    cache_iterator_type after(cache_->lower_bound(t));
      
    if (after == cache_->begin())
      return after;
      
    if (after == cache_->end())
      return --cache_iterator_type(after);
      
    cache_iterator_type before(after);
    --before;
    return abs(before->first - t) < abs(after->first - t) ? before : after;
  }

  optional<cache_iterator_type> advance(const tangent_vector& tangent, const double& from_t, const double& to_t, size_t recursion = 1) const
  {
    const point& origin = tangent.context();
      
    if (recursion > max_recursion_)
      return optional<cache_iterator_type>();
      
    //cout << "  Geodesic: " << from_t << " -> " << to_t << endl;
      
    shared_ptr<atlas> a(atlas_);
      
    for (set<shared_ptr<atlas::chart> >::const_iterator i = a->charts.begin(); i != a->charts.end(); ++i)
      if (origin[*i] && tangent[*i])
      {
        optional<cache_iterator_type> result(advance_on_chart(*i, *origin[*i], *tangent[*i], from_t, to_t));
        if (result)
            return result;
      }
      
    optional<cache_iterator_type> halfway(advance(tangent, from_t, (from_t + to_t) / 2, recursion + 1));
    return halfway ? advance((*halfway)->second, (from_t + to_t) / 2, to_t, recursion + 1) : optional<cache_iterator_type>();
  }
    
  optional<cache_iterator_type> advance_on_chart(const shared_ptr<atlas::chart>& c, const nvector<double>& x, const nvector<double>& dx, const double& from_t, const double& to_t) const
  {
    bulirsch_stoer<vector<2, nvector<double> >, modified_midpoint_stepper> solver((geodesic_callback(c)), from_t, make_vector(x, dx));
      
    if (!solver.step(to_t))
      return optional<cache_iterator_type>();
      
    point dest(atlas_.lock(), c, solver.y()[0]);
    tangent_vector tv(dest, c, solver.y()[1]);
      
    return optional<cache_iterator_type>(cache_->insert(cache_value_type(to_t, tv)).first);
  }
    
  const static size_t max_recursion_ = 7;

  const static double positive_infinity =  1e300;
  const static double negative_infinity = -1e300;
};        
        \end{codeblock}

    \section[Spherical polar coordinate transformation]{Generalised spherical polar coordinate transformation}
        \label{A:grwb-code-generalised-spherical}

        \begin{codeblock}
template <typename T> inline nvector<T> to_polar(const nvector<T>& x)
{
  nvector<T> r(x);
  T cosine(abs(x));
  const T zero_t(zero(cosine));
    
  r[0] = cosine;
  for (size_t i(1); i < x.size() - 1; ++i)
    if (cosine != zero_t)
    {        
      r[i] = asin(x[i - 1] / cosine);
      cosine *= cos(r[i]);
    }
    else
      r[i] = zero_t;

  r[x.size() - 1] = atan2(x[x.size() - 2], x[x.size() - 1]);

  return r;
}

template <typename T> inline nvector<T> from_polar(const nvector<T>& x)
{
  nvector<T> r(x);
  T cosine(x[0]);

  for (size_t i(0); i < x.size() - 1; ++i)
  {
    r[i] = cosine * sin(x[i + 1]);
    cosine *= cos(x[i + 1]);
  }
  r[x.size() - 1] = cosine;

  return r;
}

template <typename T> inline nvector<T> from_polar_with_radius(const nvector<T>& x, const T& radius)
{
  nvector<T> x2(x.size() + 1, unchanging(radius));
  for (size_t i(0); i < x.size(); ++i)
    x2[i + 1] = x[i];
  return from_polar(x2);
}

template <typename T> inline nvector<T> to_polar_without_radius(const nvector<T>& x)
{
  const nvector<T> polar(to_polar(x));
  return nvector<T>(polar.size() - 1, polar.begin() + 1);
}
        \end{codeblock} 
            
    \section{Connecting geodesic}
        \label{A:grwb-code-connecting-geodesic}
        
        \begin{codeblock}
namespace connecting_geodesic_detail
{
  class geodesic_shooter
  {
  public:
    geodesic_shooter(const point& a, const point& b, const shared_ptr<atlas::chart>& c)
      : a_(a),
    b_(b),
    chart_(c)
    {
    }
      
    optional<double> operator()(const nvector<double>& v) const
    {
      geodesic geo(tangent_vector(a_, chart_, from_polar_with_radius(v, 1.)));
      optional<pair<double, double> > r(min_euclidean_separation(geo, b_));
      return r ? optional<double>(r->second) : optional<double>();
    }
      
  private:
    const point& a_;
    const point& b_;
    const shared_ptr<atlas::chart>& chart_;
  };
}

inline optional<geodesic> connecting_geodesic(const point& a, const point& b)
{
  const shared_ptr<atlas::chart>& c(a.valid_chart());
  connecting_geodesic_detail::geodesic_shooter shooter(a, b, c);

  function<optional<double> (const nvector<double>&)> shooter_function(shooter);
  powell_minimiser<double, nvector<double> > pm(shooter_function);

  const nvector<double> va(*a[c]), vb(*b[c]);
  optional<pair<nvector<double>, double> > r(pm(to_polar_without_radius(vb - va)));

  if (!r)
    return optional<geodesic>();

  const nvector<double> v(from_polar_with_radius(r->first, 1.));
  const double scale(min_euclidean_separation(geodesic(tangent_vector(a, c, v)), b)->first);

  return optional<geodesic>(geodesic(tangent_vector(a, c, v * scale)));
}        
        \end{codeblock}
        
    \section{Connecting null geodesic}
        \label{A:grwb-code-connecting-null-geodesic}
        
        \begin{codeblock}
namespace connecting_null_geodesic_detail
{
  class null_geodesic_shooter
  {
  public:
    null_geodesic_shooter(const function<optional<point> (const double&)>& curve, const point& a, const shared_ptr<atlas::chart>& c, const nvector<nvector<double> >& tangent_basis, const double& null_guess, const double& guess)
      : guess_(make_vector(null_guess, guess)),
    curve_(curve),
    a_(a),
    chart_(c),          
    basis(tangent_basis)
    {
    }
      
    optional<double> operator()(const nvector<double>& v) const
    {
      const nvector<double> v2(from_polar_with_radius(v, 1.));
      nvector<double> v3(v2.size() + 1, unchanging(1.));
      for (size_t i(0); i < v2.size(); ++i)
         v3[i + 1] = v2[i];
        
      geodesic geo(tangent_vector(a_, chart_, basis * v3));
      optional<pair<nvector<double>, double> > r(min_euclidean_separation(geo, curve_, guess_));
        
      if (!r)
         return optional<double>();
        
      guess_ = r->first;
      return optional<double>(r->second);
    }

    const nvector<double>& last_guess() const
    {
      return guess_;
    }
      
  private:
    mutable nvector<double> guess_;
    const function<optional<point> (const double&)>& curve_;
    const point& a_;
    const shared_ptr<atlas::chart>& chart_;
    const nvector<nvector<double> >& basis;
  };
}

inline optional<pair<double, geodesic> > connecting_null_geodesic(const function<optional<point> (const double&)>& curve, const point& a, const double& guess)
{
  const static nvector<nvector<double> > polar_basis(make_vector(make_vector(0.01, 0.), make_vector(0., 0.01)));

  const shared_ptr<atlas::chart>& c(a.valid_chart());
  const nvector<nvector<double> > basis(orthonormal_tangent_basis(a, c));
    
  const point b(*curve(guess));

  const nvector<double> va(*a[c]), vb(*b[c]);
  const nvector<double> spacelike(vb.size() - 1, (inverse(basis) * (vb - va)).begin() + 1);
  const nvector<double> spacelike_polar(to_polar_without_radius(spacelike));

  const connecting_null_geodesic_detail::null_geodesic_shooter shooter(curve, a, c, basis, abs(spacelike), guess);
  const function<optional<double> (const nvector<double>&)> shooter_function(shooter);
  const powell_minimiser<double, nvector<double> > pm(shooter_function);

  optional<pair<nvector<double>, double> > r(pm(spacelike_polar, polar_basis));
  if (!r)
    return optional<pair<double, geodesic> >();

  const nvector<double> vr(from_polar_with_radius(r->first, 1.));
  nvector<double> vr2(vr.size() + 1, unchanging(1.));
  for (size_t i(0); i < vr.size(); ++i)
    vr2[i + 1] = vr[i];
  const nvector<double> solution(basis * vr2);   

  const nvector<double> scales(min_euclidean_separation(geodesic(tangent_vector(a, c, solution)), curve, shooter.last_guess())->first);

  return optional<pair<double, geodesic> >(make_pair(scales[1], geodesic(tangent_vector(a, c, solution * scales[0]))));    
}

inline optional<pair<double, geodesic> > connecting_null_geodesic(const point& a, const function<optional<point> (const double&)>& curve, const double& guess = 0.)
{
  return connecting_null_geodesic(curve, a, guess);
}        
        \end{codeblock} \clearemptydoublepage
    \chapter{Numerical experiment code listing}
    \label{A:karim-code}

    This appendix lists the code, written by the author, for the numerical experiment described in Chapter~\ref{C:karim-grwb}.  The code fragments listed in that chapter were adapted from portions of this code; they were simplified for clarity, and any code not directly relevant to the discussion was removed.

    The comments at the beginning of Appendix~\ref{A:grwb-code} also apply to the code listed here.
    
    \begin{codeblock}
template <class T> class geodesic_interferometer : public numerical_experiment<T>
{
public:
  explicit geodesic_interferometer(const shared_ptr<T>& _);

private:
  void reflect(const double& r, const double& interferometer_speed, const double& arm_length);
};

template <class T> class karim_interferometer : public numerical_experiment<T>
{
public:
  explicit karim_interferometer(const shared_ptr<T>& _);

private:
  void reflect(const double& r, const double& interferometer_speed, const double& arm_length);
};

template <class T> inline karim_interferometer<T>::karim_interferometer(const shared_ptr<T>& _)
  : numerical_experiment<T>(_)
{
  add_distortion("Spherical to Orthonormal");
  add_distortion("Linear");
  nvector<nvector<double> >& lin(*dynamic_pointer_cast<linear_distortion>(distortions.back())->matrix);
  lin[0][0] = 0.;
  lin[0][3] = 1.;

  reflect(10., 0.2, 3.);
}

template <class T> inline geodesic_interferometer<T>::geodesic_interferometer(const shared_ptr<T>& _)
  : numerical_experiment<T>(_)
{
  add_distortion("Spherical to Orthonormal");
  add_distortion("Linear");
  nvector<nvector<double> >& lin(*dynamic_pointer_cast<linear_distortion>(distortions.back())->matrix);
  lin[0][0] = 0.;
  lin[0][3] = 1.;

  reflect(4., 0.2, 2.);
}

template <class T> inline void geodesic_interferometer<T>::reflect(const double& r, const double& interferometer_speed, const double& arm_length)
{
  const double two_arm_length(2. * arm_length);

  const shared_ptr<atlas::chart>& c(any_chart());

  const nvector<double> coordinate_direction(make_vector(1., 0., 0., interferometer_speed / r));

  const point origin(this->atlas(), c, make_vector(0., r, half_pi, half_pi));
  const tangent_vector tangent(normalise(tangent_vector(origin, c, coordinate_direction)));
  const cached_worldline beam_splitter(coordinate_line(tangent, c));

  cout << endl << "Beam-splitter origin = " << *origin[c] << ", tangent = " << *tangent[c] << "." << endl;
  cout << "Interferometer speed: " << interferometer_speed << endl;
  cout << "Interferometer arm length: " << arm_length << "." << endl;
  output << r << endl;
  output << interferometer_speed << endl;
  output << arm_length << endl;

  const tangent_vector radial_mirror_direction(orthonormalise(tangent_vector(origin, c, make_vector(0., 1., 0., 0.)), tangent));
  const tangent_vector theta_mirror_direction(orthonormalise(orthonormalise(tangent_vector(origin, c, make_vector(0., 0., 1., 0.)), tangent), radial_mirror_direction));
  const tangent_vector phi_mirror_direction(orthonormalise(orthonormalise(orthonormalise(tangent_vector(origin, c, make_vector(0., 0., 0., 1.)), tangent), radial_mirror_direction), theta_mirror_direction));
  const vector<5, tangent_vector> directions(make_vector(radial_mirror_direction, -radial_mirror_direction, theta_mirror_direction, phi_mirror_direction, -phi_mirror_direction));
    
  double max_affine_length(0.);

  for (const tangent_vector* i(directions.begin()); i != directions.end(); ++i)
  {
    const point mirror_origin(*geodesic(*i)(arm_length));
    const tangent_vector mirror_tangent(normalise(tangent_vector(mirror_origin, c, coordinate_direction)));  
    const cached_worldline mirror(coordinate_line(mirror_tangent, c));

    const geodesic outward_ray(connecting_null_geodesic(origin, mirror, arm_length)->second);
    const point reflection(*outward_ray(1.));
      
    const pair<double, geodesic> con(*connecting_null_geodesic(reflection, beam_splitter, two_arm_length));
    const double& affine_length(con.first);
    const geodesic& inward_ray(con.second);

    plot(mirror, 0., affine_length);
    plot(outward_ray, 0., 1.);
    plot(inward_ray, 0., 1.);

    cout << "  Mirror direction = " << *(*i)[c] << ", time experienced by beam splitter = " << affine_length << "." << endl;
    cout << "    Outward ray tangent = " << *(*outward_ray.tangent(0.))[c] << endl;
    cout << "    Inward ray tangent = " << *(*inward_ray.tangent(0.))[c] << endl;      
    output << affine_length << endl;

    if (affine_length > max_affine_length)
      max_affine_length = affine_length;
  }

  plot(beam_splitter, 0., max_affine_length);
}

template <class T> inline void karim_interferometer<T>::reflect(const double& r, const double& interferometer_speed, const double& arm_length)
{
  const double two_arm_length(2. * arm_length);

  const shared_ptr<atlas::chart>& c(any_chart());

  const nvector<double> coordinate_direction(make_vector(1., 0., 0., interferometer_speed / r));

  const point origin(this->atlas(), c, make_vector(0., r, half_pi, half_pi));
  const tangent_vector tangent(normalise(tangent_vector(origin, c, coordinate_direction)));
  const cached_worldline beam_splitter(coordinate_line(tangent, c));

  cout << endl << "Beam-splitter origin: " << *origin[c] << ", tangent: " << *tangent[c] << "." << endl;
  cout << "Interferometer speed: " << interferometer_speed << endl;
  cout << "Interferometer arm length: " << arm_length << "." << endl;
  output << r << endl;
  output << interferometer_speed << endl;
  output << arm_length << endl;

  const tangent_vector radial_mirror_direction(origin, c, make_vector(0., 1., 0., 0.));
  const tangent_vector theta_mirror_direction(origin, c, make_vector(0., 0., 1. / r, 0.));
  const tangent_vector phi_mirror_direction(origin, c, make_vector(0., 0., 0., 1. / r));
  const vector<5, tangent_vector> directions(make_vector(radial_mirror_direction, -radial_mirror_direction, theta_mirror_direction, phi_mirror_direction, -phi_mirror_direction));

  double max_affine_length(0.);

  for (const tangent_vector* i(directions.begin()); i != directions.end(); ++i)
  {
    const point mirror_origin(*coordinate_line(*i, c)(arm_length));
    const tangent_vector mirror_tangent(normalise(tangent_vector(mirror_origin, c, coordinate_direction)));
    const cached_worldline mirror(coordinate_line(mirror_tangent, c));

    const geodesic outward_ray(connecting_null_geodesic(origin, mirror, arm_length)->second);
    const point reflection(*outward_ray(1.));
      
    const pair<double, geodesic> con(*connecting_null_geodesic(reflection, beam_splitter, two_arm_length));
    const double& affine_length(con.first);
    const geodesic& inward_ray(con.second);

    plot(mirror, 0., affine_length);
    plot(outward_ray, 0., 1.);
    plot(inward_ray, 0., 1.);

    cout << "  Mirror direction = " << *(*i)[c] << ", time experienced by beam splitter = " << affine_length << "." << endl;
    cout << "    Outward ray tangent = " << *(*outward_ray.tangent(0.))[c] << endl;
    cout << "    Inward ray tangent = " << *(*inward_ray.tangent(0.))[c] << endl;      
    output << affine_length << endl;

    if (affine_length > max_affine_length)
      max_affine_length = affine_length;
  }

  plot(beam_splitter, 0., max_affine_length);
}
    \end{codeblock} \clearemptydoublepage

\end{document}